\documentclass[aps,prd,twocolumn,preprintnumbers,nofootinbib,superscriptaddress,amsmath]{revtex4-2}

\usepackage{graphicx}
\usepackage{amsmath}
\usepackage{amssymb}
\usepackage{amsfonts}
\usepackage[colorlinks]{hyperref}
\usepackage{url}
\usepackage[dvipsnames]{xcolor}
\usepackage{bm}
\usepackage{array}
\usepackage{placeins}
\usepackage{bbold}
\usepackage{algorithm}
\usepackage{algpseudocode}
\usepackage[normalem]{ulem}
\usepackage{xspace}
\usepackage{orcidlink}
\usepackage{subcaption}
\usepackage[font=small]{caption}
\usepackage{ragged2e}

\definecolor{dodgerblue}{HTML}{1E90FF}
\definecolor{crimson}{HTML}{DC143C}
\definecolor{auwien}{HTML}{8B64DD}

\hypersetup{
    colorlinks=true,
    linkcolor=auwien,
    filecolor=auwien,
    citecolor = auwien,      
    urlcolor=auwien,
}

\newcommand{\ssout}[1]{}

\def\d{\mathrm{d}}
\newcommand{\rmi}{\mathrm{i}}
\newcommand{\rme}{\operatorname{e}}
\newcommand{\pyEFPE}{\texttt{pyEFPE}\xspace}
\newcommand{\STfour}{\texttt{SpinTaylorT4}\xspace}
\newcommand{\Ftwo}{\texttt{TaylorF2}\xspace}
\newcommand{\XP}{\texttt{IMRPhenomXP}\xspace}
\newcommand{\FtwoEcc}{\texttt{TaylorF2Ecc}\xspace}
\allowdisplaybreaks[4]
\newcommand{\sub}[1]{_{\text{#1}}}

\newcommand{\uvec}[1]{\bm{\hat{#1}}}

\newcommand{\ord}[1]{\mathcal{O} \left( #1 \right)}
\newcommand{\D}{\mathcal{D}}
\newcommand{\av}[1]{\left\langle #1 \right\rangle}

\begin{document}
\makeatother

\preprint{IFT-UAM/CSIC-25-12}

\title{Improved post-Newtonian waveform model for inspiralling precessing-eccentric compact binaries}

\author{Gonzalo Morras \orcidlink{0000-0002-9977-8546}}
\email{gonzalo.morras@uam.es}
\affiliation{Instituto de F\'isica Te\'orica UAM/CSIC, Universidad Aut\'onoma de Madrid, Cantoblanco 28049 Madrid, Spain}
\affiliation{School of Physics and Astronomy and Institute for Gravitational Wave Astronomy, University of Birmingham, Edgbaston, Birmingham, B15 2TT, United Kingdom}

\author{Geraint Pratten \orcidlink{0000-0003-4984-0775}}
\email{g.pratten@bham.ac.uk}
\affiliation{School of Physics and Astronomy and Institute for Gravitational Wave Astronomy, University of Birmingham, Edgbaston, Birmingham, B15 2TT, United Kingdom}

\author{Patricia Schmidt \orcidlink{0000-0003-1542-1791}}
\email{p.schmidt@bham.ac.uk}
\affiliation{School of Physics and Astronomy and Institute for Gravitational Wave Astronomy, University of Birmingham, Edgbaston, Birmingham, B15 2TT, United Kingdom}


\date{\today}

\begin{abstract}

The measurement of spin-precession and orbital eccentricity in gravitational-wave (GW) signals is a key priority in GW astronomy, as these effects not only provide unique insights into the astrophysical formation and evolution of compact binaries but also, if neglected in waveform models, could introduce significant biases in parameter estimation, searches, and tests of General Relativity. Despite the growing potential of upcoming LIGO-Virgo-KAGRA observing runs and future detectors to measure eccentric-precessing signals, accurately and efficiently modeling them remains a significant challenge.
In this work, we present \pyEFPE, a frequency-domain post-Newtonian (PN) waveform model for the inspiral of precessing-eccentric compact binaries.
\pyEFPE improves upon previous models by introducing analytical expressions for the Fourier mode amplitudes, enhancing the numerical stability of the multiple scale analysis framework, and adding recently derived PN corrections, critical to accurately describe signals in GW detectors. Additionally, we simplify the numerical implementation and introduce a scheme to interpolate the polarization amplitudes, achieving a speedup of up to $\sim \ord{15}$ in the waveform computations, making the model practical for data analysis applications.
We thoroughly validate \pyEFPE by comparing it to other waveform models in the quasi-circular and eccentric-spin-aligned limits, finding good agreement. Additionally, we demonstrate \pyEFPE's capability to analyze simulated GW events, accurately recovering the parameters of signals described by both \pyEFPE and \XP.
While \pyEFPE still lacks important physical effects, such as higher-order PN corrections, higher-order modes, mode asymmetries, tidal interactions or the merger-ringdown phase, it represents a significant step towards more complete waveform models, offering a flexible and efficient framework that can be extended in future work to incorporate these effects.

\end{abstract}
\maketitle

\section{Introduction}
\label{sec:intro}

One of the most pressing questions in modern relativistic astrophysics is to understand the formation and evolutionary processes of stellar-mass black holes.
Even though the LIGO~\cite{LIGOScientific:2014pky}, Virgo~\cite{VIRGO:2014yos}, and KAGRA~\cite{KAGRA:2018plz} detectors are making hundreds of gravitational-wave (GW) detections~\cite{LIGOScientific:2018mvr,LIGOScientific:2020ibl,KAGRA:2021vkt}, there are still numerous open questions on the interpretation of the underlying astrophysical population~\cite{Mandel:2018hfr,Mapelli:2021hdk,Gerosa:2021mno,Mandel:2021smh,KAGRA:2021duu}. 
Among all the observables, spin-precession~\cite{Rodriguez:2016vmx,Gerosa:2018wbw,Banerjee:2023ycw} and eccentricity~\cite{Naoz:2016klo,Samsing:2017xmd,Tagawa:2020jnc,Zevin:2021rtf} are two of the most exciting as they are thought to be comparatively clean tracers for the underlying astrophysical formation channel. 

Whilst the majority of current observations are consistent with quasi-circular (non-eccentric) binaries, there is tentative evidence that some binary black hole events may have non-zero eccentricity~\cite{Gayathri:2020coq,Gamba:2021gap,Romero-Shaw:2022xko,Gupte:2024jfe}, based on analyses using aligned-spin eccentric models.
However, it has recently been noted that the complex interplay between spin precession and residual eccentricity can introduce systematic biases in astrophysical inference~\cite{Fumagalli:2023hde,Fumagalli:2024gko}, see also~\cite{Romero-Shaw:2022fbf}, emphasizing the need for waveform models that account for both effects.
Neglecting either spin precession~\cite{Chatziioannou:2014bma,Pratten:2020igi} or eccentricity~\cite{Favata:2021vhw} can lead to biases in the inferred parameters of GW signals. In particular, ignoring eccentricity has been shown to induce significant errors in the estimation of the chirp mass due to their intrinsic degeneracy.
Developing waveform models that incorporate both effects is therefore essential, not only to improve astrophysical inference, but also to address challenges in incorporating additional physics into GW searches~\cite{McIsaac:2023ijd,Phukon:2024amh,Gadre:2024ndy,Schmidt:2024jbp} and to conduct precision tests of General Relativity~\cite{Saini:2022igm,Narayan:2023vhm,Shaikh:2024wyn,Bhat:2024hyb}.

Furthermore, one of the key science goals for LISA is to trace the origin and evolution of massive black holes~\cite{Sesana:2021jfh,LISA:2022yao,LISA:2024hlh} with mass $M \sim 10^4 - 10^9 M_{\odot}$. 
Whilst the population properties of these binaries are highly uncertain, accurate constraints on their masses, spins, and orbital eccentricity will provide crucial insight into black hole seed formation scenarios, galaxy assembly, and the evolution of large scale structure. 
For example, spin precession could be coupled to the properties of the host galaxy, with binaries in gas-rich environments potentially undergoing a secular alignment of the black hole spin via the Bardeen-Petterson effect~\cite{Bardeen:1975bpe,Bogdanovic:2007hp}, itself driven by Lense-Thirring precession~\cite{Lense:1918zz}.
Similarly, orbital eccentricity is thought to be sensitive to the surrounding environment, with several mechanisms being able to generate non-negligible eccentricity through binary-disk interactions~\cite{Armitage:2005xq,Macfadyen:2006jx,Dotti:2008za,Hayasaki:2008eu,Cuadra:2008xn,Haiman:2009pvg,Zrake:2020zkw}, 

It should therefore be apparent that the rapid development of waveform models that accurately capture both spin-precession and eccentricity is one of the key challenges in modern GW astronomy. 
In the past few years, there has been significant progress in constructing aligned-spin eccentric waveform models. This includes recent progress using post-Newtonian (PN) calculations~\cite{Boetzel:2017zza,Moore:2019xkm,Boetzel:2019nfw,Cho:2021oai,Paul:2022xfy,Henry:2023tka,Sridhar:2024zms}, the effective-one-body (EOB) formalism~\cite{Chiaramello:2020ehz,Nagar:2021gss,Khalil:2021txt,Ramos-Buades:2021adz,Albanesi:2022xge,Albanesi:2022ywx,Placidi:2021rkh,Nagar:2024oyk,Nagar:2024dzj,Gamboa:2024imd,Gamboa:2024hli}, and NR informed models~\cite{Islam:2021mha,Paul:2024ujx}.
However, only limited progress has been made 
in the joint modeling of precession and eccentricity~\cite{Csizmadia:2012wy,Ireland:2019tao,vandeMeent:2017bcc,Klein:2010ti,Klein:2018ybm,Klein:2021jtd,Arredondo:2024nsl,Liu:2023ldr,Gamba:2024cvy}. 

In this work, we build upon a series of studies that, over the past few years, have established the foundations for constructing Efficient Fully Precessing Eccentric (EFPE) waveforms~\cite{Klein:2018ybm,Klein:2021jtd,Arredondo:2024nsl}. 
Ref.~\cite{Klein:2018ybm} developed much of the formalism for these frequency-domain post-Newtonian waveform models for inspiralling precessing-eccentric compact binaries. Building on this, Ref.~\cite{Klein:2021jtd} introduced a more efficient method to model spin-precession dynamics using a multiple scale analysis (MSA) approach~\cite{Bender:1999msa,Klein:2013qda,Chatziioannou:2013dza,Kesden:2014sla,Gerosa:2015tea,Chatziioannou:2017tdw,Gerosa:2023xsx}. Most recently, Ref.~\cite{Arredondo:2024nsl} formalized the description of the eccentricity correction to the waveform amplitudes, allowing to extend the domain of validity of EFPE waveforms to moderate eccentricities $(e \lesssim 0.8)$. With respect to these works, in \pyEFPE we introduce several key improvements
\begin{enumerate}
    \item In Sec.~\ref{sec:NewtAmp} we derive closed analytical expressions for the Newtonian Fourier mode amplitudes, greatly simplifying the amplitude computation with respect to Ref.~\cite{Arredondo:2024nsl}, while improving the accuracy and speed.
    \item In Sec.~\ref{sec:MSA} we derive self-consistent expressions for the MSA introduced in Ref.~\cite{Klein:2021jtd} and improve its numerical stability, making the \pyEFPE waveform more stable and less prone to failures.
    \item In Sec.~\ref{sec:RR} and in appendix~\ref{sec:appendix:PN_formulas}, we revisit the PN equations, fixing typos and incorporating the recently derived 2.5PN and 3PN aligned-spin eccentric contributions from Ref.~\cite{Henry:2023tka} into the evolution equations. 
    \item In Sec.~\ref{sec:waveform} we provide a detailed description of the numerical implementation of the \pyEFPE waveform, demonstrating how to use the Runge-Kutta method to solve the equations of motion and derive frequency-domain waveforms in a simpler and more efficient manner than in Ref.~\cite{Klein:2018ybm}. We also describe how to interpolate the amplitudes to speed up the waveform by a factor up to $\sim \ord{15}$ at a minimal cost of accuracy.
    \item The \pyEFPE waveform is publicly available in~\cite{pyEFPE_repo}, making it the first EFPE model widely accessible to the community. Furthermore, its \texttt{Python} implementation is designed to be easy to understand, use, and modify.
\end{enumerate}

As for the rest of the paper, in Sec.~\ref{sec:FourierTransform}, we explain how \pyEFPE uses the SUA approximation~\cite{Klein:2014bua} to analytically compute the waveform in the frequency domain, fixing a typo in Refs.~\cite{Klein:2018ybm,Klein:2021jtd,Arredondo:2024nsl}. In Sec.~\ref{sec:validate}, we thoroughly test and validate \pyEFPE, verifying that it reproduces the expected phenomenology of inspiral waveforms with spin-precession and orbital eccentricity, performing mismatch comparisons with other waveform models in the quasi-circular and eccentric-spin-aligned limits, and demonstrating its capability to estimate the parameters of simulated GW signals. Finally, in Sec.~\ref{sec:conclusion} we conclude, summarizing our findings and outlining potential directions for future work.

Unless otherwise specified, in this paper we use geometric units ($G=c=1$), and write vectors in boldface, adding a hat in the case of unit vectors. Additionally, angular momenta are expressed as dimensionless quantities by scaling them with the total mass squared, $M^2$, such that, for example $\bm{L} = \bm{L}^\mathrm{physical}/M^2$ or $\bm{S}_i = \bm{S}_i^\mathrm{physical}/M^2$.

\section{Gravitational waves from precessing eccentric binaries}
\label{sec:NewtAmp}

In this section, we derive the GW emission of a precessing eccentric binary, assuming the system's dynamics are known. 
While these dynamics are thoroughly studied in later sections, starting with the GW emission provides a broad overview of the problem and helps identify the parts of the system requiring modeling. Specifically, we focus on the leading-order (Newtonian) quadrupolar GW emission, e.g.~\cite{Mishra:2015bqa}, incorporating precession effects in the waveform by performing a time-dependent rotation of the waveform modes~\cite{Schmidt:2010it,OShaughnessy:2011pmr,Boyle:2011gg,Schmidt:2012}.

\subsection{Newtonian eccentric orbit}
\label{sec:NewtAmp:orbit}

At Newtonian order, the orbit of an eccentric binary in the center-of-mass frame can be described with the Keplerian parametrization~\cite{Damour:1985ecc,Colwell:1993book}

\begin{subequations}
\label{eq:NewtonianOrbit}
\begin{align}
    r(u) & = a (1 - e \cos{u}) \, ,\label{eq:NewtonianOrbit:r} \\
    \phi_N & = v(u) \equiv 2 \arctan\left[\left(\frac{1+e}{1-e}\right)^{1/2} \tan{\frac{u}{2}}\right] \, , \label{eq:NewtonianOrbit:v} \\
    \ell & \equiv n (t - t_0) = u - e \sin{u} \, , \label{eq:NewtonianOrbit:l}
\end{align}
\end{subequations}

\noindent where the relative separation vector is given by $\bm{x} = r(\cos{\phi_N},\sin{\phi_N},0)$, $a$ is the semi-major axis, $e$ the eccentricity, $n = 2 \pi/P$ the mean motion, where $P$ is the orbital period, and $t_0$ is a constant of integration; the auxiliary variables $u$, $v$ and $\ell$ are the eccentric, true and mean anomalies, respectively. 

When PN effects are taken into account, the orbit will no longer be described by Eq.~\eqref{eq:NewtonianOrbit}, and small perturbations appear. However, the orbit can be written in a similar way, using the ``quasi-Keplerian'' parametrization~\cite{Damour:1985ecc,Damour:1988mr,Schaefer:1993qk,Wex:1995qk,Memmesheimer:2004cv}. Here, we will ignore small periodic corrections to the orbit and only keep the secular effects. In particular, the equation for the orbital phase $\phi$ will be modified by the periastron advance $k$~\cite{Damour:2004bz,Arun:2007rg} to read: 

\begin{align}
    \phi & = (1+k)v = (1+k) \ell + (1+k)(v-\ell) \nonumber \\
    & \approx \underbrace{(1+k) \ell}_{\lambda} + \underbrace{v-\ell}_{W} \, .
    \label{eq:phi_lW}
\end{align}
In the first term, the periastron advance $k$ cannot be neglected, even though it is formally of 1PN order, as $\ell$ becomes very large for long observation times and $k \ell$ induces a significant change in the orbital phase. 
In the second term, the contribution of $k$ to $\phi$ is subdominant since $|v-\ell|\leq \pi$ and therefore $k(v-\ell)$ is a small periodic phase change of order 1PN. 

The periastron advance can be seen explicitly by looking at the argument of periastron $\delta \lambda = \phi -v$, which tracks the phase of the periastron of the elliptic orbit, and can be computed as

\begin{equation}
    \delta\lambda = \lambda - \ell = k \ell \, ,
    \label{eq:dlambda_def}
\end{equation}

\noindent which grows linearly in time and is directly proportional to the periastron advance $k$.

\subsection{Waveform of a spin-precessing binary}
\label{sec:NewtAmp:Waveform}

To describe the waveform of a spin-precessing binary we follow the descriptions and conventions of Ref.~\cite{Arredondo:2024nsl}. 
The GW polarizations, $h_{+,\times}$, can be decomposed in terms of spin-weighted spherical harmonics~\cite{Thorne:1980ru,Kidder:2007}, i.e.,

\begin{equation}
  h_+ - i h_\times = \sum_{l=2}^\infty \sum_{m=-l}^l h^{lm} ~{}_{-2} Y^{lm}(\Theta\sub{p}, \Phi\sub{p}) \, ,
  \label{eq:hlm_def}
\end{equation}

\noindent where $(\Theta\sub{p}, \Phi\sub{p})$ are the spherical angles of the wave propagation vector as measured in the inertial binary source frame, and ${}_{-2} Y^{lm}$ are the spin-weighted spherical harmonics of spin weight $-2$. 

When the spins of the component objects $\bm{S}_i$ are not aligned with the orbital angular momentum $\bm{L}$, the system undergoes spin induced precession~\cite{Barker:1975ae, Apostolatos:1994pre} in which $\bm{L}$, and therefore the orbital plane, precesses around the total angular momentum $\bm{J} = \bm{L} + \bm{S}_1 + \bm{S}_2$. This spin-induced orbital precession greatly complicates the structure of the GW modes $h_{l m}$ of Eq.~\eqref{eq:hlm_def}~\cite{Apostolatos:1994pre}. However, the modes can be mostly simplified by transforming from the inertial to a ``co-precessing'' frame that is instantaneously aligned with the orbital angular momentum and then, using the dynamics, rotating the co-precessing frame modes $H_{l m}$ back to the inertial frame in which the $h^{lm}$ are defined (Eq.~\eqref{eq:hlm_def})~\cite{Schmidt:2010it,OShaughnessy:2011pmr,Boyle:2011gg,Schmidt:2012}. 
Note that in the co-precessing frame, precession effects are reduced but not entirely eliminated~\cite{Schmidt:2012,Boyle:2014ioa,Ramos-Buades:2020noq}. 
The rotation of the modes is given by \cite{Schmidt:2010it}

\begin{equation}
  h^{l m'} = \sum_{m=-l}^l D^l_{m' m}(\phi_z,\theta_L,\zeta) H^{l m} \, ,
  \label{eq:twist-up}
\end{equation}

\noindent where $D^l_{m'm}(\phi_z,\theta_L,\zeta)$ are the Wigner $D$-matrices and $\phi_z$, $\theta_L$ and $\zeta$ are the three Euler angles that describe the rotation from the co-precessing to the inertial frame. In particular, $\theta_L$ is the angle between $\uvec{L}$ and $\uvec{J}$, $\phi_z$ is the angle of the projection of $\bm{L}$ onto the plane perpendicular to $\uvec{J}$ and $\zeta$ is the third Euler angle, fixed by the minimal-rotation condition $\dot{\zeta} = - \dot{\phi}_z \cos{\theta_L}$~\cite{Buonanno:2002fy,Boyle:2011gg}.

Therefore, the problem has now simplified to describing the GW modes $H^{lm}$ of an approximately non-precessing system, which will be functions of the eccentric orbit. Using the quasi-Keplerian parametrization (Eqs.~(\ref{eq:NewtonianOrbit},\ref{eq:phi_lW})), these modes can be expressed as~\cite{Arredondo:2024nsl}

\begin{align}
    H^{l m}(t) &= h_0 \rme^{-\rmi m\phi(t)} K^{l m}[u(t)] \nonumber\\
    &\equiv h_0 \hat{H}^{lm}(t) \, ,
    \label{eq:Hlm_def}
\end{align}

\noindent where 
\begin{equation}
    h_0 \equiv 4\sqrt{\frac{\pi}{5}} \frac{M\nu}{d_L} (M \omega)^{2/3} \, ,
\end{equation}
\noindent with $d_L$ being the luminosity distance to the binary, $M = m_1 + m_2$ the total mass, $\omega$ the mean orbital angular velocity, and $\nu = m_1 m_2/M^2$ the symmetric mass ratio.
Since $H^{l m}$ are the GW modes in the co-precessing frame, ignoring small mode asymmetries~\cite{Boyle:2014ioa,Ramos-Buades:2020noq}, they satisfy
\begin{align}
    H^{l -m} = (-1)^l (H^{l m})^{*}  \, .
    \label{eq:Hl-m_from_Hlm}
\end{align}
\noindent Putting together Eqs.~(\ref{eq:hlm_def},\ref{eq:Hlm_def}), the waveform polarizations in the inertial frame are given by
\begin{equation}
    h_+ - \rmi h_\times = \sum_{l=2}^\infty \sum_{m=-l}^l A_{l,m} \hat{H}^{lm} \, ,
    \label{eq:hpmihc}
\end{equation}
\noindent where we have defined
\begin{equation}
    A_{l,m}(t) \equiv h_0 \sum_{m'=-l}^l {}_{-2}Y^{l m'}(\Theta, \Phi)  D^l_{m'm}(\phi_z,\theta_L,\zeta) \, ,
    \label{eq:Almmp}
\end{equation}

\noindent and $(\Theta, \Phi)$ are the spherical angles of the binary as measured in the co-precessing frame. Note that, given the rotation formula for spherical harmonics, $A_{l,m}(t)$ is proportional to the inertial frame spherical harmonics. Since $h_+$ and $h_\times$ are real valued, using Eq.~\eqref{eq:hpmihc} and its complex conjugate, together with the mode symmetry of Eq.~\eqref{eq:Hl-m_from_Hlm} we can separate the two polarizations as
\begin{equation}
  h_{+,\times}(t) = \sum_{l=2}^\infty \sum_{m=-l}^l \mathsf{A}^{+,\times}_{l,m} \hat{H}^{lm} \, ,
  \label{eq:hpc_AH}
\end{equation}
\noindent where we have defined 
\begin{subequations}
\begin{align}
  & \mathsf{A}^+_{l,m} = \frac{1}{2} \left[ A_{l,m} + (-1)^l (A_{l,-m})^* \right] \, , \\
  &\mathsf{A}^\times_{l,m} = \frac{\rmi}{2} \left[ A_{l,m} - (-1)^l (A_{l,-m})^* \right] \, .
\end{align}
\label{eq:Apc_lm_def}
\end{subequations}
\noindent Note that the $\mathsf{A}^{+,\times}_{l,m}$ amplitudes also satisfy a mode symmetry similar to Eq.~\eqref{eq:Hl-m_from_Hlm}, i.e.

\begin{equation}
    \mathsf{A}^{+,\times}_{l,-m} = (-1)^l (\mathsf{A}^{+,\times}_{l,m})^{*}
    \label{eq:Apc_lm_mode_symtry}
\end{equation}
Using $(D^l_{m'-m})^{*} = (-1)^{m'+m}D^l_{-m'm}$, we can simplify Eq.~\eqref{eq:Apc_lm_def} as

\begin{subequations}
\begin{align}
    & \mathsf{A}^{+,\times}_{l,m} = h_0 \sum_{m'=-l}^{l} \mathsf{P}^{+,\times}_{l,m,m'}(\Theta, \Phi)  D^l_{m'm}(\phi_z,\theta_L,\zeta) \, , \label{eq:Apc_lm_simp:def} \\
    & \mathsf{P}^{+}_{l,m,m'} = \frac{1}{2}\left[{}_{-2}Y^{l m'} + (-1)^{l + m + m'} ({}_{-2}Y^{l -m'})^{*}\right] , \label{eq:Apc_lm_simp:Pp} \\
    & \mathsf{P}^{\times}_{l,m,m'} = \frac{\rmi}{2}\left[ {}_{-2}Y^{l m'} - (-1)^{l + m + m'} ({}_{-2}Y^{l -m'})^{*}\right] , \label{eq:Apc_lm_simp:Pc}
\end{align}
\label{eq:Apc_lm_simp}
\end{subequations}

\noindent where the $ \mathsf{P}^{+,\times}_{l,m,m'}(\Theta, \Phi)$ terms remain constant throughout the evolution and therefore need to only be computed once at initialization.

To compute the GW polarizations as a function of time we need to evaluate the co-precessing GW modes $\hat{H}^{l m}(\ell, u(\ell))$. Naively, this requires numerically solving the transcendental Eq.~\eqref{eq:NewtonianOrbit:l} to obtain the eccentric anomaly $u$ as a function of the mean anomaly $\ell$. However, this can be avoided by writing the GW modes as a Fourier series in the mean anomaly, which will also prove useful when transforming the GW signal to the frequency domain. Therefore, we follow \cite{Arredondo:2024nsl} and write

\begin{equation}
    \hat{H}^{l m} = \rme^{-\rmi m \lambda} \sum_{p=-\infty}^\infty N_p^{l m} \rme^{-\rmi p \ell} \, ,
    \label{eq:Hlm_Nlmp}
\end{equation}

\noindent where we have separated the mean orbital phase term $\rme^{-\rmi m \lambda}$, since it is not $2 \pi$-periodic in $\ell$~\cite{Damour:2004bz}, and $N_p^{l m}$ are the Fourier series coefficients, which are defined by

\begin{equation}
    N_p^{l m} = \frac{1}{2\pi}\int_{-\pi}^\pi \left(\rme^{\rmi m \lambda} \hat{H}^{l m}\right) \rme^{\rmi p \ell} \d \ell\, .
    \label{eq:Nlm_p}
\end{equation}
From the mode symmetry of Eq.~\eqref{eq:Hl-m_from_Hlm} and Eq.~\eqref{eq:Hlm_Nlmp}, we can deduce that 
\begin{align}
    N^{l -m}_p = (-1)^l (N^{l m}_{-p})^{*}
    \label{eq:Nl-m_from_Nlm}.
\end{align}
Substituting Eq.~\eqref{eq:Hlm_Nlmp} in Eq.~\eqref{eq:hpc_AH}, and writing the eccentric anomaly in terms of the argument of periastron $\delta\lambda = \lambda - \ell$, we can finally express the GW waveform as~\cite{Arredondo:2024nsl}
\begin{equation}
  h_{+,\times}(t) = \sum_{l=2}^\infty \sum_{m=-l}^l \sum_{n=-\infty}^\infty \mathcal{A}^{+,\times}_{l,m,n}(t) \rme^{-\rmi(n\lambda + (m-n)\delta\lambda)} \, ,
  \label{eq:hpc_decomp}
\end{equation}
\noindent where 
\begin{equation}
  \mathcal{A}^{+,\times}_{l,m,n}(t) = N^{lm}_{n-m}(t) \mathsf{A}^{+,\times}_{l,m}(t)  \, .
  \label{eq:Almn}
\end{equation}
\noindent In the argument of the exponential of Eq.~\eqref{eq:hpc_decomp}, we have explicitly separated the contributions of the rapidly evolving mean orbital phase $\lambda$ from the slowly evolving argument of periastron $\delta\lambda$. Finally, the spin-precession effects are captured by the slow time variation of the coefficients $\mathcal{A}^{+,\times}_{l,m,n}(t)$.

\subsection{Newtonian Fourier Mode Amplitudes}
\label{sec:NewtAmp:Np}

In this subsection, and in \pyEFPE, we consider the amplitudes to leading order in the post-Newtonian (PN) expansion, also called the Newtonian order. 
The only GW modes that contribute at this Newtonian order are the ones with $l=2$, $m=\{0, \pm 2\} $, given by~\cite{Mishra:2015bqa,Arredondo:2024nsl}

\begin{subequations}
\label{eq:H2m_Newtonian}
\begin{align}
    \hat{H}^{20} &= \sqrt{\frac{2}{3}} \frac{e\cos u}{1-e\cos u} \, , \label{eq:H2m_Newtonian:H20} \\
    \hat{H}^{22} &= \frac{2\rme^{-2\rmi\phi}}{1-e\cos u} \! \left( \!\frac{1-e^2+\rmi e\sqrt{1-e^2} \sin u}{1- e \cos u} - \frac{e}{2}\cos u \! \right) \! , \label{eq:H2m_Newtonian:H22}
\end{align}
\end{subequations}

\noindent where $\hat{H}^{2 -2}$ can be obtained from Eq.~\eqref{eq:H2m_Newtonian:H22} by using the property of Eq.~\eqref{eq:Hl-m_from_Hlm}. For a generic function of $u$, we can write its Fourier transform with respect to $\ell$ (Eq.~\eqref{eq:Nlm_p}) as the following integral over $u$

\begin{align}
    \frac{1}{2 \pi} \int_{-\pi}^\pi & f(u(\ell)) \rme^{\rmi p \ell} \d\ell = \frac{1}{2 \pi} \int_{-\pi}^\pi f(u) \rme^{\rmi p \ell(u)} \frac{\d\ell}{\d u} \d u \nonumber \\
    & = \frac{1}{2 \pi} \int_{-\pi}^\pi (1 -e \cos{u}) f(u) \rme^{\rmi p (u - e \sin{u})} \d u.
    \label{eq:fu_furier}
\end{align}

The coefficients $N^{20}_p$ can be computed by substituting the corresponding mode (Eq.~\eqref{eq:H2m_Newtonian:H20}) in Eq.~\eqref{eq:fu_furier}

\begin{align}
    N_p^{2 0} & =  \frac{1}{2 \pi} \int_{-\pi}^\pi \frac{e}{\sqrt{6}} \left(\rme^{i u} + \rme^{-i u} \right) \rme^{\rmi p (u - e \sin{u})} \d u  \nonumber \\
    & = \frac{e}{\sqrt{6}} (J_{p +1}(p e) + J_{p-1}(p e)) \, ,
    \label{eq:N20_p_ini}
\end{align}

\noindent where $J_n(z)$ is the Bessel function of integer order $n$~\cite{Abramowitz_and_Stegun},  defined as

\begin{equation}
    J_n(z) = \frac{1}{2 \pi} \int_{-\pi}^\pi \rme^{\rmi ( n u - z \sin{u})} \d u  \quad (n \in \mathbb{Z}) \, .
    \label{eq:BesselJ_def}
\end{equation}
In the $p=0$ case, we note that the integral of Eq.~\eqref{eq:N20_p_ini} could have been trivially computed to yield $N^{20}_0 = 0$, while in the case $p\neq 0$ we use the recurrence relations of the Bessel functions~\cite{Abramowitz_and_Stegun} to write:

\begin{align}
    N_p^{2 0} = 
    \begin{cases}
        0 &, \; p = 0\\
        \sqrt{\frac{2}{3}} J_p(p e) &, \; p \neq 0
     \end{cases}
    \, ,
    \label{eq:N20_p}
\end{align}

\noindent in agreement with the expression found in~\cite{Arredondo:2024nsl}. 

Computing $N^{22}_p$ is more difficult, since substituting Eq.~\eqref{eq:H2m_Newtonian:H20} in Eq.~\eqref{eq:fu_furier} leads to an integral that seems a-priori very complicated. To simplify this integral we begin by using Eq.~\eqref{eq:NewtonianOrbit:v} for $v(u)$ together with basic trigonometric relations to write
\begin{equation}
    \rme^{-\rmi v(u)} = \frac{\cos{u} - e - \rmi \sqrt{1-e^2}\sin{u}}{1-e\cos{u}} \, .
    \label{eq:expiv}
\end{equation}

In Eq.~\eqref{eq:H2m_Newtonian:H22} for $\hat{H}^{2 2}$, we use Eq.~\eqref{eq:phi_lW} to expand $\phi = \lambda + v - \ell$ together with Eq.~\eqref{eq:expiv} to write
\begin{align}
    \hat{H}^{22} = &\frac{2\rme^{-2\rmi\lambda} \rme^{2\rmi\ell}}{1-e\cos u} \left( \frac{\cos{u} - e - \rmi \sqrt{1-e^2}\sin{u}}{1-e\cos{u}} \right)^2\nonumber\\
    &\times \left(\frac{1-e^2+\rmi e\sqrt{1-e^2} \sin u}{1- e \cos u} - \frac{e}{2}\cos u \right) \, .
    \label{eq:H22_simplified}    
\end{align}

\noindent Note that we can always write Eq.~\eqref{eq:H22_simplified} as

\begin{align}
    \hat{H}^{22} = &\rme^{-2\rmi\lambda} \rme^{2\rmi\ell} \frac{\d F^{22}}{\d\ell} \, ,
    \label{eq:H22_as_der}    
\end{align}

\noindent where the derivative with respect to $\ell$ is computed as
\begin{equation}
    \frac{\d}{\d\ell} = \frac{\d u}{\d \ell} \frac{\d}{\d u} = \frac{1}{1-e\cos{u}} \frac{\d}{\d u} \, .
    \label{eq:dl_du}
\end{equation}

\noindent The expression of $F^{22}$ can be computed by integration of Eq.~\eqref{eq:H22_as_der}, where we find

\begin{align}
    F^{22}(u) = & \frac{\rmi}{1-e\cos{u}} \Big(-\sqrt{1-e^2} + \rmi e \sin{u}  \nonumber \\
    & +\sqrt{1-e^2}\cos{2 u} - \rmi \left(1 - \frac{e^2}{2}\right)\sin{2 u} \Big) \, .
    \label{eq:F22_def}    
\end{align}

\noindent Substituting Eq.~\eqref{eq:H22_as_der} in Eq.~\eqref{eq:Nlm_p}, we have

\begin{align}
     N_p^{2 2} & = \frac{1}{2\pi}\int_{-\pi}^\pi \frac{\d F^{22}}{\d\ell} \rme^{\rmi (p+2) \ell} \d \ell \nonumber \\
     & = \frac{-\rmi (p+2)}{2\pi} \int_{-\pi}^\pi  F^{22} \rme^{\rmi (p+2) \ell} \d \ell \, ,
    \label{eq:N22_p_ez}    
\end{align}

\noindent where we have used the formula for the Fourier series of the derivative of a function. If we use the same trick as in Eq.~\eqref{eq:fu_furier} to write the integral in terms of $u$, the combination $(1 - e\cos{u})F^{22}(u)$ appears in the integrand. As can be seen in Eq.~\eqref{eq:F22_def}, this is a finite series of trigonometric functions. Therefore, the integral in Eq.~\eqref{eq:N22_p_ez} can be computed in terms of Bessel functions in the same way as was done in Eq.~\eqref{eq:N20_p}. Finally, the Fourier coefficients $N^{2 2}_p$ are given by:

\begin{align}
    N^{2 2}_{j - 2} =  j &\Bigg(\!-\sqrt{1-e^2} J_j(j e) + \frac{e}{2} (J_{j+1}(j e)- J_{j-1}(j e))  \nonumber \\
    &+ \frac{1}{2} \left(\sqrt{1-e^2} + \left(1 -\frac{e^2}{2}\right)\right) J_{j-2}(j e) \nonumber \\
    &+ \frac{1}{2} \left(\sqrt{1-e^2} - \left(1 -\frac{e^2}{2}\right)\right) J_{j+2}(j e) \Bigg) \, .
    \label{eq:N22_j}    
\end{align}
We note that Eq.~\eqref{eq:N22_j} provides a straightforward closed-form analytical expression for the $l=m=2$ Newtonian Fourier mode amplitudes. 
This stands in stark contrast to the formulation in Ref.\cite{Arredondo:2024nsl}, where the same quantity is expressed as intricate infinite nested sums. In real applications, these sums must be truncated, leading to less accurate results.
In App.~\ref{sec:appendix:mode_amp_comparison}, we demonstrate the level of agreement between the exact expressions in Eq.~\eqref{eq:N22_j} and the nested sums in~\cite{Arredondo:2024nsl}, finding good agreement.
The Bessel functions, which appear in the Fourier mode amplitudes of Eqs.~(\ref{eq:N20_p}, \ref{eq:N22_j}), have very efficient implementations~\cite{BesselFunctionsCode}, available in widely used software packages~\cite{Scipy:2020}. Consequently, in the form presented in this section, the Newtonian Fourier mode amplitudes are simple and fast to compute exactly.

\subsection{Fourier modes to be included}
\label{sec:NewtAmp:Number}

In order to make the waveform model as efficient as possible, in Eq.~\eqref{eq:hpc_decomp} we will want to include as few Fourier modes as possible to describe the strain with a given tolerance.
To do this, we start by using the orthogonality relations of the tensor spherical harmonics ${}_{-2}Y^{lm}(\Theta, \Phi)$ and the Wigner $D$-matrices $D^l_{m' m}(\phi_z,\theta_L,\zeta)$, i.e.

\begin{subequations}
\label{eq:Orthogonality}
\begin{align}
   \int \frac{\d \Omega}{4 \pi} {}_{-2}Y^{l_1 m_1} ({}_{-2}Y^{l_2 m_2})^{*} & = \delta_{l_1 l_2} \delta_{m_1 m_2}  \, ,\label{eq:Orthogonality:Y} \\
   \int \frac{\d \Omega}{4 \pi} \int_0^{2\pi} \frac{\d\zeta}{2\pi} D^{l_1}_{m'_1 m_1} (D^{l_2}_{m'_2 m_2})^{*} & = \frac{\delta_{l_1 l_2} \delta_{m_1' m_2'} \delta_{m_1 m_2}}{2 l+1} \, 
\end{align}
\end{subequations}

\noindent together with the GW waveform of Eq.~\eqref{eq:hpmihc}, to prove that
\begin{align}
    \av{|h_+|^2 + |h_\times|^2} & = \sum_{l=2}^\infty \sum_{m=-l}^l |H^{l m}|^2 \nonumber \\
    & = \sum_{l=2}^\infty \left( |H^{l 0}|^2 + 2\sum_{m=1}^l |H^{l m}|^2 \right) \, ,
    \label{eq:expecth2}    
\end{align}

\noindent where we have used the mode symmetry of Eq.~\eqref{eq:Hl-m_from_Hlm}. To find out how many Fourier modes have to be included, we compute the average value of $\av{ |h_+|^2 + |h_\times|^2}$ over one orbital cycle, i.e.

\begin{align}
    \Vert h \Vert^2 &= \int_{-\pi}^{\pi} \frac{\d\ell}{2 \pi} \av{|h_+|^2 + |h_\times|^2} \nonumber \\
    & = \sum_{l=2}^\infty \left( \Vert H^{l 0}\Vert^2 + 2\sum_{m=1}^l \Vert H^{l m}\Vert^2 \right) \, ,
    \label{eq:normh}
\end{align}

\noindent where we have defined
\begin{align}
    \Vert \hat{H}^{l m}\Vert^2 & = \!\! \int_{-\pi}^{\pi} \! \frac{\d \ell}{2 \pi} |H^{l m} (\ell)|^2 = \sum_{p=-\infty}^\infty |N^{l m}_p|^2 \, ,
    \label{eq:normHlm}
\end{align}

\noindent and in the second equality of Eq.~\eqref{eq:normHlm} we have substituted the Fourier series of $\hat{H}^{l m}$, defined in Eq.~\eqref{eq:Hlm_Nlmp}. In real applications, we include a finite number of terms in the Fourier series of $\hat{H}^{l m}$, inducing an error in the strain which we will want to keep under control. 

Restricting ourselves to the Newtonian amplitudes, we can compute the norms of $\Vert \hat{H}^{2 0} \Vert$ and $\Vert \hat{H}^{2 2}\Vert$ in closed form using Eq.~\eqref{eq:H2m_Newtonian}, together with Eq.~\eqref{eq:NewtonianOrbit:l}

\begin{subequations}
\label{eq:normH2m_Newtonian}
\begin{align}
    \Vert\hat{H}^{20}\Vert^2 &= \frac{2}{3}\int_{-\pi}^{\pi} \frac{\d u}{2\pi} \frac{e^2\cos^2 u}{1-e\cos u} = \frac{2}{3} \left(\frac{1}{\sqrt{1-e^2}} - 1\right) \, , \label{eq:normH2m_Newtonian:H20} \\
    \Vert\hat{H}^{20}\Vert^2 &= 4 \int_{-\pi}^{\pi} \frac{\d u}{2 \pi}  \frac{\left|\frac{1-e^2+\rmi e\sqrt{1-e^2} \sin u}{1- e \cos u} - \frac{e}{2}\cos u \right|^2}{1-e\cos u}  \nonumber\\
    & = \frac{5}{\sqrt{1-e^2}} - 1\, . \label{eq:normH2m_Newtonian:H22}
\end{align}
\end{subequations}

\begin{figure}[t!]
\centering  
\includegraphics[width=0.5\textwidth]{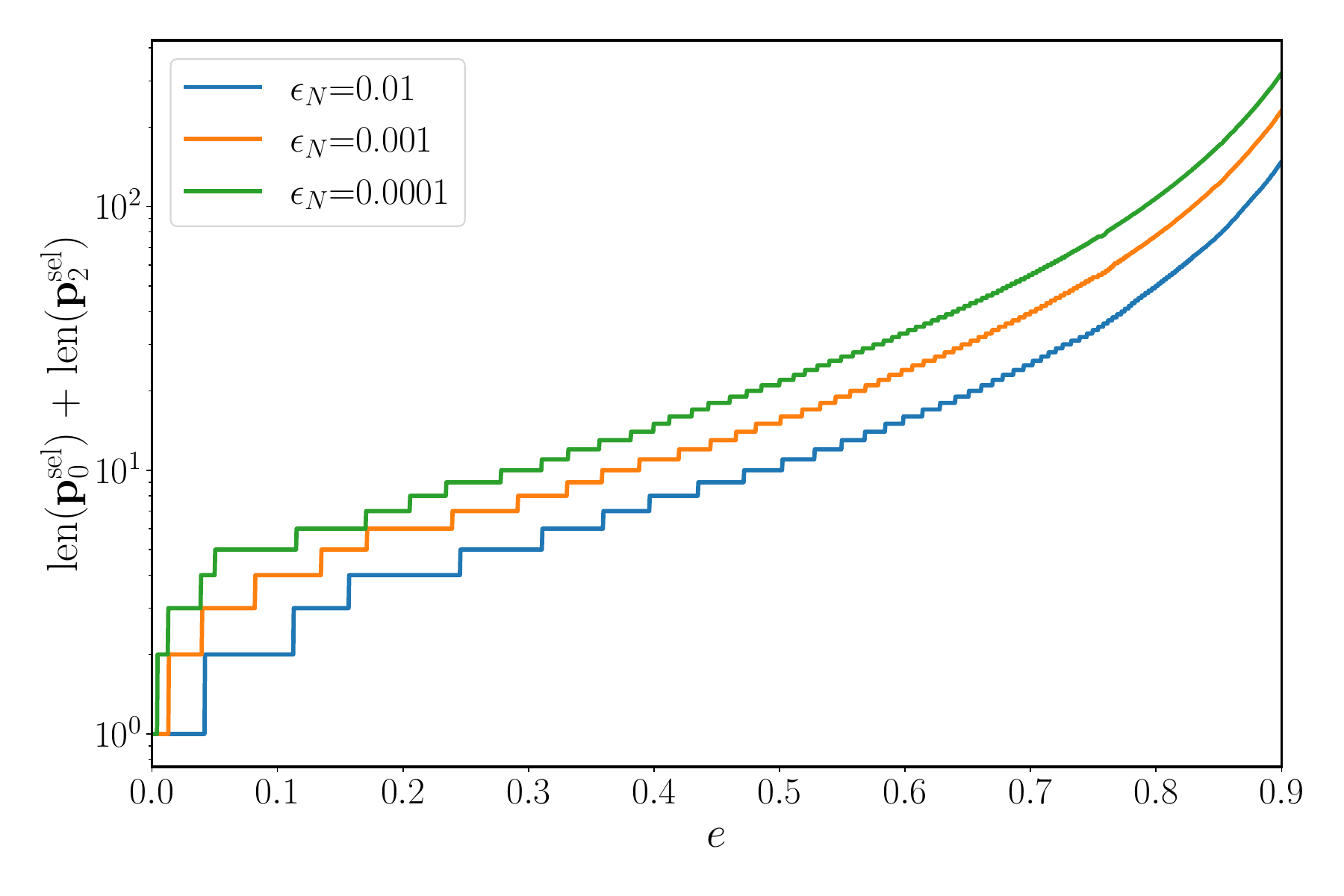}
\caption{\justifying Number of Fourier modes $N^{2 m}_p$ needed to represent the strain $h$ with different tolerances as a function of eccentricity. The number of modes is given by adding the length of the vectors $\bm{p}_0^\mathrm{sel}$ and $\bm{p}_2^\mathrm{sel}$, defined in Eq.~\eqref{eq:Newtonian_orders_needed}.}
\label{fig:Newtonian_orders_needed}
\end{figure}

Therefore, when only considering the Newtonian amplitudes, using Eqs.~(\ref{eq:normh},\ref{eq:normH2m_Newtonian}) we have that

\begin{align}
    \Vert h \Vert^2 & = \frac{8}{3} \left( \frac{4}{\sqrt{1-e^2}} - 1 \right) \label{eq:normh_Newtonian:closed}, \\
    & = 2 \left( \sum_{p_0 = 1}^\infty |N_{p_0}^{2 0}|^2 + \sum_{p_2 = -\infty}^\infty |N_{p_2}^{2 2}|^2 \right),
    \label{eq:normh_Newtonian:sums}  
\end{align}

\noindent where we have used that $N_{-p}^{2 0} = N_{p}^{2 0}$. To minimize the number of terms that have to be included in the sums of Eq.~\eqref{eq:normh_Newtonian:sums}, we order the mode amplitudes $\{|N_p^{2 0}|^2, \,  |N_p^{2 2}|^2\}$ from larger to smaller and take a sufficient number of them such that the error on $\Vert h \Vert^2$ is smaller than a given tolerance $\epsilon_N$, i.e.

\begin{equation}
    \frac{\Vert h \Vert^2 - 2 \left(\sum_{p_0 \in \bm{p}_0^\mathrm{sel}} |N_{p_0}^{2 0}|^2 + \sum_{p_2 \in \bm{p}_2^\mathrm{sel}} |N_{p_2}^{2 2}|^2 \right)}{\Vert h \Vert^2} < \epsilon_N \, ,
    \label{eq:Newtonian_orders_needed}
\end{equation}

\noindent where $\bm{p}_0^\mathrm{sel}$ and $\bm{p}_2^\mathrm{sel}$ represent the $(l,m) = (2, 0)$ and $(l,m) = (2, \pm 2)$ modes with largest norms that have to be selected. In Fig.~\ref{fig:Newtonian_orders_needed} we show, as a function of eccentricity $e$, how many Fourier modes are needed to represent the strain with different tolerances. We can observe that for small eccentricities, it is enough to include only one mode, which corresponds to the usual $N^{2 2}_0$ mode, that is, the only $N^{l m}_p$ that does not vanish when $e$ goes to 0. As the eccentricity increases, the number of modes we have to include also increases, becoming infinite in the limit $e \to 1$. We also observe that, at a given eccentricity, smaller tolerances require more modes to be included.

In App.~\ref{sec:appendix:epsilonN_logL} we give an interpretation of the amplitude tolerance $\epsilon_N$ in the context of data analysis applications, by showing that $\epsilon_N$ is closely related to the log-likelihood error via
\begin{equation}
    \Delta \log\mathcal{L} \lesssim \rho_\mathrm{opt}^2 \epsilon_N \, , \label{eq:DeltalogL_approx}
\end{equation}
\noindent where $ \rho_\mathrm{opt}$ is the optimal signal-to-noise ratio (SNR) of the signal being studied. 
As discussed in App.~\ref{sec:appendix:epsilonN_logL}, for most data analysis applications one requires the log-likelihood error to be $\lesssim 1$, implying that
\begin{equation}
    \epsilon_N \lesssim \frac{1}{\rho_\mathrm{opt}^2} \, . \label{eq:sensible_epsilonN_main}
\end{equation}
Therefore, from Eq.~\eqref{eq:sensible_epsilonN_main}, we can estimate the amplitude tolerance $\epsilon_N$ required to analyze a signal with a certain optimal SNR $\rho_\mathrm{opt}$.

\section{Spin precession description}
\label{sec:MSA}

In this section we closely follow Ref.~\cite{Klein:2021jtd} to model the spin-precession (SP) of compact objects in eccentric orbits using a multiple scale analysis (MSA) approach. While the core idea and procedure are based on Ref.~\cite{Klein:2021jtd}, we fix a few typos therein and rewrite equations in a simpler form, improving their numerical stability and precision to make \pyEFPE more stable and less prone to failures.
Whilst we could numerically evolve the full system of equations, e.g.~\cite{Buonanno:2002fy}, resulting in a more accurate description of the precession dynamics, it can become computationally prohibitive for the low mass binaries seen in ground-based detectors~\cite{Colleoni:2023czp} and the long-lived binaries expected in LISA~\cite{LISA:2024hlh}. 
These limitations become even more pronounced when simultaneously accounting for precession and eccentricity. 
This motivates our detailed exploration of the MSA system of equations. 

The MSA approximation uses that the radiation reaction (RR) time-scale is much longer than the SP time-scale to analytically solve the SP equations in the absence of RR, and then we add the RR by varying some of the constants of the SP solution.
In the absence of RR, including leading PN order spin-orbit and spin-spin interactions, the SP equations are given by~\cite{Barker:1979pre,Klein:2018ybm,Klein:2021jtd}:

\begin{subequations}
\label{eq:raw_prec_eqs}
\begin{align}
\D \uvec{L} &= - y^6 \left( \bm{\Omega}_1 + \bm{\Omega}_2 \right) \, , \label{eq:raw_prec_eqs:L} \\
\D \bm{s}_1 &= \mu_2 y^5 \bm{\Omega}_1 \, ,  \label{eq:raw_prec_eqs:s1} \\
\D \bm{s}_2 &= \mu_1 y^5 \bm{\Omega}_2 \, ,  \label{eq:raw_prec_eqs:s2}
\end{align}
\end{subequations}

\noindent where we have used the following definitions \cite{Klein:2021jtd}

\begin{subequations}
\label{eq:raw_prec_eqs_defs}
\begin{align}
\D &= \frac{M}{\left(1-e^2 \right)^{3/2}} \frac{d}{dt} \, , \\
y &= \frac{(M \omega)^{1/3}}{\sqrt{1 - e^2}} \, , \label{eq:raw_prec_eqs_defs:y} \\
\mu_i &= \frac{m_i}{M} \, , \\
\bm{s}_i &= \frac{\bm{S}_i}{\mu_i} \, , \\
L &= \frac{\nu}{y} \, , \\
\bm{\Omega}_i &=  \left[ \frac{1}{2}\mu_i + \frac{3}{2} \left(1 - y \uvec{L} \cdot \bm{s} \right) \right] \uvec{L} \times \bm{s}_i + \frac{1}{2} y \bm{s}_j \times \bm{s}_i \, ,
\end{align}
\end{subequations}

\noindent and $m_i$ are the individual masses, $\bm{S}_i$ are the individual spins, $\bm{L}$ is the Newtonian angular momentum, $M = m_1 + m_2$ is the total mass, $e$ is the orbital eccentricity, $y$ is a PN parameter related to the norm of $\bm{L}$, $\omega$ is the mean orbital angular velocity, $\mu_i$ are the dimensionless mass parameters, 
$\nu = \mu_1 \mu_2$ is the symmetric mass ratio, $\bm{s}_i$ are the reduced individual spins, and $\bm{s} = \bm{s}_1 + \bm{s}_2$ is the total reduced spin.

The SP equations of Eq.~\eqref{eq:raw_prec_eqs} contain seven conserved quantities. The norm of the orbital angular momentum $L$ and the three components of the total angular momentum vector $\bm{J} = \bm{L} + \mu_1 \bm{s}_1 + \mu_2 \bm{s}_2$ are conserved only when ignoring RR, while the norm of the spin vectors $s_1$ and $s_2$ and the effective spin parameter $\chi\sub{eff}$ are also conserved in the presence of RR. Here $\chi\sub{eff}$ is given by~\cite{Damour:2001tu,Racine:2008qv,Ajith:2009bn}
\begin{equation}
\chi\sub{eff} = \uvec{L} \cdot \bm{s} \, .
\label{eq:chi_eff_0}    
\end{equation}

Taking into account these seven constants of motion, the SP equations of Eq.~\eqref{eq:raw_prec_eqs} have only two dynamical variables left. We can go to a non-inertial frame where the $z$-axis is aligned with $\bm{J}$, and where the orbital angular momentum $\bm{L}$ is perpendicular to the $y$-axis, with $\uvec{L} \cdot \uvec{x} \geq 0$ \cite{Schmidt:2017btt}, i.e.

\begin{equation}
    \uvec{L} = \sin\theta_L \uvec{x} + \cos\theta_L \uvec{z} \, ,
    \label{eq:L_coprec}
\end{equation}

\noindent with $\theta_L \in [0,\pi]$. In this $J$-aligned frame, the angular momenta can all be expressed in terms of one variable, which as in Ref.~\cite{Klein:2021jtd} we choose to be the reduced aligned-spin difference

\begin{align}
\delta\chi &= \uvec{L} \cdot \left( \bm{s}_1 - \bm{s}_2 \right) \, .
\label{eq:dchi_def}
\end{align}

Using the SP equations of Eq.~\eqref{eq:raw_prec_eqs}, the derivative of $\delta\chi$ can be written as

\begin{align}
    \D\delta\chi & = (\D \uvec{L}) \cdot \left( \bm{s}_1 - \bm{s}_2 \right) + \uvec{L} \cdot \left( \D\bm{s}_1 - \D \bm{s}_2 \right) \nonumber \\
    & = 3 (1 - y \chi\sub{eff}) \left( \uvec{L} \times \bm{s}_1 \right) \cdot \bm{s}_2 \, .
    \label{eq:Ddchi_raw}
\end{align}

\noindent Using  the orbital angular momentum in the frame of Eq.~\eqref{eq:L_coprec}, and the conserved quantities previously described, we can write Eq.~\eqref{eq:Ddchi_raw} as~\cite{Klein:2021jtd}

\begin{equation}
\left( \D \delta \chi \right)^2 = \frac{9}{4} A^2 y^{11} \left( \delta \mu  \delta \chi^3 + B \delta \chi^2 + C \delta \chi + D \right), \label{eq:Dchidsq}
\end{equation}
\noindent where 
\begin{equation}
    \delta\mu = \mu_1 - \mu_2 \, ,
    \label{eq:dmu_ref}
\end{equation}
\noindent and the coefficients of the cubic polynomial take the following values

\begin{widetext}
\begin{subequations}
\label{eq:Ddchi_coefs_raw}
\begin{align}
A & = 1 - y \chi\sub{eff},  \label{eq:Ddchi_coefs_raw:A} \\
B &= \frac{y}{2 \nu^2} \left[ - 2 \nu \left(J^2 - L^2 - L \chi\sub{eff} \right) + \delta\mu \left( S_1^2 - S_2^2 \right) - \delta\mu^2 \left( 2 L^2 +  S_1^2 + S_2^2 \right) \right] ,  \label{eq:Ddchi_coefs_raw:B} \\
C &= \frac{y}{2 \nu^2} \left\{ \left(1 + \delta\mu^2 \right) \chi\sub{eff} \left( S_1^2 - S_2^2 \right) + 2 \delta \mu \left[ 2 L \left(J^2 - L^2 - L \chi\sub{eff} \right) - (2L + \chi\sub{eff}) \left( S_1^2 + S_2^2 \right) - \nu L \chi\sub{eff}^2 \right] \right\},  \label{eq:Ddchi_coefs_raw:C} \\
D &= \frac{y}{2 \nu^2} \big\{ -2  \left(J^2 - L^2 - L \chi\sub{eff} \right) \left[J^2 - L^2 - L \chi\sub{eff} - 2 \left(S_1^2 + S_2^2 \right) - \nu \chi\sub{eff}^2\right] \nonumber\\
&+ \left(S_1^2 - S_2^2 \right) \left[ \delta\mu \chi\sub{eff}^2 - 2 \left(S_1^2 - S_2^2 \right)  \right] - \chi\sub{eff}^2 \left( S_1^2 + S_2^2 \right) \big\} \, . \label{eq:Ddchi_coefs_raw:D}
\end{align}
\end{subequations}
\end{widetext}

During the inspiral, the PN parameter $y$ is small ($y \ll 1$) and thus $J \sim L \gg S_{1,2}$. This leads to large numerical cancellations in Eq.~\eqref{eq:Ddchi_coefs_raw}. To mitigate against this, we define $\Delta_{J^2}$ such that

\begin{equation}
    J^2 = \left(L + \frac{1}{2} (\chi\sub{eff} + \delta X_0)\right)^2 + S_{1\perp,0}^2 + S_{2\perp,0}^2 + \nu \Delta_{J^2} \, ,
    \label{eq:DJ2_def}
\end{equation}

\noindent where we have defined

\begin{subequations}
\label{eq:DJ2_supp_def}
\begin{align}
\delta X_0 & = \delta\mu \, \delta\chi_0 \, , \\
S_{i\perp,0}^2 & = \left\Vert \bm{S}_{i\perp,0}\right\Vert^2 = \left\Vert \bm{S}_{i,0} - (\uvec{L}_0 \cdot \bm{S}_{i,0})\uvec{L}_0 \right\Vert^2 \, ,
\end{align}
\end{subequations}

\noindent and the subscript ``0'' denotes that the value at the initial time is taken. Since $\delta X_0$ and $S_{i \perp,0}$ are constants by definition, and $J$, $L$ and $\chi_\mathrm{eff}$ are constants of motion of the SP equations, in the absence of RR, $\Delta_{J^2}$ is also constant, taking the value

\begin{equation}
    \Delta_{J^2} = 2 \bm{s}_{1\perp,0} \cdot \bm{s}_{2\perp,0} \, .
    \label{eq:DJ20_def}
\end{equation}
However, when adding the effects of RR, we will see in Sec.~\ref{sec:RR} that $\Delta_{J^2}$ slowly varies on the RR time-scale. 
Substituting Eq.~\eqref{eq:DJ2_def} into Eqs.~\eqref{eq:Ddchi_coefs_raw}, we arrive at the following simpler expressions that avoid large numerical cancellations:

\begin{subequations}
\label{eq:Ddchi_coefs}
\begin{align}
B &= -\frac{\delta\mu^2}{y} - y\chi\sub{eff}^2 - \delta X_0 - b_\perp \, ,  \label{eq:Ddchi_coefs:B} \\
\delta\mu \, C &= \delta X_0 \left[2 \left(\frac{\delta\mu^2}{y} + y\chi\sub{eff}^2 \right) - \delta X_0 \right] -c_\perp ,  \label{eq:Ddchi_coefs:C} \\
\delta\mu^2 D &= -\delta X_0^2 \left[\frac{\delta\mu^2}{y} + y\chi\sub{eff}^2 - \delta X_0  -  b_\perp \right] + \delta X_0 c_\perp + d_\perp  . \label{eq:Ddchi_coefs:D}
\end{align}
\end{subequations}

\noindent where we separate in $b_\perp$, $c_\perp$ and $d_\perp$ the part of the coefficients that vanishes in the aligned spin case, where precession should not be present. Explicitly, these are given by

\begin{subequations}
\label{eq:Ddchi_coefs_perp}
\begin{align}
b_\perp &= y \left(s_{1\perp,0}^2 + s_{2\perp,0}^2 + \Delta_{J^2} \right) \, ,  \label{eq:Ddchi_coefs_perp:b} \\
c_\perp &= - 2\delta\mu \left[ \delta\mu \Delta_{J^2} + (s_{1\perp,0}^2 - s_{2\perp,0}^2) y \chi\sub{eff} \right],  \label{eq:Ddchi_coefs_perp:c} \\
d_\perp &= y \delta\mu^2 \left(4 s_{1\perp,0}^2 s_{2\perp,0}^2 - \Delta_{J^2}^2 \right) \, . \label{eq:Ddchi_coefs_perp:d}
\end{align}
\end{subequations}

The differential equation of Eq.~\eqref{eq:Dchidsq} can be solved analytically in terms of the roots of the cubic polynomial on the right hand side. That is, if we write~\cite{Klein:2021jtd}

\begin{align}
    \left( \D \delta \chi \right)^2 & = - \frac{1}{y} \left( \frac{3}{2} A y^{6} \right)^2 \nonumber\\
& \quad \times ( \delta \chi - \delta\chi_+ ) ( \delta \chi - \delta\chi_- ) ( \delta \chi_3 - \delta\mu \delta\chi ) \, ,   \label{eq:Dchidsq_poly} \\
    \delta\chi_- &\leq \delta\chi_+ \leq \frac{\delta\chi_3}{\delta\mu} \, , \label{eq:Dchidsq_roots}
\end{align}

\noindent then the solution of this equation is~\cite{Klein:2021jtd}

\begin{align}
 \delta \chi &= \delta \chi_- + (\delta \chi_+ - \delta \chi_-) \text{sn}^2 (\psi\sub{p}; m) \, , \label{eq:dchi_sol}\\
\end{align}

\noindent where 

\begin{subequations}
\label{eq:m_Dpsip_defs}
\begin{align}
 m &= \frac{\delta\mu (\delta \chi_+ - \delta \chi_-)}{\delta \chi_3 - \delta\mu \delta \chi_-} \, , \\
 \D \psi\sub{p} &= \frac{3 A y^6}{4} \sqrt{ \frac{1}{y} (\delta \chi_3 - \delta \mu \delta \chi_-)} \label{eq:Dpsi_p}\, ,
\end{align}    
\end{subequations}

\noindent and $\text{sn}(\psi\sub{p}; m) = \sin(\mathrm{am}(\psi\sub{p}; m))$ is the Jacobi elliptic sine function, with $\mathrm{am}(\psi\sub{p}; m)$ being the Jacobi amplitude. We use the same conventions for the elliptic functions and integrals as in Ref.~\cite{Klein:2021jtd}. To find the roots of the cubic polynomial, we start computing the coefficients of its depressed cubic as

\begin{subequations}
\label{eq:depressed_cubic}
\begin{align}
p & = \frac{1}{y^2} \left( \frac{B^2}{3} - \delta \mu C \right) \nonumber \\
& = \frac{1}{y^2} \left( 3 p_\parallel^2 + 2 p_\perp \right) \, , \label{eq:depressed_cubic:p} \\
q &= \frac{1}{y^3} \left( \frac{2 B^3}{27} - \frac{B \delta \mu C}{3}  + \delta\mu^2 D  \right) \nonumber \\
& = \frac{1}{y^3} \left(-2 (p_\parallel^2 + p_\perp) p_\parallel + d_\perp \right)\, ,\label{eq:depressed_cubic:q}
\end{align}
\end{subequations}

\noindent where it can be shown that $p>0$. 
To avoid numerical instabilities, we separate in $p_\parallel$ the terms that are non-zero in the aligned spin case and in $p_\perp$ the ones that vanish. 
Explicitly, they are given by
\begin{subequations}
\label{eq:depressed_cubic_extra}
\begin{align}
p_\parallel & = \frac{1}{3} \left(\frac{\delta\mu^2}{y} + y\chi\sub{eff}^2  - 2 \delta X_0 + b_\perp \right) \, , \label{eq:depressed_cubic_extra:ppar} \\
p_\perp &= \delta X_0 b_\perp + \frac{1}{2} c_\perp \, . \label{eq:depressed_cubic_extra:pperp}
\end{align}
\end{subequations}

Similarly to~\cite{Klein:2021jtd}, the roots of the cubic polynomial of Eq.~\eqref{eq:Dchidsq_poly} can be written in terms of
\begin{subequations}
\label{eq:Ydefs}
\begin{align}
    Y_3 &= 2 \sqrt{\frac{p}{3}} \cos\left[ \frac{\arg(G)}{3} \right] \, , \\
    Y_\pm &= 2 \sqrt{\frac{p}{3}} \cos\left[ \frac{\arg(G) \mp 2 \pi}{3} \right] \, , \\
    G &= -\frac{q}{2} + i \left[ \left( \frac{p}{3} \right)^3 - \left( \frac{q}{2} \right)^2 \right]^{1/2} \nonumber \\
      &= -\frac{q}{2} + \frac{i}{y^3} 
         \Bigg[ \frac{p_\perp^2 (p_\parallel^2 + 8 p_\perp)}{27} + (p_\parallel^2 + p_\perp) p_\parallel d_\perp 
         - \frac{d_\perp^2}{4} \Bigg]^{1/2} \, , \label{eq:Ydefs:G_simplified} \\
    dY &= \frac{B}{3 y} \, ,
\end{align}
\end{subequations}
\noindent such that
\begin{subequations}    
\begin{align}
 \delta \chi_3 &= y \left( Y_3 - dY \right) \, , \\
 \delta \chi_\pm &= \frac{y}{\delta\mu} \left( Y_\pm - dY \right) \, .
\end{align}
\end{subequations}

\begin{figure}[t!]
\centering  
\includegraphics[width=0.5\textwidth]{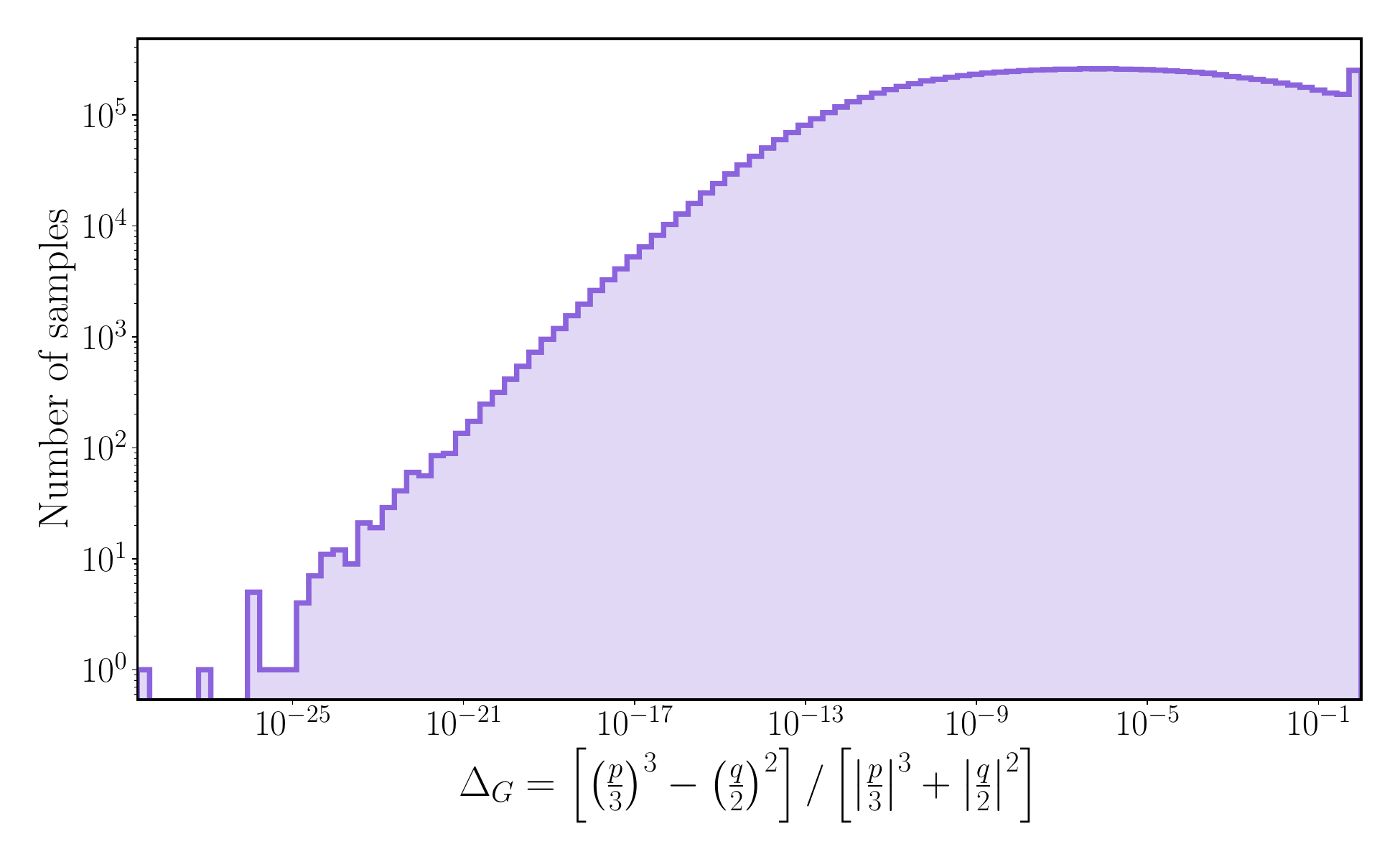}
\caption{\justifying Histogram of $\Delta_G$ computed for $10^7$ random samples drawn from a distribution isotropic in spin orientations, uniform in dimensionless spin magnitudes $\chi_i = s_i/\mu_i \in [0,1]$, uniform in mass ratio $q = m_2/m_1 \in [0, 1]$ and log-uniform in the PN parameter $y \in [0.001, 6^{-1/2}]$.}
\label{fig:argG_cancelation}
\end{figure}

\noindent Note that in Eq.~\eqref{eq:Ydefs:G_simplified} we have simplified the square root term to avoid the large numerical cancellations. From Eq.~\eqref{eq:depressed_cubic_extra} we expect that $|p_\parallel| \gg |p_\perp|$ and $|p_\parallel| \gg d_\perp$, and thus $\left(\frac{p}{3} \right)^3 \sim \left( \frac{q}{2} \right)^2 \sim \ord{p_\parallel^6}$. However, when we compute the difference between these two terms in Eq.~\eqref{eq:Ydefs:G_simplified}, we get $\left(\frac{p}{3} \right)^3 - \left( \frac{q}{2} \right)^2 \sim \ord{p_\parallel^3 d_\perp}$ which is much smaller, and points to a large numerical cancellation happening. To illustrate this point, we show in Fig.~\ref{fig:argG_cancelation} the level of cancellation by plotting the distribution of 
\begin{equation}
    \Delta_G = \left[\left(\frac{p}{3} \right)^3 - \left( \frac{q}{2} \right)^2\right]/\left[\left| \frac{p}{3} \right|^3 + \left|\frac{q}{2} \right|^2 \right] \, ,
    \label{eq:DeltaG}
\end{equation}
for random angular momenta and masses, as described in the figure caption. For $\Delta_G \lesssim 10^{-12}$, which as we observe in Fig.~\ref{fig:argG_cancelation} is quite common, we expect the numerical error to have a large impact when using 64 bit floating precision numbers.

Using the quantities defined in Eq.~\eqref{eq:Ydefs}, we can simplify Eq.~\eqref{eq:m_Dpsip_defs} as
\begin{subequations}
\label{eq:m_Dpsip_simple}
\begin{align}
 \D \psi\sub{p} &= \frac{3 A y^6}{4} \sqrt{ Y_3 - Y_- },  \label{eq:Dpsip_Y} \\
 m &= \frac{Y_+ - Y_-}{Y_3 - Y_-}  = \frac{\sin\left[ \frac{\arg\left(G \right)}{3}\right]}{\cos\left[\frac{\arg\left(G \right)}{3} - \frac{\pi}{6}\right]}. \label{eq:ellip_m_of_argG}
\end{align}    
\end{subequations}
It will also be convenient to write the solution of $\delta\chi$ of Eq.~\eqref{eq:dchi_sol} as
\begin{equation}
    \delta \chi = \delta\chi\sub{av} - \delta\chi\sub{diff} \left(1 - 2 \text{sn}^2 (\psi\sub{p} , m)\right) \, ,
    \label{eq:dchi_sol_avdiff}
\end{equation}
where we have defined
\begin{subequations}
\label{eq:dchiavdiff}
\begin{align}
\delta\chi\sub{av}   &= \frac{\delta\chi_+ + \delta\chi_-}{2} \, , \label{eq:dchiavdiff:dchiav} \\
\delta\chi\sub{diff} & = \frac{\delta\chi_+ - \delta\chi_-}{2} \, . \label{eq:dchiavdiff:dchidiff}
\end{align}
\end{subequations}
These variables can be written as:
\begin{subequations}
\label{eq:dchiavdiff_explicit}
\begin{align}
\delta\chi\sub{av}   &= \delta \chi_0 + \frac{p_\parallel - y \sqrt{p/3} \cos\left[ \frac{\arg\left(G \right)}{3}\right]}{\delta \mu} \nonumber \\
& = \delta \chi_0 + \frac{p_\parallel^2 \sin^2\left[ \frac{\arg\left(G \right)}{3}\right] - (2 p_\perp /3) \cos^2\left[ \frac{\arg\left(G \right)}{3}\right]}{\delta \mu \left(p_\parallel + y \sqrt{p/3} \cos\left[ \frac{\arg\left(G \right)}{3}\right] \right)} \, , \label{eq:dchiavdiff_explicit:dchiav} \\
\delta\chi\sub{diff} & = \frac{y}{\delta\mu}\sqrt{p} \sin\left[ \frac{\arg\left(G \right)}{3}\right]  \, , \label{eq:dchiavdiff_explicit:dchidiff}
\end{align}
\end{subequations}

\noindent where in the second line of Eq.~\eqref{eq:dchiavdiff_explicit:dchiav} we have written $\delta\chi_\mathrm{av}$ in a way that avoids large numerical cancellations that can happen when $p_\parallel \gg |p_\perp|$. 

To go from the non-inertial precession frame, where the solution of Eq.~\eqref{eq:Dchidsq} was found, to the inertial frame, we perform a set of rotations described by the three Euler angles $\phi_z, \theta_L$ and $\zeta$ also introduced in Eq.~\eqref{eq:twist-up}. The Euler angles can be computed as

\begin{subequations}
\label{eq:EulerAngles}
\begin{align}
\D \phi_z &= \frac{1}{\sin^2\theta_L} \left(\D \uvec{L}\right) \cdot \left( \uvec{J} \times \uvec{L} \right) \, , \label{eq:EulerAngles:Dphiz} \\
\cos \theta_L &= \uvec{J} \cdot \uvec{L} = \frac{1}{2J} \left( 2 L + \chi\sub{eff} + \delta \mu \delta \chi \right) \, , \label{eq:EulerAngles:costhL}\\
\D \zeta & = - \cos \theta_L \D \phi_z \label{eq:EulerAngles:Dzeta} \, ,
\end{align}
\end{subequations}
\noindent where the final equation corresponds to the minimal rotation condition~\cite{Buonanno:2002fy,Boyle:2011gg}.
We can compute Eq.~\eqref{eq:EulerAngles:Dphiz} by substituting all the angular momenta in terms of the solution of Eq.~\eqref{eq:dchi_sol_avdiff}, such that
\begin{subequations}
\label{eq:Dphiz}
\begin{align}
\D \phi_z &= \frac{J y^6}{2} + \frac{\D \psi\sub{p}}{\nu \sqrt{Y_3 - Y_-}} \left[ \frac{N_+}{D_+} + \frac{N_-}{D_-} \right] \, , \\
N_\pm &= (J \pm L) \left(J \pm L \pm 2 \nu \chi\sub{eff} \right) - \delta \mu \left(S_1^2 - S_2^2\right) \, , \\
D_\pm &= 2 J (1 \pm \cos{\theta_L}) = 2(J \pm L) \pm \chi\sub{eff} \pm \delta\mu \, \delta\chi  \nonumber \\
&= B_\pm - C_\pm \left[1 - 2 \text{sn}^2 (\psi\sub{p}, m) \right] \, , \label{eq:Dphiz:Dpm} \\
B_\pm & = 2 (J \pm L) \pm \chi\sub{eff} \pm\delta\mu \, \delta\chi\sub{av} \, , \\
C_\pm & = \pm \, \delta\mu \, \delta\chi\sub{diff} \, .
\end{align}
\end{subequations}
Eq.~\eqref{eq:Dphiz} can be analytically integrated. To do so, we first show how to integrate a general function in terms of our solution, i.e. we study how to integrate $g(\mathrm{sn}^2 (\psi\sub{p}(t), m))$ in
\begin{align}
    G(t_0,t_f) &= \int_{t_0}^{t_f} g(\mathrm{sn}^2 (\psi\sub{p}(t), m)) \d t \nonumber\\
    & = \int_{\psi\sub{p}(t_0)}^{\psi\sub{p}(t_f)} g(\mathrm{sn}^2 (\psi\sub{p}, m)) \frac{\d t}{\d \psi\sub{p}} \d\psi\sub{p} \, .
    \label{eq:MSA_general_int_start}
\end{align}
When ignoring radiation reaction, the derivative $\d\psi\sub{p}/\d t$ is a constant that can be taken outside of the integral. Furthermore, we can write the integral in terms of the Jacobi amplitude $\varphi = \mathrm{am}(\psi\sub{p}; m)$, i.e.
\begin{equation}
    \psi\sub{p} = \int_0^\varphi \frac{\d \theta}{\sqrt{1 - m \sin^2\theta}} \rightarrow \d\psi\sub{p} = \frac{\d \varphi}{\sqrt{1 - m \sin^2\varphi}} \, ,
    \label{eq:JacobiAmp_d}
\end{equation}
\noindent simplifying the integral of Eq.~\eqref{eq:MSA_general_int_start} as 

\begin{equation}
     G(t_0,t_f) = \frac{1}{\d \psi\sub{p}/\d t} \int_{\varphi(t_0)}^{\varphi(t_f)} \frac{g(\sin \varphi)}{\sqrt{1 - m \sin^2\varphi}}  \d\varphi \, .
    \label{eq:MSA_general_int_JA}
\end{equation}

Also, we define the precession average of a function, as the average of the function over one precession cycle, i.e.

\begin{equation}
    \av{g} = \frac{1}{T\sub{p}} G(t_0, t_0 + T\sub{p}) \, ,
    \label{eq:prec_avg_def}
\end{equation}

\noindent where $T\sub{p}$ is the precession period, given by 

\begin{equation}
    T\sub{p} = \frac{2 K(m)}{\d \psi\sub{p}/\d t} \, ,
    \label{eq:T_precesion}
\end{equation}

\noindent and $K(m)$ is the Jacobi complete elliptic integral of the first kind. Therefore, the precession average of any function can be computed as

\begin{equation}
    \av{g} = \frac{1}{2 K(m)} \int_{-\pi/2}^{\pi/2} \frac{g(\sin\varphi)}{\sqrt{1 - m \sin^2\varphi}} \d\varphi
    \label{eq:prec_avg_simple}
\end{equation}

With these equations, we can integrate $\D\phi_z$ in Eq.~\eqref{eq:Dphiz} analytically, separating the result in a secular part $\phi_{z,0}$ and periodic part $\delta\phi_z$, i.e.

\begin{subequations}
\label{eq:phiz_sol}
\begin{align}
\phi_z &= \phi_{z,0} + \delta\phi_z, \\
\D \phi_{z,0} &= \av{\D \phi_z} \nonumber\\
&= \frac{J y^6}{2} + \frac{3 ( 1 - y \chi\sub{eff}) y^6}{4 \nu K(m)} \big\{ P_+ + P_- \big\}, \label{eq:phiz_sol:Dphiz0} \\
\delta \phi_z &= \int_{t_0}^{t_0 +\delta t}  \D \phi_z - \av{\D \phi_z}  \, \d t \nonumber\\
&= \frac{1}{\nu \sqrt{Y_3 - Y_-}} \big\{ \delta P_+(\hat{\psi}\sub{p}) + \delta P_-(\hat{\psi}\sub{p})\big\} , \label{eq:phiz_sol:dphiz}
\end{align}
\end{subequations}

\noindent where we have defined

\begin{subequations}
\label{eq:P_pm_def}
\begin{align}
P_\pm & = \frac{N_\pm}{B_\pm - C_\pm}  \Pi\left( \frac{-2 C_\pm}{B_\pm - C_\pm}, m \right) \, ,  \\
\delta P_\pm(\hat{\psi}\sub{p}) & = \frac{N_\pm}{B_\pm - C_\pm} \Bigg\{ \Pi \left[ \frac{- 2 C_+}{B_+ - C_+} ; \text{am}(\hat{\psi}\sub{p}; m) ; m \right] \nonumber\\ 
& \qquad - \frac{\Pi\left( \frac{- 2 C_+}{B_+ - C_+}, m \right)}{K(m)} \hat{\psi}\sub{p} \Bigg\} \, ,
\end{align}
\end{subequations}

\noindent where $\Pi(n, m)$ is the complete elliptic integral of the third kind and $\Pi(n; \phi; m)$ is the incomplete elliptic integral of the third kind. $t_0$ represents the start of a precession cycle and $\delta t \in (0, T\sub{p}]$ tracks the time within the precession cycle. Consequently, the variable $\hat{\psi}\sub{p}$ tracks the phase within the precession cycle and satisfies the following conditions
\begin{subequations}
\label{eq:hatpsip_def}
\begin{align}
&\psi\sub{p} - \hat{\psi}\sub{p} = 2 n K(m) \text{, for some }  n \in \mathbb{Z} \, , \\
&-K(m) < \hat{\psi}\sub{p} \leq K(m) \, .
\end{align}
\end{subequations}

We can perform the same steps for the Euler angle $\zeta$, writing Eq.~\eqref{eq:EulerAngles:Dzeta} in terms of $\delta\chi$ and integrating it analytically, separating the secular part $\zeta_0$ from the periodic part $\delta\zeta$. Doing this we obtain:

\begin{subequations}
\label{eq:zeta_sol}
\begin{align}
\zeta = & \zeta_0 + \delta\zeta, \\
\D \zeta_0 = & -\frac{\left(2L + \chi\sub{eff} + \delta\mu \av{\delta\chi} \right) y^6}{4} \nonumber \\ 
& - \frac{3 (L + \nu \chi\sub{eff})(1 - y \chi\sub{eff}) y^6}{2\nu} \nonumber\\
& + \frac{3 ( 1 - y \chi\sub{eff}) y^6}{4 \nu K(m)} \big\{ P_+ - P_- \big\}, \\
\delta\zeta = &  \frac{y \sqrt{Y_3 - Y_-}}{3 (1 - y \chi\sub{eff})} \left\{ E \left[ \text{am} (\hat{\psi}\sub{p}; m) ; m \right] - \frac{E(m)}{K(m)} \hat{\psi}\sub{p} \right\} \nonumber\\
&+ \frac{1}{\nu \sqrt{Y_3 - Y_-}} \big\{\delta P_+(\hat{\psi}\sub{p}) - \delta P_-(\hat{\psi}\sub{p}) \big\} \, . \label{eq:zeta_sol:dzeta}
\end{align}
\end{subequations}

\noindent where we have introduced the precession average of $\delta\chi$, given by

\begin{align}
    \av{\delta\chi} &= \delta\chi\sub{av} - \frac{2\delta\chi\sub{diff}}{m} \left[ \frac{E(m)}{K(m)} - 1 + \frac{m}{2} \right] \, ,
    \label{eq:dchi_prec_avg}
\end{align}

\noindent and $E(m)$ is the complete elliptic integral of the second kind, while $E(\phi; m)$ is the incomplete elliptic integral of the second kind. Notice that for small $m$

\begin{equation}
    \frac{2}{m} \left[ \frac{E(m)}{K(m)} - 1 + \frac{m}{2} \right] = - \frac{m}{8} + O(m^2) \, ,
    \label{eq:m_factor_dchi_prec_avg_series}
\end{equation}

\noindent and this term vanishes in the $m \to 0$ limit. 
In Eq.~\eqref{eq:P_pm_def} we observe that we can have indeterminations in $P_{\pm}$ and $\delta P_{\pm}$, which appear in the secular and periodic parts of both $\phi_z$ (Eq.~\eqref{eq:phiz_sol}) and $\zeta$ (Eq.~\eqref{eq:zeta_sol}). These indeterminations happen under two conditions. First, when $B_+ - C_+ = 0$, where one can show $N_+ = 0$ and there is a $0/0$ indetermination. Second, when $B_- + C_- = 0$, $\Pi(-2C_-/(B_- - C_-), m)$ diverges, and one can show $N_- = 0$, leading to a $0 \times \infty$ indetermination. Both scenarios happen because we have $D_{\pm} = 0$ at some point in the precessional cycle. From Eq.~\eqref{eq:Dphiz:Dpm}, this occurs when the total and orbital angular momenta, $J$ and $L$, get aligned ($\cos{\theta_L} = \pm 1$). This alignment corresponds to a Gimbal lock, which makes the Euler angles $\phi_z$ and $\zeta$ interdependent. To prevent these indeterminations, which would result in computational errors, we impose a minimum value of $D_\pm$, ensuring $D_\pm \geq \epsilon_D$. From Eq.~\eqref{eq:Dphiz:Dpm} this corresponds to imposing the constraint

\begin{equation}
    B_\pm \mp  C_\pm =  B_\pm - \delta\mu \delta\chi\sub{diff} \geq  \epsilon_D  \, .
\end{equation}

To implement this constraint in the code, we choose to update $B_\pm$ as

\begin{equation}
    B_\pm \leftarrow  \max\left(B_\pm, \delta\mu \delta\chi\sub{diff}+\epsilon_D\right)  \, ,
\end{equation}

\noindent keeping the values of the rest of the variables in Eq.~\eqref{eq:P_pm_def} constant.  
This regularization avoids singularities and indeterminations by effectively turning off precession when the total and orbital angular momenta align during the precessional cycle.

\section{Addition of Radiation Reaction effects}
\label{sec:RR}

In Sec.~\ref{sec:NewtAmp} and Sec.~\ref{sec:MSA} we have explored the system ignoring RR, i.e. ignoring how the binary evolves due to the emission of gravitational waves. Taking RR into account, the PN parameter $y$ of Eq.~\eqref{eq:raw_prec_eqs_defs:y} will increase as the system becomes more compact, while the eccentricity $e$ will decrease, as the system circularizes, reaching a residual eccentricity induced by spin~\cite{Klein:2018ybm,Klein:2010ti}

\begin{equation}
    e\sub{min}^2 = \frac{5y^4}{304} \left\Vert \bm{s}_{1\perp} - \bm{s}_{2\perp}  \right\Vert^2 \, .
    \label{eq:emin}
\end{equation}

The post-Newtonian evolution equations for the PN parameter $y$ and the squared eccentricity $e^2$, the mean orbital phase $\lambda$ and the argument of periastron $\delta\lambda$ can be expressed in the following way:

\begin{subequations}
\label{eq:RR_eqs_no_SP}
\begin{align}
 \D y &= \nu y^9 \sum_{n \geq 0} a_n \left(y, e^2, \uvec{L}, \bm{s}_1, \bm{s}_2 \right) y^n \, , \\
 \D e^2 &= - \nu y^8 \sum_{n \geq 0} b_n \left(y, e^2, \uvec{L}, \bm{s}_1, \bm{s}_2 \right) y^n \, , \\
 \D \lambda &= y^3 \, , \\
 \D \delta\lambda &= \frac{k y^3}{1 + k}, \quad k = y^2 \sum_{n \geq 0} k_n\left(y, e^2, \uvec{L}, \bm{s}_1, \bm{s}_2 \right) y^n  \label{eq:RR_eqs_no_SP:dl}\, ,
\end{align}
\end{subequations}

\noindent where $k$ is the periastron advance of Eq.~\eqref{eq:phi_lW}. The coefficients $a_n$, $b_n$ and $k_n$ are obtained from Ref.~\cite{Klein:2018ybm} to 3PN order in the non-spinning part and 2PN in the fully spinning part. We correct some errors in the 2PN and 3PN non-spinning instantaneous terms pointed out in Ref.~\cite{Ebersold:2019kdc} associated with an incorrect transformation from ADM to harmonic coordinates. We also fix a typo in the coefficient multiplying $(\uvec{L} \cdot \bm{s}_2)^2$ of $b_2$, where an $111/4$ should be replaced by an $111/2$~\cite{Henry:2023tka}. As it is well known in the literature~\cite{Cutler:1992tc}, detectors are highly sensitive to the phase of the binary, and describing it at 2PN can induce strong biases in the recovered parameters. To alleviate this, we add the state-of-the-art 2.5PN and 3PN spin-spin and spin-orbit corrections for aligned-spin eccentric binaries derived in Ref.~\cite{Henry:2023tka}. The explicit expressions for the evolution equations used can be found in appendix~\ref{sec:appendix:PN_formulas}.

Since no analytical solution is known for the differential equations of Eq.~\eqref{eq:RR_eqs_no_SP}, we have to numerically solve them using, for example, Runge-Kutta methods~\cite{Press:2007nr,Scipy:2020}. 
However, we observe that the PN coefficients depend not only on $y$ and $e^2$, but also on the angular momenta $\uvec{L}, \bm{s}_1, \bm{s}_2$, which vary on the SP timescale. Since $\D\psi\sub{p}, \D\phi_z, \D\zeta \sim O(y^5)$, while $\D\log(y), \D e^2 \sim \ord{y^8} $, the angular momenta vary much more rapidly than $y$ and $e^2$. That is, the SP time-scale is much shorter than the RR time-scale. To avoid having to integrate Eqs.~\eqref{eq:RR_eqs_no_SP} on the shorter (and thus more computationally expensive) SP time-scale, we replace the quantities in Eq.~\eqref{eq:RR_eqs_no_SP} that depend on the SP time-scale by their average over one precession cycle, which can analytically be computed with the solution of Sec.~\ref{sec:MSA}. These averages vary on the RR time-scale, and the differential equations become numerically cheaper to integrate.

\subsection{Spin couplings averaging}
\label{sec:RR:spin_avg}

The 1.5PN and 2.5PN spin-orbit couplings of App.~\ref{sec:appendix:PN_formulas} can be written in terms of $\chi_\mathrm{eff}$ and $\delta \chi$. Since $\chi_\mathrm{eff}$ is constant, we only have to compute $\av{\delta\chi}$, which was already done in Eq.~\eqref{eq:dchi_prec_avg}. Meanwhile, the 2PN spin-spin couplings can be expressed as linear combinations of

\begin{subequations}
\begin{align}
    \sigma_i^{(1)} &= \bm{s}_i^2, \\
    \sigma_i^{(2)} &= \left(\bm{\hat{L}} \cdot \bm{s}_i\right)^2, \\
    \sigma_i^{(3)} &= \left|\bm{\hat{L}} \times \bm{s}_i\right|^2 \cos 2\psi_i,
\end{align}
\end{subequations}

\noindent where $i \in \{0, 1, 2\}$, $\bm{s}_0 = \bm{s}_1 + \bm{s}_2$ and $\psi_i$ denotes the angle subtended by the periastron line and the projection of $\bm{s}_i$ onto the orbital plane. We have that $\sigma_{1,2}^{(1)} = s_{1,2}^2$ and $\sigma_0^{(2)} = \chi\sub{eff}^2$ are conserved, and, as was found in Ref.~\cite{Klein:2021jtd}, we have that 
\begin{equation}
    \av{\sigma_i^{(3)}} = 0 \, \quad \text{for $i=\{0,1,2\}$}\, .
\end{equation}

The precession average for $\sigma_0^{(1)}$ can be computed in terms of our solution as

\begin{align}
    & \av{\sigma_0^{(1)}} = \av{\left( \bm{s}_1 + \bm{s}_2 \right)^2}  \nonumber\\ & = \av{\frac{J^2 - L^2 - \delta\mu \left(\mu_1 s_1^2 - \mu_2 s_2^2\right) - L \left(\chi\sub{eff} + \delta\mu \delta\chi \right) }{\nu}} \nonumber \\
    & = \chi\sub{eff}^2 + s_{1\perp,0}^2 + s_{2\perp,0}^2 + \Delta_{J^2}^2 + \delta\mu \left(\frac{\delta\chi_0 - \av{\delta\chi}}{y}\right).
\end{align}

Finally, the square of the component of the spins aligned with the orbital angular momentum $\sigma_{1,2}^{(2)}$ can be written as

\begin{align}
    \av{\sigma_{1,2}^{(2)} } &= \av{\left(\bm{\hat{L}} \cdot \bm{s}_{1,2}\right)^2} = \av{\left(\frac{\chi_\mathrm{eff} \pm \delta \chi}{2}\right)^2} \nonumber\\
    & = \frac{\chi_\mathrm{eff}^2 \pm 2 \chi_\mathrm{eff}\av{\delta\chi} + \av{\delta\chi^2}}{4}  \, ,
\end{align}

\noindent where we have used that $\chi_\mathrm{eff}$ is conserved, $\av{\delta\chi}$ can be computed from Eq.~\eqref{eq:dchi_prec_avg} and we can compute $\av{\delta\chi^2}$ using Eqs.~(\ref{eq:dchi_sol_avdiff},\ref{eq:prec_avg_simple}), obtaining

\begin{equation}
    \av{\delta\chi^2} = \av{\delta\chi}^2 + \left(\frac{1}{2} - F_\sigma(m) \right) \delta\chi_\mathrm{diff}^2 \, ,
    \label{eq:dchi2_prec_avg}
\end{equation}

\noindent where we have defined the function

\begin{equation}
    F_\sigma(m) = \frac{4}{m^2} \bigg[ \frac{E(m)^2}{K(m)^2}+ \frac{2 E(m)}{3 K(m)} (m - 2) + \frac{m^2}{8} - \frac{m}{3} + \frac{1}{3} \bigg] .
    \label{eq:F_sigma_def}
\end{equation}

We note that in Eq.~\eqref{eq:F_sigma_def} there is a difference in the sign of $F_\mathrm{\sigma}$ with respect to Ref.~\cite{Klein:2021jtd} due to a typo therein. For small values of $m$ (i.e. $m \ll 1$) we have

\begin{equation}
    F_\sigma(m) = \frac{m^2}{256} + O(m^3)
    \label{eq:F_sigma_low_m}
\end{equation}

\noindent and this term vanishes in the $m \to 0$ limit. In App.~\ref{sec:appendix:PN_formulas:SpinAvg-SS} we show how the fully spinning spin-spin coefficients are simplified when using the expressions for $\av{\sigma_i^{(j)}}$ derived in this section.

\subsection{Total angular momentum}
\label{sec:RR:J}

Through radiation reaction, the total angular momentum varies due to the radiation of orbital angular momentum, i.e. \cite{Klein:2021jtd}

\begin{align}
\D \bm{J} =  (\D L) \uvec{L} =  \D \left( \frac{\nu}{y} \right) \uvec{L} = - \frac{\nu \D y}{y^2} \uvec{L} \, .
\label{eq:DvecJ}
\end{align}
We can multiply both sides of Eq.~\eqref{eq:DvecJ} by $\bm{J}$ and use that $\bm{J}\cdot \uvec{L} = J \cos{\theta_L}$ (Eq.~\eqref{eq:EulerAngles:costhL}) to write 

\begin{align}
\D J^2 = - \frac{L \D y}{y}  \left( 2 L + \chi\sub{eff} + \delta \mu \delta \chi \right) \, .
\label{eq:DJ2_SP}
\end{align}
Using Eq.~\eqref{eq:DJ2_def} to write $J^2$ in terms of $\Delta_{J^2}$ (Eq.~\eqref{eq:DJ2_def}) we can further simplify this equation as

\begin{align}
\D \Delta_{J^2} = \delta \mu \frac{\delta\chi_0 - \delta \chi}{y^2} \D y \, ,   
\label{eq:DDJ2_SP}
\end{align}

\noindent where $\delta\chi$ varies in the SP time-scale, following Eq.~\eqref{eq:dchi_sol_avdiff}. To avoid the computationally costly integration of Eq.~\eqref{eq:DDJ2_SP} on the SP time-scale, we separate in a similar way as in Eq.~\eqref{eq:phiz_sol} the secular and periodic parts of $\Delta_{J^2}$, i.e.

\begin{subequations}
\label{eq:DDJ2_prec_avg}
\begin{align}
\Delta_{J^2} & = \Delta_{J^2,0} + \delta \Delta_{J^2} \, , \label{eq:DDJ2_prec_avg:separation}\\
\frac{\D \Delta_{J^2,0}}{\D y} & = \delta \mu \frac{\delta\chi_0 - \av{\delta \chi}}{y^2}, \,  \label{eq:DDJ2_prec_avg:secular}\\
\delta \Delta_{J^2} & = \frac{4\nu \sqrt{Y_3 - Y_-}}{3 (1 - y \chi\sub{eff}) } y^2 \left[ \left( \frac{32}{5} + \frac{28}{5} e^2 \right) + O(y) \right] \nonumber\\
&\times\left\{ E[ \text{am}( \hat{\psi}\sub{p}; m); m] - \frac{E(m)}{K(m)} \hat{\psi}\sub{p} \right\} \, . \label{eq:DDJ2_prec_avg:periodic}
\end{align}    
\end{subequations}

Analyzing Eq.~\eqref{eq:DDJ2_prec_avg:secular}, we have that $\D \Delta_{J^2,0}/\D y \sim O(y^0)$. Therefore, $\Delta_{J^2,0}$ is slowly varying through RR and thus simple to numerically integrate. Furthermore, since from the initial conditions $\Delta_{J^2,0}(t_0) \sim O(y^0)$ (Eq.~\eqref{eq:DJ20_def}), we have that $\Delta_{J^2,0} \sim O(y^0)$ all throughout the evolution. On the other hand, given that $\sqrt{Y_3 - Y_-}\sim O(y^{-1})$ and that the term in brackets containing the elliptic integrals is $\ord{m} \sim O(y^2)$, the periodic part $\delta \Delta_{J^2} \sim O(y^3)$ is much smaller than the secular part $\Delta_{J^2,0}$. 

Using that $\delta \Delta_{J^2} \approx 2 J_0 \delta J /\nu$ to compare Eq.~\eqref{eq:DDJ2_prec_avg:periodic} with the corresponding equations in Ref.~\cite{Klein:2021jtd}, we observe that there is a factor of 2 missing in Ref.~\cite{Klein:2021jtd} due to a typo therein.

\subsection{Effects of the evolution of the elliptic parameter}
\label{sec:RR:psip}

Through radiation reaction the parameter $m$ of Eq.~\eqref{eq:ellip_m_of_argG} evolves with time. Since the period of the spin-precession solution (Eq.~\eqref{eq:dchi_sol_avdiff}) is $2 K(m)$ in $\psi\sub{p}$, the value of $\psi\sub{p}$ accumulated at different times corresponds to a different number of precession cycles. Because of this, when including RR, Eq.~\eqref{eq:hatpsip_def} is not valid to find the phase within the precession cycle $\hat{\psi}\sub{p}$. To solve this, we define a new phase $\bar{\psi}_p$ that is proportional to the accumulated number of cycles. In particular,
\begin{equation}
    \D\bar{\psi}_p = \frac{\pi}{2 K(m)} \D \psi\sub{p} \, .
    \label{eq:barpsip_def}
\end{equation}
\noindent From this, we can recover the correct phase of the Jacobi elliptic functions via

\begin{equation}
    \hat{\psi}\sub{p}(t) = \frac{2 K[m(t)]}{\pi} \hat{\bar{\psi}}_p(t) \, ,
    \label{eq:hatpsip_hatbarpsip}
\end{equation}

\noindent where the angle $\hat{\bar{\psi}}_p$ satisfies the following conditions
\begin{subequations}
\label{eq:hatbarpsip_def}
\begin{align}
&\bar{\psi}_p - \hat{\bar{\psi}}_p = n \pi \text{, for some }  n \in \mathbb{Z} \, , \\
&-\pi < \hat{\bar{\psi}}_p \leq \pi \, ,
\end{align}
\end{subequations}

\noindent and therefore tracks the phase within the precession cycle also when including RR.

\section{Fourier Transform approximation}
\label{sec:FourierTransform}

In the previous sections we laid out the foundations to obtain the time-domain GW polarizations $h_{+,\times}(t)$ emitted by a binary that includes the effects of spin-precession and orbital eccentricity. However, for data-analysis applications such as parameter estimation (PE) or searches, we usually need the frequency-domain polarizations $\tilde{h}_{+,\times}(f)$, obtained with the Fourier transform 

\begin{equation}
    \tilde{h}(f) = \int_{-\infty}^\infty \d t \, h(t) \rme^{2 \pi \rmi f t} \, ,
    \label{eq:FourierTransform_def}
\end{equation}
\noindent where for notational simplicity we have dropped the subscripts $+,\times$. In principle, the Fourier Transform of Eq.~\eqref{eq:FourierTransform_def} can be computed numerically by, for example, using the Fast Fourier Transform (FFT). In practice however, using the FFT to compute $\tilde{h}(f)$ is computationally costly and, unless the strain data $h(t)$ is properly conditioned~\cite{Harris:1978win,LIGOScientific:2019hgc,Talbot:2021igi}, it can introduce its own sources of error. Therefore we seek to approximate the frequency-domain strain analytically.

Our starting point is the expression of Eq.~\eqref{eq:hpc_decomp} for the time-domain GW polarizations $h_{+,\times}(t)$. For notational simplicity we again drop all subscripts and superscripts, i.e.

\begin{equation}
    h(t) = A(t) \rme^{-\rmi \phi(t)} \, ,
    \label{eq:simple_ht}
\end{equation}

\noindent where here $A(t) \equiv \mathcal{A}^{+,\times}_{l,m,n}(t)$ is an amplitude that varies in the spin-precession time scale (i.e. $\dot{A}/A \sim \ord{y^5}$) and $\phi(t) \equiv n\lambda(t) + (m-n)\delta\lambda(t)$ represents the phase of the binary, with $\dot{\lambda}\sim\ord{y^3}$ and $\dot{\delta\lambda} \sim \ord{y^5}$. For non-precessing systems one usually neglects the very slow evolution of $A(t)$, and computes the Fourier transform of Eq.~\eqref{eq:simple_ht} by using the Stationary Phase Approximation (SPA)~\cite{Sathyaprakash:1991mt,Cutler:1994ys,Droz:1999qx}. 
To compute the SPA, we first find the stationary time $t_0$, as
\begin{equation}
    2 \pi f = \dot{\phi}(t_0),
    \label{eq:tSPA_def}
\end{equation}
and Taylor expand the phase around this time 

\begin{align}
    \phi(t) & \approx \phi(t_0) + (t - t_0) \dot{\phi}(t_0) + \frac{1}{2} (t-t_0)^2 \ddot{\phi}(t_0)  \nonumber\\
    & = \phi(t_0) + 2 \pi f (t - t_0) + \frac{1}{2} \frac{(t-t_0)^2}{T_0^2} \, ,
    \label{eq:phiSPA_taylor}    
\end{align}
where we have defined

\begin{equation}
    T_0 = \frac{1}{\sqrt{\left|\ddot{\phi}(t_0)\right|}} \, .
    \label{eq:TSPA_def}
\end{equation}
Substituting this approximation in Eq.~\eqref{eq:FourierTransform_def}, neglecting the time-dependence of the amplitude (i.e. $A(t) \approx A(t_0)$), and analytically computing the Fresnel integral that appears, we obtain the usual SPA approximation

\begin{equation}
    \tilde{h}_\mathrm{SPA}(f) = \sqrt{2 \pi} T_0 A(t_0) \rme^{\rmi (2 \pi f t_0 - \phi(t_0) - \pi/4)} \, .
    \label{eq:hf_SPA}
\end{equation}
However, in Ref.~\cite{Klein:2014bua} it was noted that neglecting the time-dependence of the amplitude is not a good approximation for precessing systems and they introduced a correction to the usual SPA which they named the Shifted Uniform Asymptotics (SUA) method, showing its improved performance over the SPA. In the SUA, Eq.~\eqref{eq:hf_SPA} becomes

\begin{equation}
    \tilde{h}_\mathrm{SUA}(f) = \sqrt{2 \pi} T_0 A^\mathrm{corr}(t_0) \rme^{\rmi (2 \pi f t_0 - \phi(t_0) - \pi/4)} \, ,
    \label{eq:hf_SUA}
\end{equation}

\noindent where the only difference is that we have made $A(t_0) \to A^\mathrm{corr}(t_0)$, defined as

\begin{equation}
    A^\mathrm{corr}(t_0) = \sum_{k=-k_\mathrm{max}}^{k_\mathrm{max}} a_{k,k_\mathrm{max}} A(t_0 + k T_0) \, ,
    \label{eq:Acorr_SUA_Klein}
\end{equation}

\noindent where the constants $a_{k,k_\mathrm{max}}$ are found solving the following linear system of equations:

\begin{subequations}
\begin{align}
    \frac{1}{2} a_{0,k_\mathrm{max}} + \sum_{k=1}^{k_\mathrm{max}} a_{k, k_\mathrm{max}} & = \frac{1}{2} & \, ,  & \; p = 0,  \\
    \sum_{k=1}^{k_\mathrm{max}} \frac{(\rmi \, k^2)^p}{(2 p-1)!!} a_{k, k_\mathrm{max}} & = \frac{1}{2} & \, , & \; 1 \leq p \leq k_\mathrm{max} \, , \label{eq:akkmax_eqs_Klein:linearsystem_k>=1}\\
    a_{-k,k_\mathrm{max}}  = a_{k, k_\mathrm{max}} & &  &
\end{align}
\label{eq:akkmax_eqs_Klein}
\end{subequations}

This differs from the linear system of Ref.~\cite{Klein:2014bua} because in order to simplify Eq.~\eqref{eq:Acorr_SUA_Klein}, we have substituted $\frac{1}{2} a_{k,k_\mathrm{max}} \to a_{k,k_\mathrm{max}}$ for $|k| \geq 1$, and we have defined $a_{-k, k_\mathrm{max}} = a_{k, k_\mathrm{max}}$. This simplification was also implicitly done in Refs.~\cite{Klein:2018ybm,Klein:2021jtd,Arredondo:2024nsl}, but all of them have a typo where the factor of $1/2$ in the right hand side of Eq.~\eqref{eq:akkmax_eqs_Klein:linearsystem_k>=1} is missing.

\section{Waveform Implementation}
\label{sec:waveform}

In this section we aim to bring together all the previous results of the paper to compute the frequency domain waveform, detailing the actual numerical implementation used in our waveform approximant.

\subsection{Solving the dynamics}
\label{sec:waveform:dynamics}

As previously discussed, to describe the dynamics of our eccentric-precessing binary system, we have used the quasi-Keplerian approximation to describe the eccentric orbits and have described with the MSA approximation how these orbits precess due to spin effects. Finally, we take into account the effect of radiation reaction as a slow perturbation of the quasi-Keplerian and MSA parameters. This leads to the following equations of motion that need to be integrated

\begin{subequations}
\begin{align}
\D y &= y^9 \sum_{n \geq 0} \av{a_n} y^n, \label{eq:EFPE_eoms1} \\
\D e^2 &= y^8 \sum_{n \geq 0} \av{b_n}  y^n, \\
\D \lambda &= y^3, \\
\D \delta\lambda &= \frac{k y^3}{1 + k}, \quad k = y^2 \sum_{n \geq 0} \av{k_n} y^n, \label{eq:EFPE_eoms4}\\
\D \Delta_{J^2,0} & = \delta \mu \frac{\delta\chi_0 - \av{\delta \chi}}{y^2} \D y , \label{eq:EFPE_eoms5} \\
\D \bar{\psi}_p &= \frac{3 (1 - y \chi\sub{eff}) y^6}{4} \frac{\pi}{2K(m)} \sqrt{ Y_3 - Y_- }, \label{eq:EFPE_eoms6}\\
\D \phi_{z,0} &= \frac{J y^6}{2} + \frac{3 ( 1 - y \chi\sub{eff}) y^6}{4 \nu K(m)} \left( P_+ + P_- \right),  \label{eq:EFPE_eoms7}\\
\D \zeta_0 &= -\frac{\left(2L + \chi\sub{eff} + \delta\mu \av{\delta\chi} \right) y^6}{4} \nonumber\\
&- \frac{3 (L + \nu \chi\sub{eff})(1 - y \chi\sub{eff}) y^6}{2\nu} \nonumber\\
&+ \frac{3 ( 1 - y \chi\sub{eff}) y^6}{4 \nu K(m)} \left(P_+ - P_- \right), \label{eq:EFPE_eoms8}
\end{align}
\label{eq:EFPE_eoms}
\end{subequations}

\noindent where all the quantities have been previously defined, and to avoid the solution to depend on the spin-precession time-scale we neglect the contribution of $\delta\Delta_{J^2}$ in all of the equations. Given some initial conditions, e.g. as will be described in Sec.~\ref{sec:waveform:ini}, the system of coupled ordinary differential equations of Eq.~\eqref{eq:EFPE_eoms} can be numerically integrated. To do this we write it as 

\begin{equation}
    \frac{\d\bm{\mathcal{V}}}{\d t} = \bm{F}(\bm{\mathcal{V}}) \, ,
    \label{eq:RungeKutta_eoms}
\end{equation}

\noindent where we have defined 

\begin{equation}
    \bm{\mathcal{V}} \equiv \{y, e^2, \lambda, \delta\lambda, \Delta_{J^2,0}, \bar{\psi}_p, \phi_{z,0}, \zeta_0 \} \, .
    \label{eq:RungeKutta_Vdef}
\end{equation}

We integrate Eq.~\eqref{eq:RungeKutta_eoms} using adaptive Runge-Kutta methods, which automatically estimate the time-steps needed to keep the error under a specified tolerance. In particular, we use an explicit Runge-Kutta of order 5(4)~\cite{DORMAND198019}, where the error is controlled assuming accuracy of the fourth-order method, but steps are taken using the fifth-order accurate formula. The result of this 5(4) Runge-Kutta can be expressed as a sequence of times $\{t_j\}$ between each of which we construct a quartic interpolation polynomial of $\bm{\mathcal{V}}$~\cite{Shampine1986}, i.e.

\begin{equation}
    \mathcal{V}_i(t) = \mathcal{V}_i(t_j) +  \sum_{k=1}^{4} Q_{i, j, k} (t - t_j)^k, \quad t_{j} \leq t \leq t_{j + 1} \, .
    \label{eq:RungeKutta_result}
\end{equation}

Writing the solution as in Eq.~\eqref{eq:RungeKutta_result} will prove useful, as it allows for easy computation of the derivatives of $\bm{\mathcal{V}}$, which are for example required when using the SPA (or SUA).

\subsection{Initial conditions}
\label{sec:waveform:ini}

As previously mentioned, to integrate the equations of motion of Eq.~\eqref{eq:EFPE_eoms} we need to give initial conditions for all of the components of $\bm{\mathcal{V}}$. However, many of these variables used in the model have a difficult astrophysical interpretation. Therefore, we initialize our system using the more physically relevant variables specified in table~\ref{table:InitialParameters} and then convert them to the variables of our model. 

\begin{table}[t!]
	\begin{tabular}{| c | c |}
        \hline
        Parameter & Description  \\ 
        \hline
        $m_1$ & Mass of primary object \\
        \hline
        $m_2$ & Mass of secondary object \\
        \hline
        $q_1 $ & \begin{tabular}{@{}c@{}}quadrupole parameter of primary \\ object ($q_1 = 1$ for a black hole) \end{tabular}  \\
        \hline
        $q_2 $ & \begin{tabular}{@{}c@{}}quadrupole parameter of secondary \\ object ($q_2 = 1$ for a black hole) \end{tabular} \\
        \hline
        $e_0$ & \begin{tabular}{@{}c@{}} Initial orbital time-eccentricity \\ in ADM coordinates \end{tabular} \\
        \hline 
        $\chi_{1x,0}, \chi_{1y,0}, \chi_{1z,0}$ & \begin{tabular}{@{}c@{}} Initial dimensionless spin vector of \\ primary object ($\bm{\chi}_1 = \bm{s}_1/\mu_1 = \bm{S}_1/\mu_1^2$ ) \end{tabular}\\
        \hline
        $\chi_{2x,0}, \chi_{2y,0}, \chi_{2z,0}$ & \begin{tabular}{@{}c@{}} Initial dimensionless spin vector of \\ secondary object ($\bm{\chi}_2 = \bm{s}_2/\mu_2 = \bm{S}_2/\mu_2^2$ ) \end{tabular}\\
        \hline
        $\iota_0$ & \begin{tabular}{@{}c@{}} Initial inclination, i.e. the angle between \\ initial orbital angular momentum $\uvec{L}_0$ \\ and vector from binary to observer $\uvec{N}$ \end{tabular}  \\
        \hline
        $d_L$ & \begin{tabular}{@{}c@{}} Initial luminosity distance \\ from binary to observer \end{tabular}   \\
        \hline
        $\lambda_0$ & Initial mean orbital phase \\
        \hline
        $\ell_{0}$ & Initial mean anomaly \\
        \hline
        $f_0^\mathrm{GW,22}$ & \begin{tabular}{@{}c@{}} Initial frequency of the $l=m=2$ \\ quasi-circular GW mode. It is twice the \\ initial orbital frequency ($f_0^\mathrm{GW,22} = 2 f_0^\mathrm{orb}$)  \end{tabular} \\
        \hline
        $f_f^\mathrm{GW,22}$ & \begin{tabular}{@{}c@{}} Final frequency of the $l=m=2$ \\ quasi-circular GW mode.\end{tabular} \\
        \hline
    \end{tabular}
    \caption{Table listing the parameters required to evaluate the model together with a short description for each.}
\label{table:InitialParameters}
\end{table}

Some of the parameters in table~\ref{table:InitialParameters} are directly used by our model, such as the component masses $m_{1,2}$, the component quadrupole parameters $q_{1,2}$, the initial eccentricity $e_0$, the luminosity distance $d_L$ and the initial mean orbital phase $\lambda_0$. Others can be easily related, such as the initial mean anomaly $\ell_0$, with the initial argument of periastron $\delta\lambda_0$ using Eq.~\eqref{eq:dlambda_def}, i.e.

\begin{equation}
    \delta\lambda_0 = \lambda_0 - \ell_0 \, ,
    \label{eq:dlambda_0}
\end{equation}

Or the initial $l=m=2$ quasi-circular GW frequency $f_0^\mathrm{GW,22} = 2 f_0^\mathrm{orb} = \omega/\pi$ with the initial PN parameter $y_0$ using Eq.~\eqref{eq:raw_prec_eqs_defs:y}, i.e.

\begin{equation}
    y_0 = \frac{\left( \pi M f_0^\mathrm{GW,22} \right)^{1/3}}{\sqrt{1 - e_0^2}} \, .
    \label{eq:y0}
\end{equation}

To apply the adaptive Runge-Kutta we also need to specify a maximum time $t_f$ up to which we integrate the equations of motion of Eq.~\eqref{eq:EFPE_eoms}. In practice however, it is more convenient to specify the endpoint of integration in terms of a maximum frequency $f_f^\mathrm{GW,22}$, usually related to the frequency up to which we are analyzing some data, or the frequency up to which we trust the PN approximation used in the inspiral. In this last case, for late times, we could hybridize with a different approximant that is better suited for the merger-ringdown stages of the binary, such as a phenomenological model~\cite{Pratten:2020ceb}, an effective one body model~\cite{Ramos-Buades:2023ehm,Nagar:2024dzj}  or a numerical relativity (NR) surrogate model~\cite{Varma:2019csw,Islam:2021mha}. In terms of the PN parameter $y$, we stop the integration when

\begin{equation}
    y = y_f = \frac{\left(\pi M f_f^\mathrm{GW,22} \right)^{1/3}}{\sqrt{1 - e^2(t_f)}} \, .
    \label{eq:yfinal}
\end{equation}

To impose this condition in an adaptive Runge-Kutta, every time a new time-step $t_{j+1}$ is proposed, we use the interpolation polynomial of Eq.~\eqref{eq:RungeKutta_Vdef} to check if $y(t)=y_f$ is satisfied at any point of the interval $t \in (t_j, t_{j+1}]$. If it is, we terminate the Runge-Kutta and define the final time as the first solution $y(t_f) = y_f$. In the quasi-circular case, we stop trusting the PN approximation for frequencies above that of the Minimum Energy Circular Orbit (MECO)~\cite{Cabero:2016ayq,Pratten:2020ceb}. This point is hard to estimate, as it requires finding where the PN energy $E_\mathrm{PN}(f)$ has its first minimum. Therefore, a stopping point that is more commonly chosen is the frequency of the Innermost Stable Circular Orbit (ISCO)~\cite{Bardeen:1972fi}, above which circular orbits of test particles around Schwarzschild are unstable, plunging to the central black hole. In terms of the PN parameter, this happens at 

\begin{equation}
    y_\mathrm{ISCO} = 6^{-1/2} \, .
    \label{eq:yISCO}
\end{equation}

As is usually done in PN approximants, the time is chosen such that $t=0$ corresponds to the coalescence time~\cite{Schmidt:2017btt}. In the eccentric case, this time is hard to analytically compute to high PN orders. Therefore, for simplicity we use the 0PN time to coalescence $\tau_c$, that can be computed as described in appendix~\ref{sec:appendix:tLO}, where in Eq.~\eqref{eq:tcoal_0PN_y0} we obtain
\begin{align}
    \tau_c(y, e) & = \frac{5}{256} \frac{M}{\nu y^8} \frac{F(e)}{\sqrt{1 - e^2}} \, ,
    \label{eq:tcoal_0PN_main}    
\end{align}
\noindent with $F(e)$ being an $\ord{1}$ function defined by Eq.~\eqref{eq:Fe_def} and shown in Fig.~\ref{fig:Fe_plot}. When we start the Runge-Kutta integration of the equations of motion, we use Eq.~\eqref{eq:tcoal_0PN_main} to set the initial time, i.e.
\begin{equation}
    t_0 = - \tau_c (y_0, e_0) \, .
    \label{eq:t0_RK_ini_guess}
\end{equation}
However, as previously described, when the Runge-Kutta integration terminates, we obtain a value of $t_f$ such that $y(t_f) = y_f$, that will slightly deviate from $-\tau_c (y_f, e_f)$, given that we ignore PN corrections in our initial guess of the coalescence time. Therefore, after finishing the Runge-Kutta integration, we shift the time of the whole solution as
\begin{equation}
    t \to t - t_f - \tau_c (y_f, e_f).
    \label{eq:RK_final_tshift}
\end{equation}
Setting the initial conditions on the angular momenta and orientation of the binary requires more choices of conventions. Throughout our description of the problem we use three different reference frames \cite{Schmidt:2017btt}

\begin{itemize}
    \item The $N$-frame: This is the inertial frame with respect to which the GW polarizations are measured. By definition it places the unit vector $\uvec{N}$ pointing from the binary to the observer in the $\uvec{Z}$ axis. To fix the remaining orientation of this frame, we choose it such that the initial orbital angular momentum $\bm{L}_0$ is on the $\uvec{X}-\uvec{Z}$ plane.
    \item The $J$-frame: This is the frame in which the binary motion is described. By definition, it places the total angular momentum $\bm{J}$ in the $\bm{z}$ axis. To fix the remaining orientation of this frame, we again choose the initial orbital angular momentum $\bm{L}_0$ to be on the $\uvec{x}-\uvec{z}$ plane. As seen in Eq.~\eqref{eq:DvecJ}, the direction of $\bm{J}$ varies in the radiation reaction time-scale, making the $J$-frame non-inertial. However, in this work we are neglecting the evolution of $\uvec{J}$ (i.e. we are doing $\uvec{J}(t) = \uvec{J}_0$), and we therefore consider the $J$-frame to be inertial. While this is usually a very good approximation, it can fail in systems exhibiting strong nutational resonances or transitional precession~\cite{Zhao:2017tro}.
    \item The $L$-frame: This is the non-inertial frame in which the effects of precession on the waveform are minimized~\cite{Schmidt:2010it,Boyle:2011gg}. The Euler Angles $(\phi_z, \theta_L, \zeta)$ are defined such that they rotate the $\bm{L}$-frame to the $\bm{J}$-frame, i.e. for any vector $\bm{v}$
    \begin{equation}
        \begin{pmatrix}
            v_{x,J} \\ v_{y,J} \\ v_{z,J}
        \end{pmatrix}_{J} = 
        \hat{R}_z(\phi_z) \hat{R}_y(\theta_L) \hat{R}_z(\zeta)
        \begin{pmatrix}
            v_{x',L} \\ v_{y',L} \\ v_{z',L}
        \end{pmatrix}_{L} \, ,
        \label{eq:L_to_J_frame}
    \end{equation}
    \noindent where $(\bm{v})_J$ and $(\bm{v})_L$ means that the vector is measured in the $J$-frame and $L$-frames respectively and $\hat{R}_{\uvec{u}}(\alpha)$ are the usual rotation matrices of an angle $\alpha$ around the rotation axis $\uvec{u}$.
\end{itemize}

The orientation of the binary is determined by the initial inclination $\iota_0$, defined as the angle between the initial orbital angular momentum $\uvec{L}_0$ and the vector from binary to observer $\uvec{N}$, therefore

\begin{equation}
    (\uvec{L}_0)_N = 
    \begin{pmatrix}
        \cos\iota_0 \\ 0 \\ \sin\iota_0
    \end{pmatrix} \, ,
    \label{eq:L0_vec_N}
\end{equation}

As is usually done in GW modeling, the input spins are given in terms of the $L$-frame dimensionless spin vectors $\bm{\chi}_i = \bm{S}_i/\mu_i^2$ of the component objects, such that $|\chi_i| \leq 1$ for sub-extremal Kerr Black Holes. The spins are specified in Cartesian components such that $\chi_{i z,0}$ is the component parallel to the orbital angular momentum (i.e. $\chi_{i z,0} = \uvec{L}_0 \cdot \bm{\chi}_{i,0}$). With our conventions, the total angular momentum in the $N$-frame is given by
\begin{align}
    & (\bm{J}_0)_N  = \hat{R}_y (\iota_0) (\bm{L}_0 + \bm{S}_{1,0} + \bm{S}_{2,0})_L \nonumber \\
    & = 
    \begin{pmatrix}
        \cos{\iota_0} & 0 & \sin\iota_0 \\ 0 & 1 & 0 \\ -\sin{\iota_0} & 0 & \cos\iota_0
    \end{pmatrix}
    \begin{pmatrix}
        S_{1 x,0} + S_{2 x, 0} \\ S_{1 y,0} + S_{2 y, 0} \\ \frac{\nu}{y} + S_{1 z,0} + S_{2 z, 0}
    \end{pmatrix} \, .
    \label{eq:J0_vec_N}
\end{align}

If we define ($\theta_{JN}, \phi_{JN}$) as the spherical angles\footnote{We can express any unit vector using two spherical angles ($\theta, \phi$) as $\uvec{v} = (\sin\theta\cos\phi, \sin\theta\sin\phi, \cos\theta)$.} of the unit vector pointing in the direction of the initial total angular momentum $\bm{J}_0$, we can bring this vector to the $z$ axis applying $\hat{R}_y(-\theta_{JN}) \hat{R}_z(-\phi_{JN})$. This almost corresponds to the rotation that transforms from the $N$ to the $J$ frames. We have to add the additional requirement that $(\uvec{L}_0)_J$ is in the $x-z$ plane of the $J$ frame. To impose this we compute the spherical angles $(\theta_{LJN},\phi_{LJN})$ of the unit vector $\hat{R}_y(-\theta_{JN}) \hat{R}_z(-\phi_{JN}) (\uvec{L}_0)_N$, and then, the rotation from the $N$ frame to the $J$ frame is given by

\begin{equation}
    \hat{R}_{NJ} = \hat{R}_z(-\phi_{LJN}) \hat{R}_y(-\theta_{JN}) \hat{R}_z(-\phi_{JN}) \, .
    \label{eq:N_to_J_frame_rot}
\end{equation}

Using this, we can define the angles $(\Theta\sub{p}, \Phi\sub{p})$ that were vaguely introduced in Eq.~\eqref{eq:hlm_def} as the spherical angles of the unit vector from the observer to the binary in the $J$ frame, i.e. $(-\uvec{N})_J$. Therefore, $(\Theta\sub{p}, \Phi\sub{p})$ are defined by

\begin{equation}
    \begin{pmatrix}
        \sin\Theta\sub{p}\cos\Phi\sub{p} \\ \sin\Theta\sub{p}\cos\Phi\sub{p} \\ \cos\Theta\sub{p}
    \end{pmatrix} 
    = \hat{R}_{NJ} 
    \begin{pmatrix}
        0 \\ 0 \\ -1
    \end{pmatrix} \, .
    \label{eq:ThetaPhi_def}
\end{equation}

To integrate the equations of motion of Eq.~\eqref{eq:EFPE_eoms}, we also have to find the value of initial value of $\Delta_{J^2,0}$. To do this, we first compute the initial value of $\Delta_{J^2}$, which is given by Eq.~\eqref{eq:DJ2_supp_def}. To find the initial value of $\Delta_{J^2,0}$ we numerically solve for it using

\begin{equation}
    \Delta_{J^2}(t_0) = \Delta_{J^2,0}(t_0) + \delta\Delta_{J^2}(t_0; \Delta_{J^2,0}(t_0)) \, ,
    \label{eq:DJ200_eqsolve}
\end{equation}

\noindent where we have made explicit that $\delta\Delta_{J^2}$ (given by Eq.~\eqref{eq:DDJ2_prec_avg:periodic}) is being evaluated assuming $\Delta_{J^2} = \Delta_{J^2,0}$, to be consistent with what is later done in the evolution equations. To evaluate $\delta\Delta_{J^2}(t_0)$ we also need the initial value of the precession phase $\psi_{\mathrm{p},0}$, this can be computed for each value of $\Delta_{J^2,0}$ assumed by solving Eq.~\eqref{eq:dchi_sol_avdiff} using $\delta\chi(t_0) = \delta\chi_0$, i.e.

\begin{equation}
    \psi_{\mathrm{p},0} = \pm F\left(\frac{1}{2}\arccos\left( \frac{\delta\chi\sub{av} - \delta\chi_0}{\delta\chi\sub{diff}}\right) ; m_0\right) \, ,
    \label{eq:am_psi_p_0}
\end{equation}

\noindent where $F(\phi; m)$ is the incomplete elliptic integral of the first kind. To choose the correct sign of $\psi_{\mathrm{p},0}$ we look at the derivative of $\D\delta\chi$, which in terms of the MSA solution is given by

\begin{equation}
    \D\delta\chi = 2 \delta\chi\sub{diff} \sin(2\mathrm{am}(\psi\sub{p}; m )) \D \psi\sub{p} \, .
    \label{eq:Ddchi_sol}
\end{equation}
Since $\D \psi\sub{p} \geq 0$ and $\delta\chi\sub{diff} \geq 0$, in order for Eq.~\eqref{eq:Ddchi_sol} to have the same sign for $\D\delta\chi$ as Eq.~\eqref{eq:Ddchi_raw}, we have to choose
\begin{equation}
    \mathrm{sign}\left(\psi_{\mathrm{p},0}\right) = \mathrm{sign}\left( \left( \uvec{L} \times \bm{s}_1 \right) \cdot \bm{s}_2 \right) \, .
    \label{eq:am_psi_p_0_sigm}
\end{equation}
Finally, to integrate the equations of motion of Eq.~\eqref{eq:EFPE_eoms} we are only missing the initial values of the average Euler angles $\phi_{z,0}$ and $\zeta_0$. Given that $\uvec{L}_0$ is on the $x-z$ plane of the $J$-frame, the initial value of $\phi_z$ satisfies $\phi_z(t_0) = 0$, and thus
\begin{equation}
    \phi_{z,0}(t_0) = - \delta\phi_{z}(t_0) \, ,
    \label{eq:phiz_0}
\end{equation}

\noindent which can be computed with Eq.~\eqref{eq:phiz_sol:dphiz}. A choice for the initial value of the Euler angle $\zeta$ corresponds to a choice of the initial $L$-frame with respect to which the input spins are measured. For simplicity, we set $\zeta(t_0) = 0$, i.e.
\begin{equation}
    \zeta_{0}(t_0) = - \delta\zeta(t_0) \, ,
    \label{eq:zeta_0}
\end{equation}
\noindent which can be computed with Eq.~\eqref{eq:zeta_sol:dzeta}. 

We note that the parameter conventions used in this work are largely consistent with those in \texttt{lalsuite}~\cite{lalsuite_code}, with the main difference being the definition of the $L$-frame. In \texttt{lalsuite}, the $x$-axis of the $L$-frame is aligned with the separation vector, which is straightforward to determine for quasi-circular binaries but more involved for eccentric systems as it requires solving the PN quasi-Kepler equation. To avoid this complexity, we instead define the $L$-frame such that $\zeta(t_0) = 0$ (Eq.~\eqref{eq:zeta_0}), initially aligning it with the $J$-frame. This choice affects the perpendicular spin components ($\chi_{ix}$, $\chi_{iy}$), but since these vary over time due to spin precession, they are not astrophysically meaningful. However, approximately conserved precessional quantities, such as $\chi_\mathrm{p}$~\cite{Schmidt:2014iyl,Gerosa:2020aiw}, are the same in both conventions.

\subsection{Computing the waveform}
\label{sec:waveform:waveform}

Using the initial conditions of Sec.~\ref{sec:waveform:ini} we first integrate the equations of motion of Eq.~\eqref{eq:EFPE_eoms} using and adaptive Runge-Kutta of order 5(4). This gives us a grid of times $\{ t_j \}_{j=0}^{N\sub{RK}}$, between each of which all the dynamical variables are described by an interpolation polynomial of the form of Eq.~\eqref{eq:RungeKutta_result}.

The first thing we do is, for each segment $[t_j, t_{j+1}]$, find the Fourier modes that have to be included in the amplitudes given a user defined tolerance $\epsilon_N$. This is found following the procedure described in Sec.~\ref{sec:NewtAmp:Number}, obtaining a set of Fourier modes with $\{(n_i, m_i)\}_{i=1}^{N_\mathrm{F}}$ (Eq.~\eqref{eq:hpc_decomp}). Given that we are only considering the Newtonian amplitudes, $m_i$ can only be $2$, $0$ or $-2$. Therefore, at each segment of the Runge-Kutta we have $N_\mathrm{F}$ modes, each having a GW phase

\begin{align}
    \phi_i(t) & = n_i \lambda(t) + (m_i - n_i) \delta\lambda(t) \nonumber\\
    & = \phi_i (t_j) + \sum_{k=1}^4 Q_{\phi,i,j,k} (t - t_j)^k , \quad t_j \leq t \leq t_{j+1} \, ,
    \label{eq:phi_modes_RK}
\end{align}

\noindent where we have defined

\begin{subequations}
    \begin{align}
        \phi_i (t_j) & = n_i \mathcal{V}_\lambda(t_j) + (m_i - n_i) \mathcal{V}_{\delta\lambda}(t_j)  \, , \\
        Q_{\phi,i,j,k} & = n_i Q_{\lambda,j,k} + (m_i - n_i) Q_{\delta\lambda,j,k} \, .
    \end{align}
\end{subequations}

Given an input frequency $f$, we then have to compute for each mode the corresponding stationary time $t^\mathrm{SPA}_i$ solving Eq.~\eqref{eq:tSPA_def}. Substituting Eq.~\eqref{eq:phi_modes_RK} in Eq.~\eqref{eq:tSPA_def}, we have that $t^\mathrm{SPA}_i$ is the solution to

\begin{equation}
    \omega = \omega_i(t_j) + \sum_{k=1}^3 Q_{\omega,i,j,k} (t^\mathrm{SPA}_i - t_j)^k , \quad t_j \leq t \leq t_{j+1} \, .
    \label{eq:omega_modes_RK}
\end{equation}

\noindent where we have defined $\omega = 2 \pi f$, $\omega_i(t_j) = Q_{\omega,i,j,1}$ and $Q_{\omega,i,j,k} = (1+k) Q_{\phi,i,j,1+k}$. In a Runge-Kutta, the time-steps are chosen such that the difference between the polynomial of Eq.~\eqref{eq:RungeKutta_result} and the true solution, remains very small. In practice, this means that each time interval is by construction sufficiently short that the solution therein is smooth and the contribution of each term in the sum of Eq.~\eqref{eq:RungeKutta_result} is much smaller than the previous. In this context, we can invert Eq.~\eqref{eq:omega_modes_RK} using series reversion~\cite{Abramowitz_and_Stegun}

\begin{align}
    t^\mathrm{SPA}_i(\omega) = t_j + \sum_{k=1}^{N_\mathrm{sr}} Q_{t,i,j,k} (\omega - \omega_i(t_j))^k \, , \nonumber \\ 
    \mathrm{with}\quad \omega_i(t_j) \leq \omega \leq \omega_i(t_{j+1}) \, ,
    \label{eq:tSPA_modes_RK}    
\end{align}

\noindent where a series reversion order of $N_j \geq 4$ is expected to give good results in our case, and the coefficients $Q_{t,i,j,k}$ are computed with usual series reversion tables~\cite{Abramowitz_and_Stegun}. The formulas used are explicitly shown in App.~\ref{sec:appendix:SeriesReversion}. 

Using Eq.~\eqref{eq:tSPA_modes_RK} we can, for any input frequency, easily and efficiently compute the SPA time corresponding to each mode that contributes at that frequency. From Eq.~\eqref{eq:phi_modes_RK}, we can see that this effectively means that we neglect the modes with $n<0$ and with $n=0, m=\{0,-2\}$, as these have monotonously decreasing GW phases and negative frequencies, which means they have no stationary time. These negative frequency modes are needed to make the time-domain polarizations real, but can be neglected when computing the frequency-domain modes.

The SPA duration scale $T^\mathrm{SPA}_i$ can be very easily computed taking the second derivative of $\phi_i(t)$ (Eq.~\eqref{eq:phi_modes_RK}) and substituting it in Eq.~\eqref{eq:TSPA_def}. With this, we can use Eq.~\eqref{eq:hpc_decomp} and the SUA approximation introduced in Sec.~\ref{sec:FourierTransform}, to compute the GW polarizations as

\begin{align}
    \tilde{h}_{+,\times}(f) = & \sum_{i} \sqrt{2 \pi} T^\mathrm{SPA}_i \rme^{\rmi \left(2 \pi f t^\mathrm{SPA}_i - \phi(t^\mathrm{SPA}_i) - \pi/4\right)} \nonumber \\
    & \times  \sum_{k=-k_\mathrm{max}}^{k_\mathrm{max}} a_{k,k_\mathrm{max}} \mathcal{A}^{+,\times}_{2,m_i,n_i}\left(t^\mathrm{SPA}_i + k T^\mathrm{SPA}_i\right) \, .
    \label{eq:hpc_numerical_exact}    
\end{align}

\noindent where we have made explicit that only $l=2$ contributes, since we are using the Newtonian amplitudes, and the sum over $i$ represents the sum over the Fourier modes that contribute at a given frequency.

The only piece missing now is how to compute the amplitudes $\mathcal{A}^{+,\times}_{2,m,n}(t)$, defined in Eq.~\eqref{eq:Almn}. Doing this is straightforward, with the solution of the Runge-Kutta (Eq.~\eqref{eq:RungeKutta_result}) we can obtain the dynamical variables of the binary at each amplitude evaluation time $t$. With these, we can exactly compute the amplitudes using the results in Secs.~\ref{sec:NewtAmp}~and~\ref{sec:MSA}. 

\subsection{Amplitude interpolation}
\label{sec:waveform:interp}

Evaluating the amplitudes in Eq.~\eqref{eq:hpc_numerical_exact} is computationally very expensive, since, as seen in Secs.~\ref{sec:NewtAmp}~and~\ref{sec:MSA} we have to compute many Bessel functions to determine the amplitudes of the Fourier modes $N^{lm}_{n-m}$ (Eqs.~\eqref{eq:N22_j} and \eqref{eq:N20_p}), elliptic functions and integrals to compute the periodic variation of the Euler angles $\delta\phi_z$ (Eq.~\eqref{eq:phiz_sol:dphiz}) and $\delta\zeta$ (Eq.~\eqref{eq:zeta_sol:dzeta}), and we have to evaluate the Wigner $D$-matrices themselves, which for $l=2$, $m = \{0, \pm 2 \}$ are given in appendix~\ref{sec:appendix:WignerD}. Furthermore, as seen in Eq.~\eqref{eq:hpc_numerical_exact}, with the SUA approximation we have to evaluate the amplitudes $2 k_\mathrm{max} + 1$ times per stationary time.

To avoid the computation of the amplitudes to dominate the runtime, we note that they are relatively slow-varying functions of time, and therefore, we can interpolate them. Writing the amplitudes of Eq.~\eqref{eq:Almn} explicitly we have

\begin{equation}
    \mathcal{A}^{+,\times}_{2,m_i,n_i} = N^{2, m_i}_{n_i - m_i}(t) \mathsf{A}^{+,\times}_{2, m_i}(t) \, .
    \label{eq:A2mn}
\end{equation}

Here, we distinguish two terms, the first term ($N^{2, m_i}_{n_i - m_i}(t)$, defined in Eq.~\eqref{eq:Nlm_p}) is related to the amplitude of each Fourier mode, and it varies on the radiation reaction time-scale. The second term ($\mathsf{A}^{+,\times}_{2, m_i}(t)$, defined in Eq.~\eqref{eq:Apc_lm_simp}) is related to how the amplitude evolves as the binary precesses and its frequency increases. Therefore, $\mathsf{A}^{+,\times}_{2, m_i}(t)$ varies on the spin-precession time-scale and, as we see in Eq.~\eqref{eq:A2mn}, it only depends on the $m_i$ of each mode, which can be $0$, $2$ or $-2$. Given these differences, it is natural to interpolate the two terms of Eq.~\eqref{eq:A2mn} separately.

We begin with $N^{2, m_i}_{n_i - m_i}(t)$. As discussed in Sec.~\ref{sec:waveform:waveform}, the Fourier modes $\{(n_i, m_i)\}_{i=1}^{N_\mathrm{F}}$ to include are determined at each interval of the Runge-Kutta and can vary between intervals. Moreover, as shown in Eqs.~\eqref{eq:N20_p}~and~\eqref{eq:N22_j}, these amplitudes depend only on the eccentricity $e^2(t)$. Given that in each segment of the Runge-Kutta, the eccentricity $e^2(t)$ is well approximated by a polynomial as in Eq.~\eqref{eq:RungeKutta_result}, we can expect that $N^{2, m_i}_{n_i - m_i}(e^2(t))$ can also be accurately represented in the same way, i.e.,

\begin{align}
    N^{2, m_i}_{n_i - m_i}(t) = \sum_{k=0}^{4} Q_{N, i, j, k} \left(t - (t_j - k_\mathrm{max} T_i^\mathrm{SPA})\right)^k \, , \nonumber \\ 
    \mathrm{with}\quad t_j - k_\mathrm{max} T_i^\mathrm{SPA} \leq t \leq t_{j+1} + k_\mathrm{max} T_i^\mathrm{SPA} \, ,
    \label{eq:N2mi_pi_inter}
\end{align}

\noindent where we have slightly expanded the Runge-Kutta segment to take into account that we have to perform the SUA time-shifts. To find the coefficients $Q_{N, i, j, k}$ we evaluate $N^{2, m_i}_{n_i - m_i}(t)$ in $N_\mathrm{fit} = 5 + N_\mathrm{extra}$ Chebyshev nodes of the second kind~\cite{Press:2007nr,Trefethen:2013apx} between $t_j - k_\mathrm{max} T_i^\mathrm{SPA}$ and $t_{j+1} + k_\mathrm{max} T_i^\mathrm{SPA}$, i.e.

\begin{equation}
    t_q^\mathrm{fit} = t_\mathrm{start} + \sin^2\left( \frac{q}{2 (N_\mathrm{fit} - 1)} \right) \Delta t, \quad q = 0, \ldots, N_\mathrm{fit} - 1,
    \label{eq:ChebyshevNodes2ndKind}
\end{equation}

\noindent where $t_\mathrm{start} = t_j - k_\mathrm{max} T_i^\mathrm{SPA}$ and $\Delta t = t_{j+1} - t_j + 2  k_\mathrm{max} T_i^\mathrm{SPA}$. We then fit the polynomial of Eq.~\eqref{eq:N2mi_pi_inter} to these points. Note that we add $N_\mathrm{extra}$ points more than strictly needed for interpolation. For small values of $N_\mathrm{extra}$ (e.g. $1$ or $2$), this makes the polynomial fitting more robust and accurate at small additional computational cost.

Next we interpolate the ``precession'' amplitudes $\mathsf{A}^{+,\times}_{2, m}(t)$. We only need to interpolate them for $m = \{0,2\}$, since using Eq.~\eqref{eq:Apc_lm_mode_symtry} we have that $\mathsf{A}^{+,\times}_{2, -2} = (\mathsf{A}^{+,\times}_{2, 2})^{*}$. These precession amplitudes are oscillating functions of time. Part of it is due to the Euler angles $(\phi_z,\theta_L, \zeta)$ oscillating in each precession cycle, having a period of $\pi$ in $\bar{\psi}_p$. The other part of the oscillation is due to the secular evolution of the Euler angles that enter the Wigner matrices, i.e. $D^l_{m', m} \propto \rme^{-\rmi (m' \phi_{z,0} + m \zeta_0) }$. This oscillation has a period of $2 \pi$ in $(m' \phi_{z,0} + m \zeta_0)$. To accurately interpolate the precession amplitudes, we need to evaluate them at least a few times per cycle. To guarantee this, we define 

\begin{subequations}
\begin{align}
    \Delta\varphi_{\mathrm{p},j}(t) & = \sum_{k=1}^4 Q_{\varphi_\mathrm{p},j,k} (t - t_j)^k , \quad t_j \leq t \leq t_{j+1} \, , \\
     Q_{\varphi_\mathrm{p},j,k} & =Q_{\bar{\psi}_\mathrm{p},j,k} + \mathrm{sign}(Q_{\phi_{z0},j,1})Q_{\phi_{z0},j,k}
\end{align}
\label{eq:phip_interp_def}
\end{subequations}

\noindent where the $\mathrm{sign}(Q_{\phi_{z0},j,1})$ factor is introduced in case $\phi_{z,0}$ is decreasing in a segment. Then, we can make sure that the amplitudes $\mathsf{A}^{+,\times}_{2, m}(t)$ are being evaluated more than $N_\mathrm{p}$ times per cycle if in each Runge-Kutta time segment we pick interpolation times $t^\mathrm{interp}_{j,q}$ such that

\begin{equation}
    \Delta\varphi_{\mathrm{p},j}(t^\mathrm{interp}_{j,q}) = \frac{\pi}{N_\mathrm{p}} q , \quad q = 0, \ldots, \left\lfloor \frac{N_\mathrm{p}}{\pi} \Delta\varphi_{\mathrm{p},j}(t_{j + 1}) \right\rfloor,
    \label{eq:Apc_interp_points}
\end{equation}

\noindent where $\lfloor \cdot \rfloor$ denotes the floor operation and Eq.~\eqref{eq:Apc_interp_points} can, in the same way as Eq.~\eqref{eq:omega_modes_RK}, be solved using the series reversion procedure described in appendix~\ref{sec:appendix:SeriesReversion}. Then, we interpolate the amplitudes $\mathsf{A}^{+,\times}_{2, m}(t)$ in the grid of points $\{ t_{j,q}^\mathrm{interp} \}$. In particular, we choose to do a cubic spline interpolation~\cite{Press:2007nr,Scipy:2020} as it will represent oscillatory functions with high accuracy without needing too many interpolation points per cycle $N_\mathrm{p}$.

\section{Testing and validation}
\label{sec:validate}

For the \pyEFPE waveform approximant to be useful in the analysis of GW detector data, it has to satisfy two conditions. First and foremost, it has to accurately describe the GW emission of the system that is being modeled, to guarantee the validity of scientific statements derived from it. However, it also has to be computationally efficient for its application to the data to be technically feasible. In this section we will test how well both conditions are satisfied.

\begin{figure}[t!]
\centering  
\includegraphics[width=0.5\textwidth]{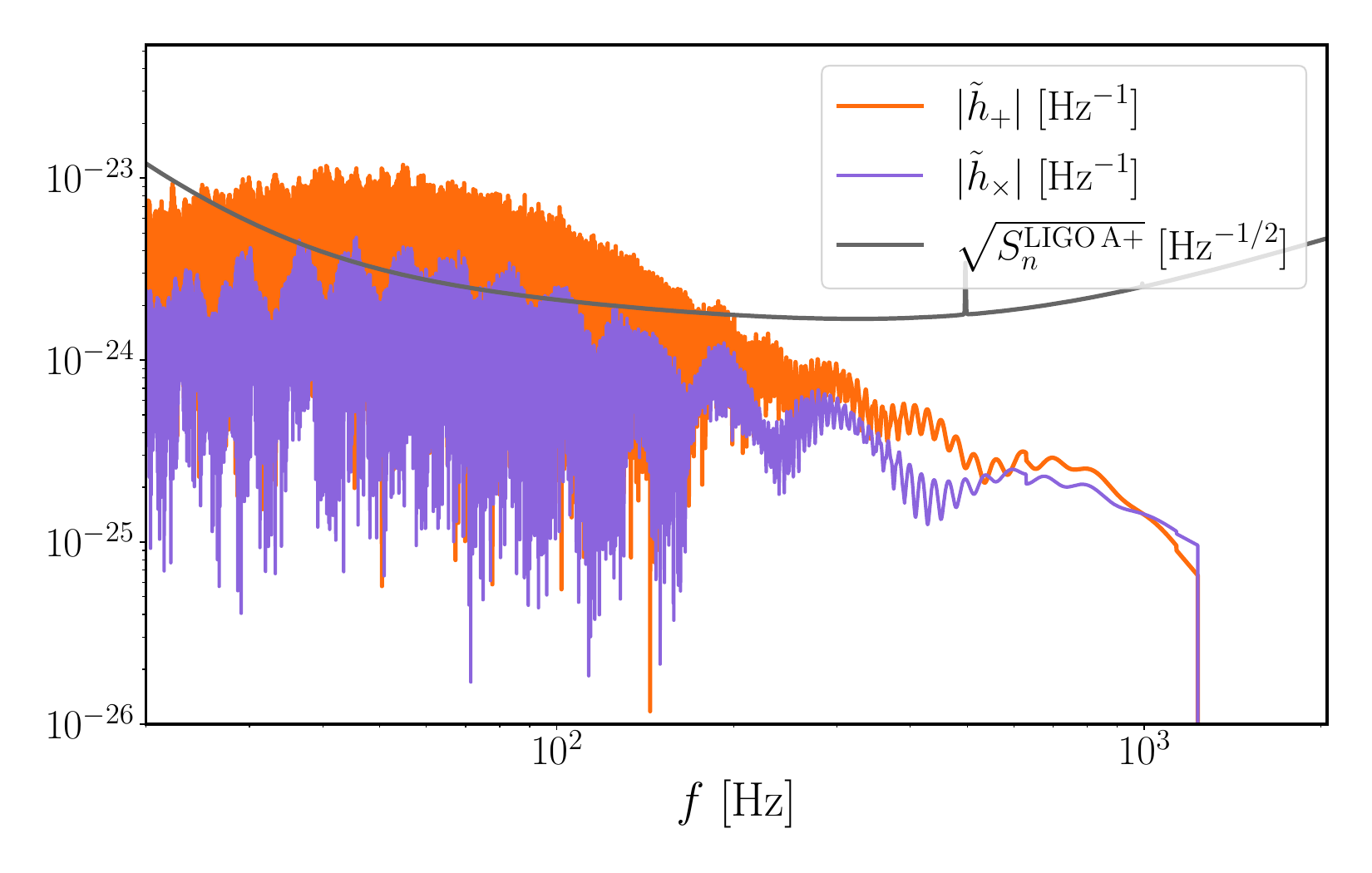}
\caption{\justifying Frequency domain polarizations for a highly-eccentric binary with masses consistent with a BNS: $m_1 = 2.4 M_\odot$, $m_2 = 1.2$, $q_1 = q_2 =1$, $e_0=0.7$, $\bm{\chi}_{1,0} = \{-0.44, -0.26, 0.48\}$, $\bm{\chi}_{2,0} = \{-0.31, 0.01, -0.84\}$, $\iota_0 = \pi/2$, $d_L = 100$~Mpc, $\lambda_0 = \ell_0 = 0$, $f_0^\mathrm{GW,22} = 10\, \mathrm{Hz}$ and $f_f^\mathrm{GW,22} = f_\mathrm{ISCO}^\mathrm{GW,22}$ (for the definition of all these parameters see table~\ref{table:InitialParameters}). The waveform is computed between $f_\mathrm{low}=20\, \mathrm{Hz}$ and $f_\mathrm{high}=4096\, \mathrm{Hz}$ with a frequency resolution of $\Delta f = 1/128\, \mathrm{Hz}$. Furthermore, the amplitude tolerance is taken to be $\epsilon_N = 10^{-3}$ (Eq.~\eqref{eq:Newtonian_orders_needed}) and for the SUA we use $k_\mathrm{max}=3$. We also show the square root of the Advanced LIGO A+ PSD~\cite{Abbott_2020,ObservingScenariosPSDs}. With this PSD, the optimal SNRs of the plus and cross polarizations are $34.8$ and $15.7$, respectively.}
\label{fig:waveform_example_fd}
\end{figure}

\begin{figure*}[t!]
\centering  
\includegraphics[width=\textwidth]{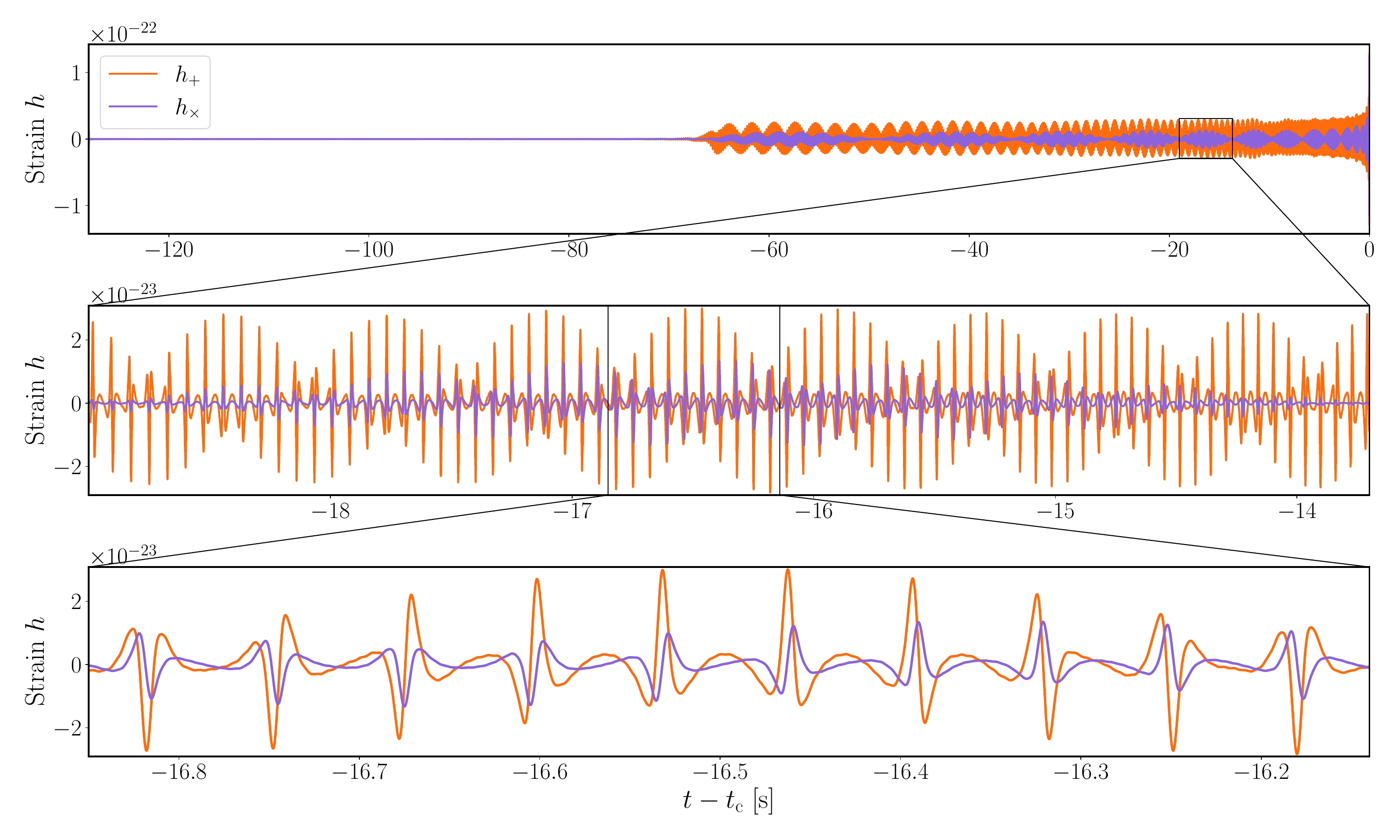}
\caption{\justifying Time-domain polarizations for the same highly eccentric BNS-like system of Fig.~\ref{fig:waveform_example_fd}. We obtain the time-domain waveform by doing the iFFT of the frequency-domain polarization shown in Fig.~\ref{fig:waveform_example_fd}, using that $\tilde{h}(-f) = \tilde{h}^{*}(f)$. In the top panel we show the full time-domain waveform, while in the middle and bottom panels we zoom into a precession cycle and a periastron advance cycle respectively. }
\label{fig:waveform_example_td}
\end{figure*}

We start by checking that \pyEFPE gives the expected waveform phenomenology. To this end, in Fig.~\ref{fig:waveform_example_fd} we show a sample waveform for a highly-eccentric system with masses consistent with a binary neutron star (BNS)\footnote{Note that while the masses are consistent with a BNS, the current model neglects tidal effects.}
, with the parameters listed in the figure caption. Given the very high initial eccentricity ($e_0 = 0.7$) and the tolerance in the amplitude of $\epsilon_N = 10^{-3}$, as seen in Fig.~\ref{fig:Newtonian_orders_needed}, at initial times (low frequencies) we have up to $40$ Fourier modes contributing. When computing the amplitude of the frequency-domain waveform, these Fourier modes interfere with each other, and because of this, the GW polarizations of Fig.~\ref{fig:waveform_example_fd} present very rapid oscillations. 
However, as the system inspirals, the eccentricity rapidly decreases~\cite{Peters:1963ux}, until only the Fourier mode with $l=m=n=2$ contributes, and we obtain the smooth amplitudes observed at later times. 
The system of Fig.~\ref{fig:waveform_example_fd} is chosen to be initially edge-on (i.e. $\iota_0 = \pi/2$) to maximize the effect of precession on the $\times$ polarization of the $l=m=2$ GW mode. 
Indeed, in Fig.~\ref{fig:waveform_example_fd} we can see how, on top of the rapid oscillations previously mentioned, the amplitude of the $\times$ polarization has a slow modulation which is due to spin-precession. 

The physics at play in the binary are clearer when looking at the time-domain polarizations. To determine these, we compute the inverse Fast Fourier Transform (iFFT) of the frequency-domain polarizations in Fig.~\ref{fig:waveform_example_fd}, obtaining the result shown in Fig.~\ref{fig:waveform_example_td}. In the top panel we observe that, even though the segment duration is $T = 128\, \mathrm{s}$ (since the frequency resolution in Fig.~\ref{fig:waveform_example_fd} is $\Delta f = 1/128\, \mathrm{Hz}$), the waveform is different from zero only for $(t - t_\mathrm{c})>-\tau_\mathrm{c}$, with $\tau_\mathrm{c} = 66.1\, \mathrm{s}$ being the coalescence time of the system from the chosen initial frequency $f_0^\mathrm{GW,22} = 10\, \mathrm{Hz}$. In the top panel of Fig.~\ref{fig:waveform_example_td} we also observe how the signal slowly evolves due to radiation reaction, increasing in frequency and amplitude and decreasing in eccenctricity. In this panel we can also clearly observe the modulation of the $\times$ amplitude due to spin-precession, that makes the inclination vary with time, and when the system is edge-on ($\iota = \pi/2$), the $\times$ polarization vanishes. To better observe this effect, in the middle panel we zoom into a precession cycle. Here we can start to discern how the waveform is composed of a series of spikes that correspond to a burst of GWs emitted each time the binary passes through periastron. We also see how the phase of each burst is different at each periastron passage, having a periodic variation due to the effect of periastron advance. In the bottom panel we zoom into one of these periastron advance cycles to more clearly see the periodic change in phase and the characteristic shape of each GW burst, that requires many Fourier modes to be described.

\subsection{Waveform comparisons}
\label{sec:validate:comp}

In this sub-section we will test the accuracy and computational speed of \pyEFPE. Given that there are no widely available waveforms that include both orbital eccentricity and spin-precession, we will check that \pyEFPE reduces to the correct limits when we turn one (or both) of these effects. Instead of working with the two polarizations, $h_+$ and $h_\times$, we compare the strains $h$ along a given direction forming an angle of $\psi$ with the polarization frame, i.e.

\begin{equation}
    h = \cos{(2\psi)} h_+ + \sin{(2\psi)} h_\times \, ,
    \label{eq:h_from_pols}
\end{equation}

\noindent where $\psi$ is the polarization angle. As is standard practice, to measure how similar two strains $h_1$ and $h_2$ are, we compute their match $\mathcal{M}$~\cite{Sathyaprakash:1991mt,Finn:1992xs} as
\begin{equation}
    \mathcal{M}(h_1, h_2) = \frac{\langle h_1 , h_2 \rangle}{\sqrt{\langle h_1, h_1 \rangle \langle h_2, h_2 \rangle}} \, ,
    \label{eq:SimpleMatchDef}
\end{equation}
\noindent where we have introduced the noise-weighted inner product $\langle \cdot, \cdot \rangle$, defined by the overlap integral \cite{Sathyaprakash:1991mt,Finn:1992xs}

\begin{equation}
    \langle a , b \rangle = 4 \mathrm{Re}\left\{ \int_{f_\mathrm{min}}^{f_\mathrm{max}} \frac{\tilde{a}^{*}(f) \tilde{b}(f)}{S_n(f)} \d f  \right\} \, ,
    \label{eq:InnerProdDef}
\end{equation}

\noindent where $S_n(f)$ is the one-sided noise Power Spectral Density (PSD) of the detector. Unless otherwise specified, we use the projected PSD for LIGO A+~\cite{Abbott_2020,ObservingScenariosPSDs}, shown in Fig.~\ref{fig:waveform_example_fd}. Given that $\langle \cdot, \cdot \rangle$ is an inner product, by the Cauchy–Schwarz inequality we have that $\mathcal{M} \leq 1$, with $\mathcal{M} = 1$ only if $\tilde{h}_1 \propto \tilde{h}_2$. Therefore, to measure the difference between two waveforms, we use the mismatch $\mathcal{MM}$, defined as

\begin{equation}
    \mathcal{MM}(h_1, h_2) = 1 - \mathcal{M}(h_1, h_2) \, .
    \label{eq:SimpleMisMatchDef}
\end{equation}

When comparing different waveform approximants, it is sometimes hard to choose both waveforms to represent the same physical system due to differences in conventions, for example in the reference orbital phase $\phi_0$, the reference time $t_0$, the polarization angle $\psi$ or rigid rotations of the in-plane spins by an angle $\phi_{S}$. Since the variables previously mentioned have small astrophysical importance, it is standard practice to minimize the mismatch over them when comparing different waveform models. Therefore, we define

\begin{equation}
    \overline{\mathcal{MM}}(h_1, h_2) = \min_{\phi_0, t_0, \psi, \phi_S} \mathcal{MM}(h_1, h_2)  \, .
    \label{eq:MinMisMatchDef}
\end{equation}

Following Ref.~\cite{Harry:2016ijz}, we analytically optimize over the polarization angle $\psi$ and numerically optimize over the reference time $t_0$ using the FFT. To numerically optimize over the reference phase $\phi_0$ and the rotation angle of the spins $\phi_S$ we use a brute force method.

As is conventional, we approximate the frequency integrals of Eq.~\eqref{eq:InnerProdDef} by summing over the Fourier frequencies assuming a segment of duration $T$ \cite{Talbot:2018cva}, i.e.

\begin{equation}
    \langle a , b \rangle = 4 \mathrm{Re}\left\{ \sum_{k= \lceil f_\mathrm{min}/\Delta f \rceil}^{\lfloor f_\mathrm{max}/\Delta f \rfloor} \frac{\tilde{a}^{*}(k \Delta f) \tilde{b}(k\Delta f)}{S_n(k \Delta f)} \Delta f  \right\} \,, 
    \label{eq:InnerProdDiscrete}
\end{equation}

\noindent where 

\begin{equation}
    \Delta f = \frac{1}{T},
    \label{eq;Df_FFT}
\end{equation}

\noindent is the frequency resolution of the FFT. The segment duration $T$ is chosen such that it contains the full signal studied. For the mismatch computations performed in this section, we use $f_\mathrm{min} = f_0^\mathrm{GW,22} = 20~\mathrm{Hz}$. To compute the duration of the signals, an upper bound is given by~\cite{Cutler:1994ys}

\begin{align}
    \tau_c &\lesssim \frac{5}{256} \left( \frac{G \mathcal{M}_c}{c^3} \right)^{-5/3} \left(\pi f_0^\mathrm{GW,22} \right)^{-8/3} 
    \, ,
    \label{eq:DurationCBC} 
\end{align}

\noindent where, as seen in Eq.~\eqref{eq:tcoal_0PN_f}, for systems with large eccentricities the duration is significantly shorter~\cite{Peters:1963ux}, and we have introduced the chirp mass $\mathcal{M}_c$,

\begin{equation}
    \mathcal{M}_c = \frac{(m_1 m_2)^{3/5}}{(m_1 + m_2)^{1/5}} \, .
    \label{eq:chirp_mass}
\end{equation}

\begin{table}[t!]
\centering
\begin{tabular}{c | c  c  }
$T~[\mathrm{s}]$ &  \multicolumn{2}{c}{$\mathcal{M}_c~[M_\odot]$} \\
{}   & Min.   & Max.  \\ 
\hline
\hline
4 & 12 & 20  \\
8 & 8 & 12  \\
16 & 5 & 8  \\
32 & 3.3 & 5  \\
64 & 2.2 & 3.3  \\
128 & 1.4 & 2.2  \\
256 & 0.95 & 1.4  \\
\hline
\end{tabular}
\caption{Segment durations $T$ studied in this section and the chirp mass range used for each one.}
\label{table:ChirpMassRanges}
\end{table}

In table~\ref{table:ChirpMassRanges} we show the different segment durations considered in this section with their corresponding chirp mass ranges. When testing the waveforms, we take random samples uniformly distributed in each chirp mass range. We always sample uniformly in mass ratio $q = m_2/m_1 \in [0.05, 1]$, the cosine of the reference inclination $\cos{\iota_0} \in [-1, 1]$, the reference phase $\phi_0 = \lambda_0 \in [0 , 2\pi]$ and the polarization angle $\psi \in [0, \pi]$. In all of the comparisons we evaluate our model assuming black holes (i.e. $q_1 = q_2 =1$) and simulating the \pyEFPE dynamics to the ISCO (i.e. $f_f^\mathrm{GW,22} = f_\mathrm{ISCO}^\mathrm{GW,22}$).
The distributions of the spins and eccentricities as well as maximum integration frequency $f_\mathrm{max}$ will be different for the different comparisons that will be considered. 

Finally, unless otherwise specified, we use an amplitude tolerance of $\epsilon_N = 10^{-3}$ (see Eq.~\eqref{eq:Newtonian_orders_needed}), for the SUA we use $k_\mathrm{max}=3$ (see Eq.~\eqref{eq:Acorr_SUA_Klein}), and interpolate the amplitudes using $N_\mathrm{extra} = 2$ (see Eq.~\eqref{eq:ChebyshevNodes2ndKind}) and $N_\mathrm{p} = 40$ (see Eq.~\eqref{eq:Apc_interp_points}). These are the default values of \pyEFPE, as they are found to be a good compromise between accuracy and computational efficiency. 
For the different waveforms we compare with, we use their publicly available \texttt{lalsuite}~\cite{lalsuite_code} implementation.

\subsubsection{Amplitude Interpolation}
\label{sec:validate:comp:interp}

We start comparing \pyEFPE with and without interpolating the amplitudes as described in Sec.~\ref{sec:waveform:interp}. For the amplitude interpolation to be useful, it has to accurately represent the exact waveform as well as speed up the evaluation of the model. To test the accuracy of the interpolation, we compute the mismatch between the exact and interpolated waveforms using Eq.~\eqref{eq:SimpleMisMatchDef}. We do not minimize the mismatch over any parameter, since, given a set of parameters, the interpolated and exact versions of \pyEFPE should represent the same system. To measure the speedup, we just compute the ratio between the runtime of \pyEFPE with the exact and interpolated amplitudes, i.e.

\begin{equation}
    \mathrm{Speedup} = \frac{T_\mathrm{R,exact}}{T_\mathrm{R,interpolated}} \, .
    \label{eq:Speedup}
\end{equation}

\begin{figure}[t!]
\centering  
\includegraphics[width=0.5\textwidth]{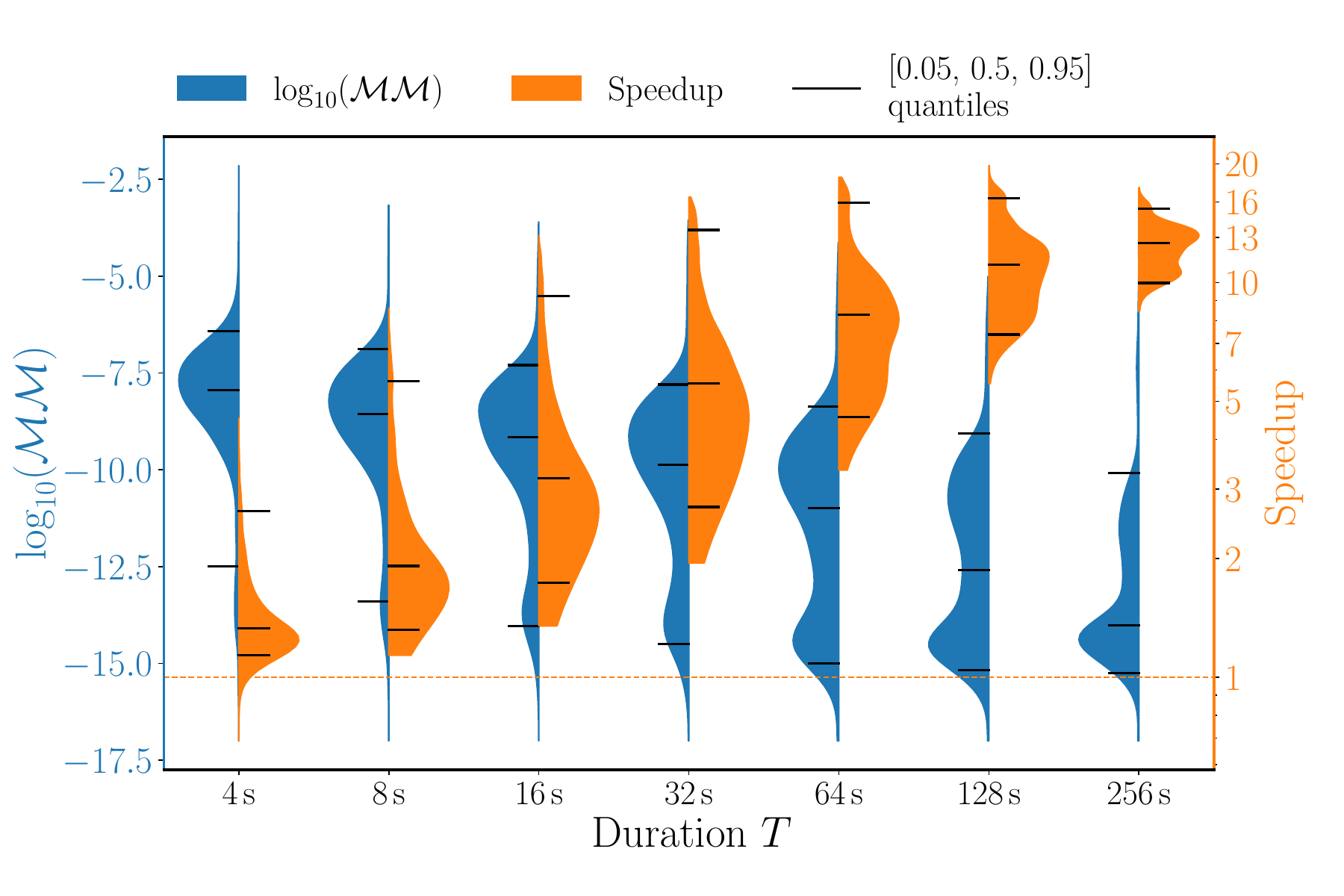}
\caption{\justifying Violin plots showing the distribution of the mismatches (left part, Eq.~\eqref{eq:SimpleMisMatchDef}) and Speedup (right part, Eq.~\eqref{eq:Speedup}) when comparing \pyEFPE with and without amplitude interpolation as a function of the segment duration $T$ or, equivalently, the chirp mass $\mathcal{M}_c$ range as specified in table~\ref{table:ChirpMassRanges}. Each distribution consists of 2000 random samples drawn from the distributions as described in Sec.~\ref{sec:validate:comp:interp}. 
}
\label{fig:pyEFPE_comparison_MM_Speedup_violins}
\end{figure}

In Fig.~\ref{fig:pyEFPE_comparison_MM_Speedup_violins} we show the distribution of mismatches and speedups for 2000 samples at each segment duration (and corresponding chirp mass range) of table~\ref{table:ChirpMassRanges}. For the distribution of the spins, we have chosen isotropic orientations and uniform spin magnitudes $\chi_i \in [0, 0.9]$, while for the eccentricity we have chosen a uniform distribution $e_0 \in [0, 0.7]$. 

In Fig.~\ref{fig:pyEFPE_comparison_MM_Speedup_violins} we observe that the mismatches are usually very small with the $95\%$ quantile being below $10^{-6}$ in all cases. However,  there is a small tail extending to mismatches of up to $\sim 10^{-2}$. These high mismatches correspond to systems where the MSA approximation develops a discontinuity that makes the interpolation of the amplitudes fail. These discontinuities can happen because of very abrupt transitional precession or because we pass through a point with $|\cos{\theta_L}| = 1$, which causes a physical discontinuity in $\D\phi_z$ and $\D\zeta$ (see Eq.~\eqref{eq:Dphiz}) that then becomes a spurious discontinuity in $\delta\phi_z$ and $\delta\zeta$ when doing the MSA (see Eq.~\eqref{eq:phiz_sol:dphiz} and Eq.~\eqref{eq:zeta_sol:dzeta}). 

\begin{figure}[t!]
\centering  
\includegraphics[width=0.5\textwidth]{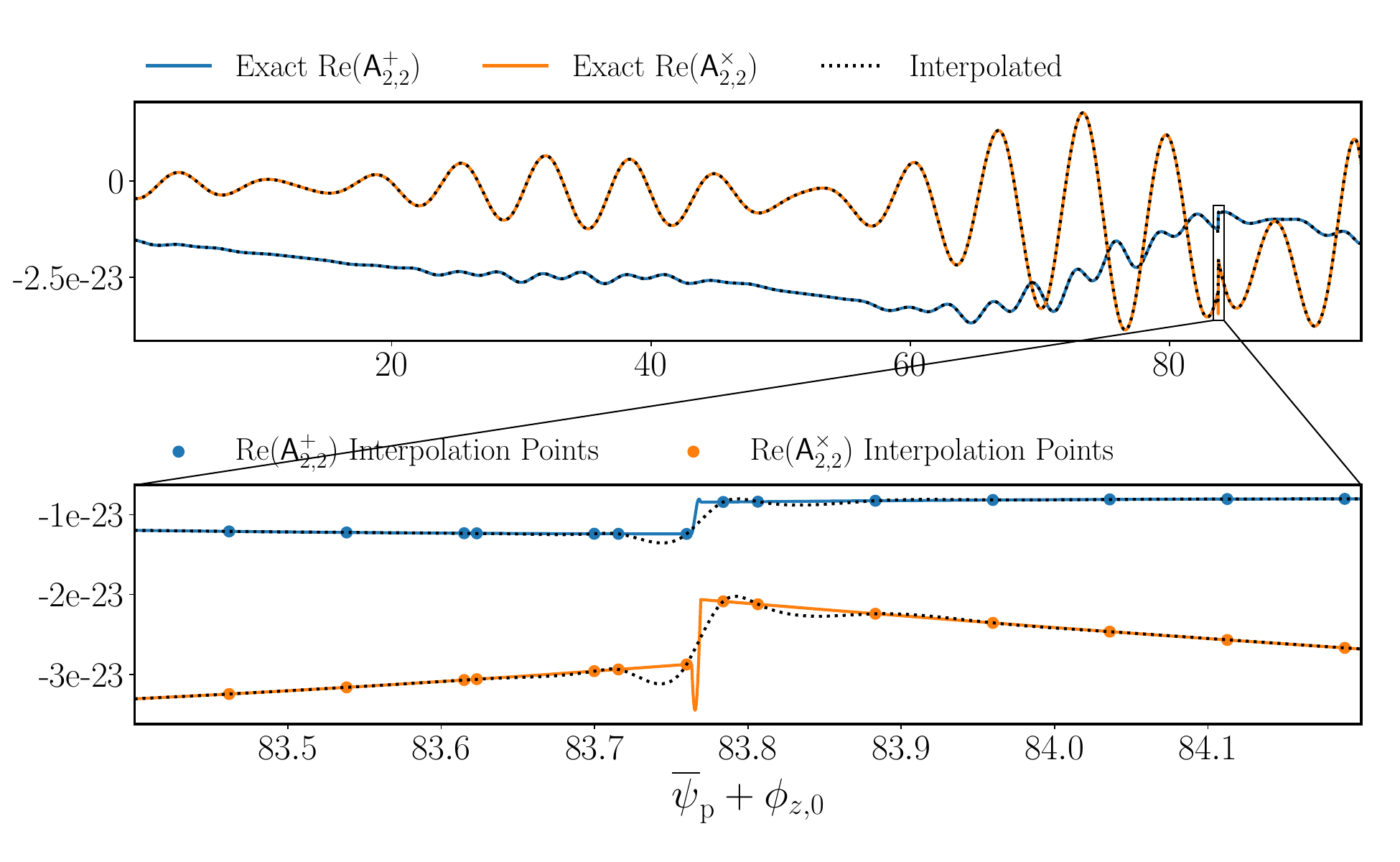}
\caption{\justifying Real part of the exact and interpolated $\mathsf{A}^{+,\times}_{2, 2}$ amplitudes (Eq.~\eqref{eq:Apc_lm_simp}) for the sample with the largest mismatch of the $T = 256\,\mathrm{s}$ distribution shown in Fig.~\ref{fig:pyEFPE_comparison_MM_Speedup_violins}, having $\mathcal{MM} = 8.8\cdot 10^{-5}$. We plot it as a function of $\overline{\psi}_\mathrm{p} + \phi_{z,0}$, since this is the variable we are doing a semi-regular grid interpolation on (see Eq.~\eqref{eq:phip_interp_def}). In the bottom panel we zoom into the amplitude discontinuity that is making the interpolation inaccurate and plot the interpolation points. The parameters of the sample shown are $m_1 = 1.656 M_\odot$, $m_2 = 1.331 M_\odot$, $e_0 = 0.596$, $\bm{\chi}_{1,0} = \{-0.092, -0.741, 0.501\}$, $\bm{\chi}_{2,0} = \{0.358, -0.118, 0.260\}$, $\iota_0 = 1.478$rad, $\lambda_0 = 1.199$rad, $\ell_0 = 0.716$rad and is located at $d_L = 10$Mpc.  }
\label{fig:amplitude_discontinuity_example}
\end{figure}

In Fig.~\ref{fig:amplitude_discontinuity_example} we show an example of such a discontinuity by plotting the exact and interpolated $\mathsf{A}^{+,\times}_{2,2}$ amplitudes for the case with the largest mismatch of the $T=256~\mathrm{s}$ samples, having the parameters listed in the caption. While it is hard to interpolate such narrow features in the amplitudes, these deviations are not a concern, since they happen where the MSA approximation fails and the model is not expected to be accurate. 

Going back to Fig.~\ref{fig:pyEFPE_comparison_MM_Speedup_violins}, we observe that the larger the segment duration, the larger the speedup, going from $\sim$ 1-2 for the $T=4\, \mathrm{s}$ set to $\sim$ 10-15 for the $T=256\, \mathrm{s}$ set. This increase is due to the fact that, even though interpolating the amplitudes consistently takes around 20 times less time than evaluating the exact expressions, shorter waveforms evaluate the amplitudes at fewer points, representing a smaller fraction of the total runtime. For the same reason, within each segment duration there is a large spread of speedups, as samples with high eccentricity have a large amount of Fourier modes contributing, needing more amplitude evaluations, and benefiting more from the interpolation speeedup. Instead, for waveforms with few amplitude evaluations, the runtime is dominated by the adaptive Runge-Kutta used to integrate the equations of motion of Eq.~\eqref{eq:EFPE_eoms}. We observe that for some of the waveforms with durations of $T=4\, \mathrm{s}$, the speedup can even be smaller than 1, due to the overhead of setting up the amplitude interpolation.

While the model's runtime heavily depends on the hardware it runs on, it is useful to give an idea of the computational cost of the waveform and determine if it is suitable for data analysis applications. A typical number of waveform evaluations $N_\mathrm{wf}$ in GW parameter estimation (PE) is $N_\mathrm{wf}\sim\ord{10^8}$. Therefore, if the mean waveform evaluation time is $T_\mathrm{wf}$ and the analysis is parallelized on $N_\mathrm{cores}$, the PE runtime is

\begin{equation}
    T_\mathrm{PE} = 1.8\,\mathrm{days} \left(\frac{N_\mathrm{wf}}{10^8} \right) \left( \frac{T_\mathrm{wf}}{0.1\,\mathrm{s}} \right) \left( \frac{64}{N_\mathrm{cores}} \right) \, ,
    \label{eq:T_PE_estimation}
\end{equation}

\noindent assuming the waveform evaluation dominates the computational cost of PE, which is usually the case. With Eq.~\eqref{eq:T_PE_estimation} we have an estimate of the waveform runtime necessary for PE to be computationally feasible,  usually requiring that $T_\mathrm{wf} \lesssim 1 \,\mathrm{s}$, for PE analyses to finish in a time-scale of weeks. 

\begin{figure}[t!]
\centering  
\includegraphics[width=0.5\textwidth]{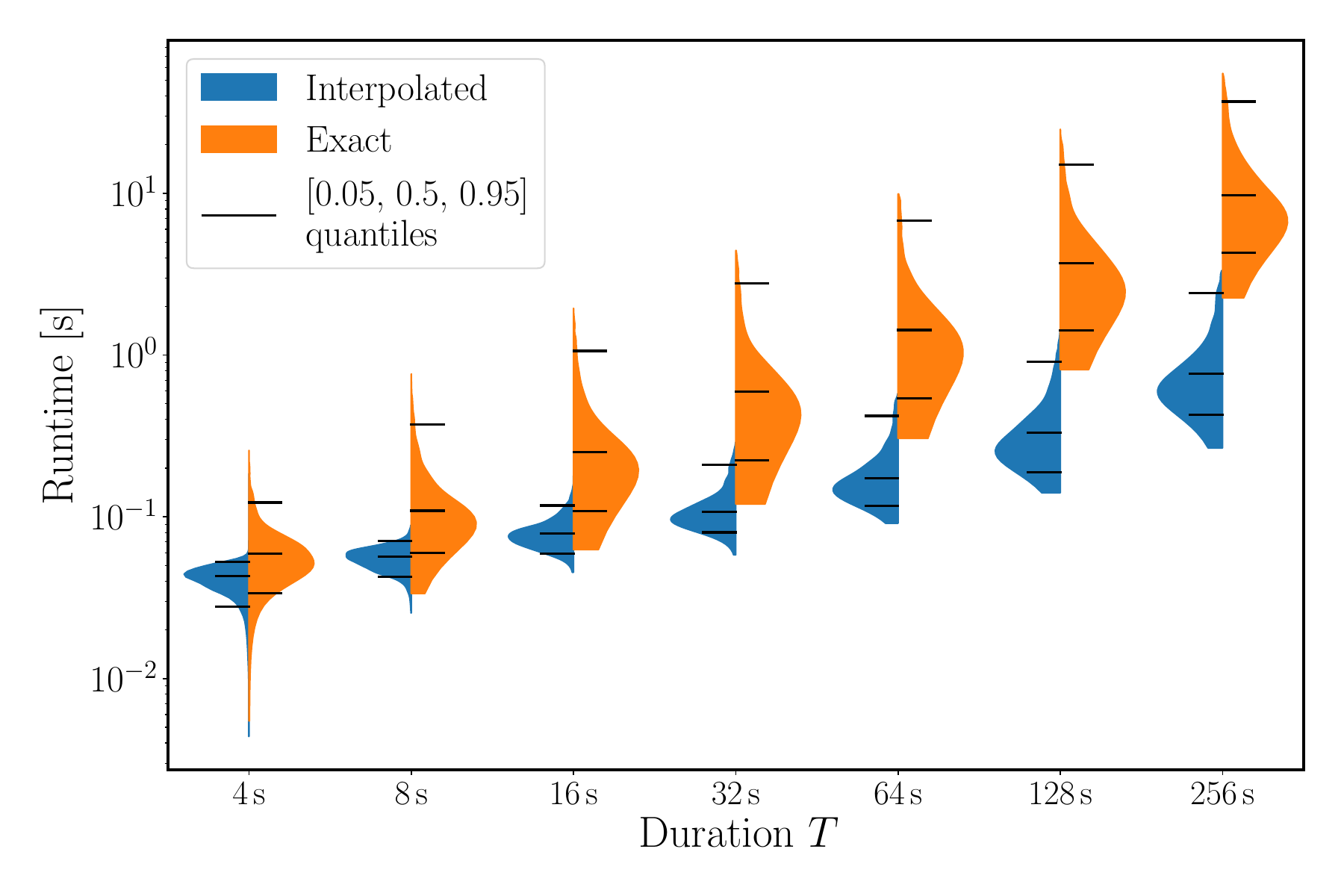}
\caption{\justifying Violin plots showing the distribution of the \pyEFPE runtime with interpolated amplitudes (left part) and exact amplitudes (right part) as a function of the segment duration $T$ or, equivalently, the chirp mass $\mathcal{M}_c$ range as specified in table~\ref{table:ChirpMassRanges}. 
The runtime distributions shown, use the same samples as in Fig.~\ref{fig:pyEFPE_comparison_MM_Speedup_violins}, and were computed on a single core of an \textit{Intel Core i7-1185G7} laptop processor.}
\label{fig:pyEFPE_comparison_runtime_violins}
\end{figure}

In Fig.~\ref{fig:pyEFPE_comparison_runtime_violins}, we show, for the same samples as in Fig.~\ref{fig:pyEFPE_comparison_MM_Speedup_violins}, the \pyEFPE runtimes on a single core of an \textit{Intel Core i7-1185G7} laptop processor. As expected, we observe that, for longer duration signals the runtime is larger, as the waveform has to be sampled at more frequencies, and for the smaller masses the Runge-Kutta integrates over longer times. For the interpolated version of \pyEFPE we see that the waveform evaluation is mostly under $\sim 0.1 \,\mathrm{s}$ for $T \leq 32\, \mathrm{s}$ and under $\sim 0.5 \,\mathrm{s}$ for $T \geq 64\, \mathrm{s}$, meaning that PE studies are computationally feasible in a time-scale of days. However, without the amplitude interpolation, each waveform evaluation can take more than $1\,\mathrm{s}$ for $T \geq 64\,\mathrm{s}$, leading to PE analyses entering time-scales of weeks to months. Therefore, the amplitude interpolation will be crucial to make PE studies computationally feasible.

\subsubsection{Comparison with quasi-circular spin-aligned models}
\label{sec:validate:comp:SAQC}

To validate \pyEFPE, we start by comparing it with other PN inspiral waveforms in the simplest scenario of quasi-circular orbits and spins aligned with the orbital angular momentum, minimizing precession effects. Therefore, we fix $e=\chi_{i\perp,0}=0$ for all samples tested and choose uniform distributions for the spin components aligned with the orbital angular momentum $\chi_{i z,0} \in [-0.9, 0.9]$. To avoid boundary effects, we perform the overlap integrals of Eq.~\eqref{eq:InnerProdDef} up to a frequency of $f_\mathrm{max} = 0.8 f_\mathrm{ISCO}$. Unless otherwise specified, for the PN models we take the same PN orders as for \pyEFPE, i.e. 3PN for the spinning and non-spinning parts of the phasing and 0PN for the amplitude.

\begin{figure}[t!]
\centering  
\includegraphics[width=0.5\textwidth]{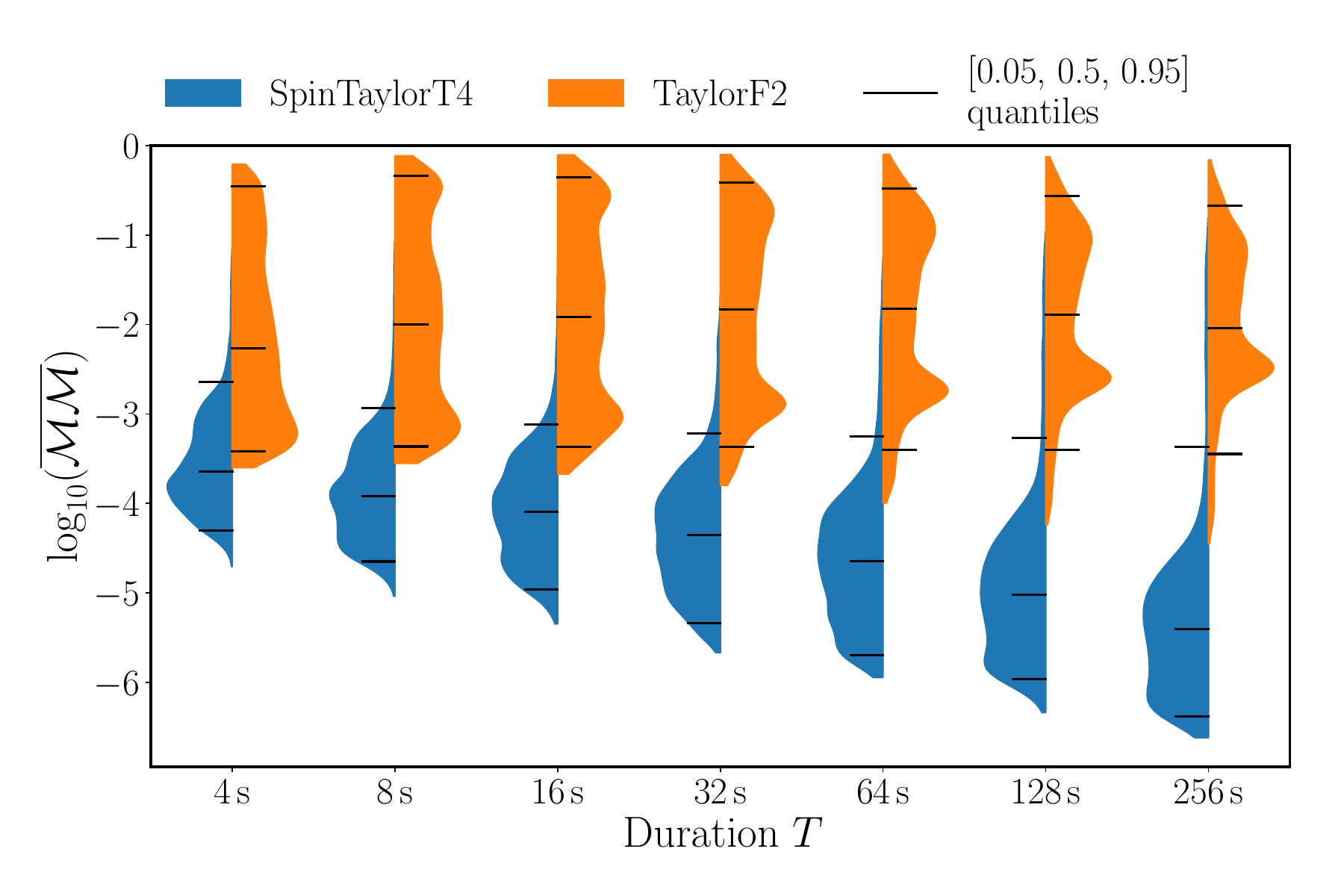}
\caption{\justifying Violin plots showing the distribution of the mismatch $\overline{\mathcal{MM}}$ (Eq.~\eqref{eq:MinMisMatchDef}) between \pyEFPE and \STfour (left part) and \Ftwo (right part) in the aligned spin case as a function of the segment duration $T$, or equivalently the chirp mass $\mathcal{M}_c$ range as specified in table~\ref{table:ChirpMassRanges}. Each distribution shown consists of 10000 random samples drawn from the distributions described in the text.}
\label{fig:MM_violins_noprec_SpinTaylorT4_TaylorF2}
\end{figure}

The first model we compare with is \STfour~\cite{lalsuite_code,Buonanno:2009zt,Sturani:2015STA,Isoyama:2020lls}, a time-domain quasi-circular precessing model that integrates the same PN evolution equations as \pyEFPE in the $e \to 0$ limit~\cite{Buonanno:2009zt}. However, to account for precession, instead of the MSA approximation it numerically solves the quasi-circular version of the precession equations of Eq~\eqref{eq:raw_prec_eqs}, taking into account corrections up to 3PN (instead of the 2PN order used by \pyEFPE). 
In Fig.~\ref{fig:MM_violins_noprec_SpinTaylorT4_TaylorF2} we observe that, in spite of their differences, the mismatches between \pyEFPE and \STfour are very small, with the bulk of the distributions being below values of $\overline{MM}\lesssim 10^{-3}$, and mismatches becoming smaller as the signals become longer. 
These mismatches are mostly due to \STfour being converted to the frequency-domain using an FFT, whereas \pyEFPE uses the SUA to obtain the frequency-domain waveform~\cite{Klein:2014bua}. 
We can also see that the mismatch distributions have very long tails extending to $\overline{MM} \sim 10^{-1}$, corresponding to systems that have
\begin{equation}
    \left(\frac{\delta\mu}{y}\right)^2 - 2 \delta \chi_0 \frac{\delta\mu}{y}  + \lvert\bm{s}_{1,0} + \bm{s}_{2,0} \rvert^2 \leq 0 \, ,
    \label{eq:MSAInstabilityCondition}
\end{equation}
\noindent for some $y \in [y_0, y_f]$. When this happens, $p_\parallel$ in Eq.~\eqref{eq:depressed_cubic_extra:ppar} becomes smaller than 0 and, even in the zero-initial perpendicular spin case where $p_\perp = q_\perp =0$, we have that $\mathrm{arg}(G) = \pi$ and the elliptic parameter in Eq.~\eqref{eq:ellip_m_of_argG} becomes $m=1$. 
The resulting effect is to make the non-precessing configuration unstable under small perturbations, as also seen in Refs.~\cite{Gerosa:2015hba,Varma:2020bon}. 
Since we have numerical errors, when Eq.~\eqref{eq:MSAInstabilityCondition} is satisfied, this instability is realized and both the \pyEFPE and \STfour waveforms become precessing. However, they do so in different ways leading to the relatively large mismatches observed. Given that in nature we never expect the perfectly aligned-spin configuration $\chi_{1\perp,0} = \chi_{2\perp,0} = 0$ to be exactly achieved, these cases are not a cause of concern.

\begin{figure}[t!]
\centering  
\includegraphics[width=0.5\textwidth]{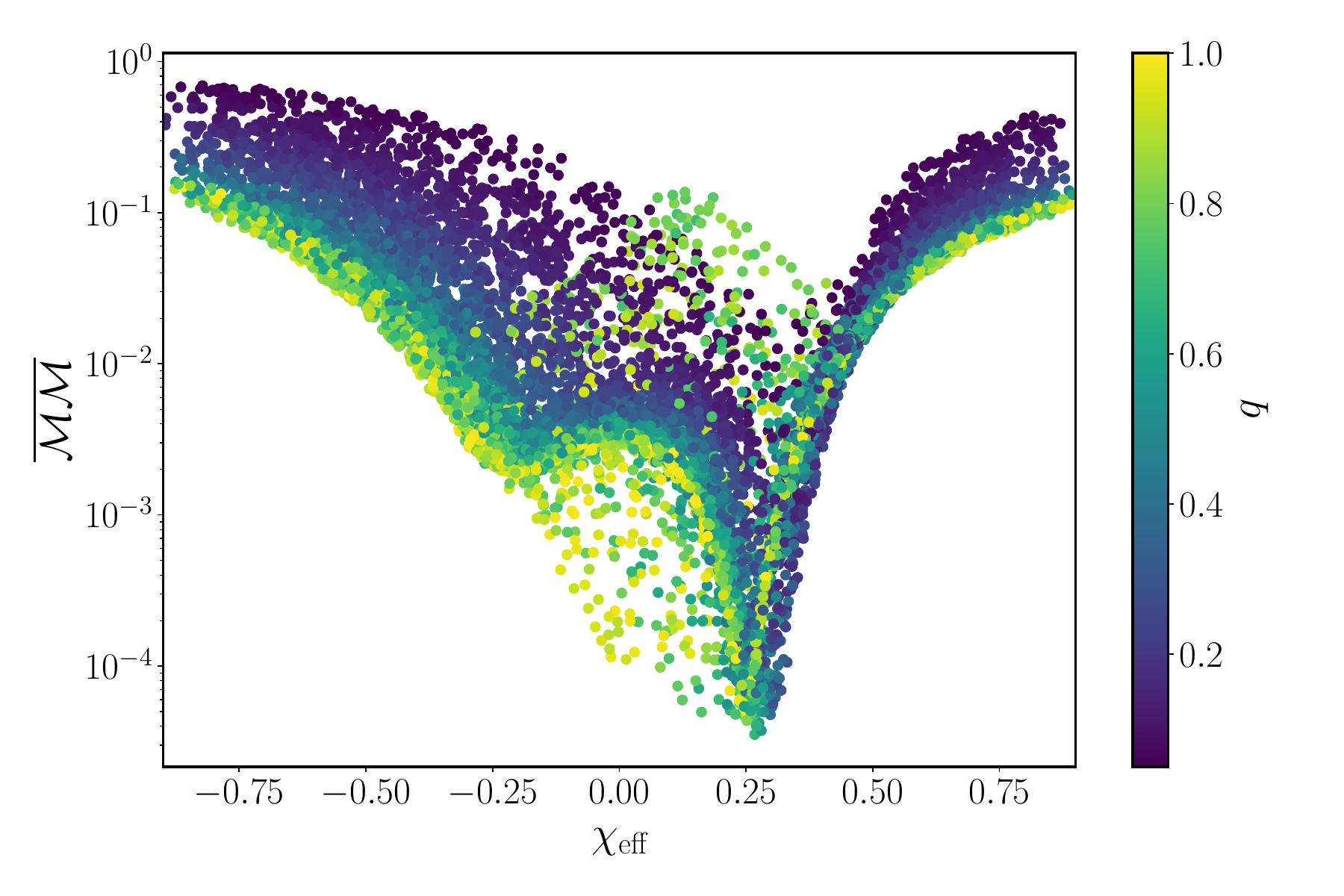}
\caption{\justifying Scatter plot of the mismatches between \pyEFPE and \Ftwo as a function of the effective inspiral spin parameter $\chi_\mathrm{eff}$ and mass ratio $q$. The samples shown correspond to the same 10000 samples in the $T=256\,\mathrm{s}$ distribution of Fig.~\ref{fig:MM_violins_noprec_SpinTaylorT4_TaylorF2}.}
\label{fig:mismatches_chi_eff_q_TaylorF2}
\end{figure}

In Fig.~\ref{fig:MM_violins_noprec_SpinTaylorT4_TaylorF2} we also compare against \Ftwo~\cite{lalsuite_code,Buonanno:2009zt,Isoyama:2020lls,Pratten:2020fqn}, the widely-used frequency-domain quasi-circular non-precessing model. We observe that in this case, the mismatches are significantly higher than in the \STfour case. This is a well known issue when comparing PN approximants~\cite{Nitz:2013mxa,Kumar:2015tha}. The difference between the \STfour family of approximants \pyEFPE belongs to, and \Ftwo is in how we write the evolution equations~\cite{Buonanno:2009zt}, having at 3PN that

\begin{subequations}
    \begin{align}
        (\D y)^\mathtt{T4} = \nu y^9 \left(a_0 + \sum_{n=2}^{6} a_n y^n \right) \, , \\
        (\D y)^\mathtt{F2} = \nu y^9 a_0 \frac{1}{1 - \sum_{n=2}^{6} c_n y^n} \, , 
    \end{align}
\end{subequations}

\noindent where the Taylor series of both agree up to $\ord{y^{15}}$. The problem is that, for extreme mass ratios (i.e. $q = m_2/m_1 \to 0$) and effective spin parameters (i.e. $|\chi_\mathrm{eff}| \to 1$), the difference between these two expressions for $\D y$ can become significant~\cite{Nitz:2013mxa}, especially at large values of $y$. This trend can clearly be seen in Fig.~\ref{fig:mismatches_chi_eff_q_TaylorF2}, where we show the mismatch between \pyEFPE and \Ftwo for the samples with $T=256\,\mathrm{s}$ as a function of $\chi_\mathrm{eff}$ and $q$. In agreement with Ref.~\cite{Nitz:2013mxa}, we observe that the mismatch grows for small mass ratios and large moduli of the effective spin parameter. On top of this ordered trend, we also observe that there are some samples with $q \sim 1$ and $\chi_\mathrm{eff} \sim 0$ having mismatches all over the place. These are samples satisfying Eq.~\eqref{eq:MSAInstabilityCondition} for which \pyEFPE dynamically acquires precession.

\subsubsection{Comparison with quasi-circular precessing models}
\label{sec:validate:comp:PrecesingQC}

Now that we have validated that \pyEFPE reduces to the expected limits in the quasi-circular spin-aligned case, we will test its behavour in the generic spin case, where the effect of spin-precession is present. Therefore, we still fix $e=0$ for all samples tested, but now use isotropic distributions in the component spin directions and uniform distributions in the component spin magnitudes $\chi_i \in [0, 0.9]$.

\begin{figure}[t!]
\centering  
\includegraphics[width=0.5\textwidth]{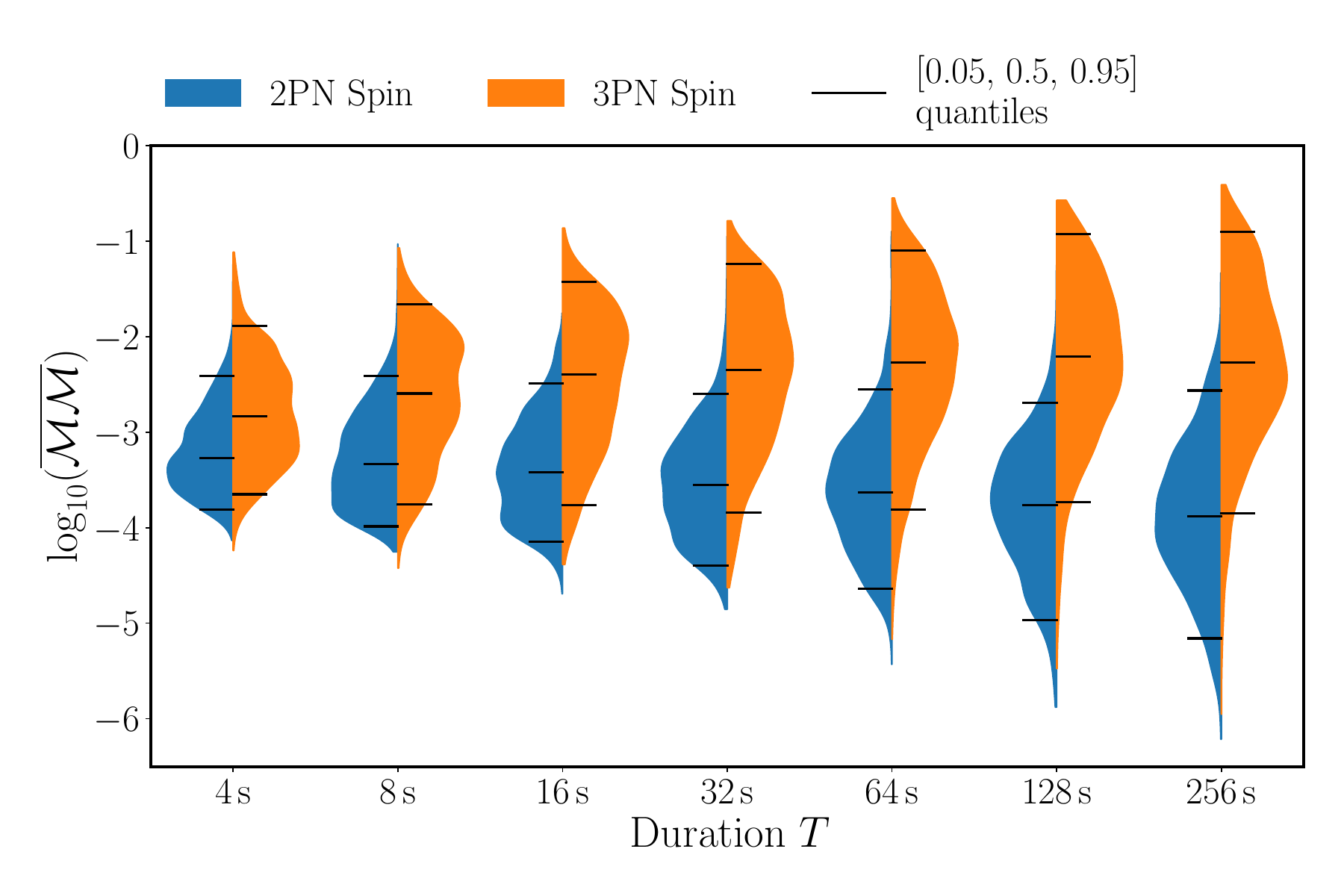}
\caption{\justifying Violin plots showing the distribution of the mismatch $\overline{\mathcal{MM}}$ (Eq.~\eqref{eq:MinMisMatchDef}) between two configurations of \pyEFPE and \STfour in the precessing spin case as a function of the segment duration $T$, or equivalently the chirp mass $\mathcal{M}_c$ range as specified in table~\ref{table:ChirpMassRanges}. The configuration ``2PN Spin'', in the left part of the violins, corresponds to taking the precession and spinning part of the evolution equations of both approximants at 2PN. The ``3PN Spin'' configuration, in the right part of the violins, corresponds to taking the precession at 2PN for \pyEFPE and 3PN for \STfour, and the spinning part of the evolution equations at 3PN for both.
Each distribution shown consists of 2000 random samples drawn from the distributions described in the text.}
\label{fig:MM_violins_prec_SpinTaylorT4_PN_spin_4_6}
\end{figure}

In Fig.~\ref{fig:MM_violins_prec_SpinTaylorT4_PN_spin_4_6} we compare \pyEFPE with \STfour in this precessing case, again integrating up to $f_\mathrm{max} = 0.8 f_\mathrm{ISCO}$, using the 0PN amplitudes and 3PN non-spinning terms in the evolution equations. However, we now consider two different cases, one where the spinning part of the evolution equations is taken to 2PN (labeled ``2PN Spin'') and other where they are taken to 3PN (labeled ``3PN Spin''). However, in the ``3PN Spin'' case, while the precession equations of \STfour include corrections up to 3PN, the \pyEFPE ones, given in Eq.~\eqref{eq:raw_prec_eqs}, are limited to 2PN, where the MSA approximation is known.

For the ``2PN Spin'' case, we find very small mismatches, similar to the ones observed in the 3PN aligned spin case of Fig.~\ref{fig:MM_violins_noprec_SpinTaylorT4_TaylorF2}, given us confidence in the fact that the MSA approximation is excellently describing the exact solution of the 2PN precession, also when including radiation reaction effects. In a similar way to the aligned spin case, the difference between \STfour and \pyEFPE is mostly due to the difference between computing the frequency domain waveform with the FFT and the SUA respectively. However, in this non-aligned spin case, there will be additional small differences between the two models coming from the residual eccentricity discussed in Eq.~\eqref{eq:emin}, which is neglected in \STfour.

For the ``3PN Spin'' case of Fig.~\ref{fig:MM_violins_prec_SpinTaylorT4_PN_spin_4_6}, the mismatches are bigger than for the ``2PN Spin'', growing for longer signals. This is due to \STfour including higher PN information in the precession equations. 
While these small corrections in the precession frequencies are not important for short signals that have few precession cycles, it can accumulate over the many precession cycles of long, leading to the mismatches observed in Fig.~\ref{fig:MM_violins_prec_SpinTaylorT4_PN_spin_4_6}. 
However, even for the longest signals studied in the ``3PN Spin'' case, the mismatches still satisfy $\overline{\mathcal{MM}} \lesssim 10^{-1}$, indicating that the use of the 2PN precession equations may be adequate for current GW detectors~\cite{Purrer:2019jcp}, though a detailed investigation on accuracy requirements should be conducted.
However, improvements to the precession description will be important for future GW detectors such as LISA~\cite{LISA:2024hlh}, the Einstein Telescope~\cite{ET:2019dnz,Branchesi:2023mws}, or Cosmic Explorer~\cite{Reitze:2019iox}, where signals are expected to be louder and longer.

\begin{figure}[t!]
\centering  
\includegraphics[width=0.5\textwidth]{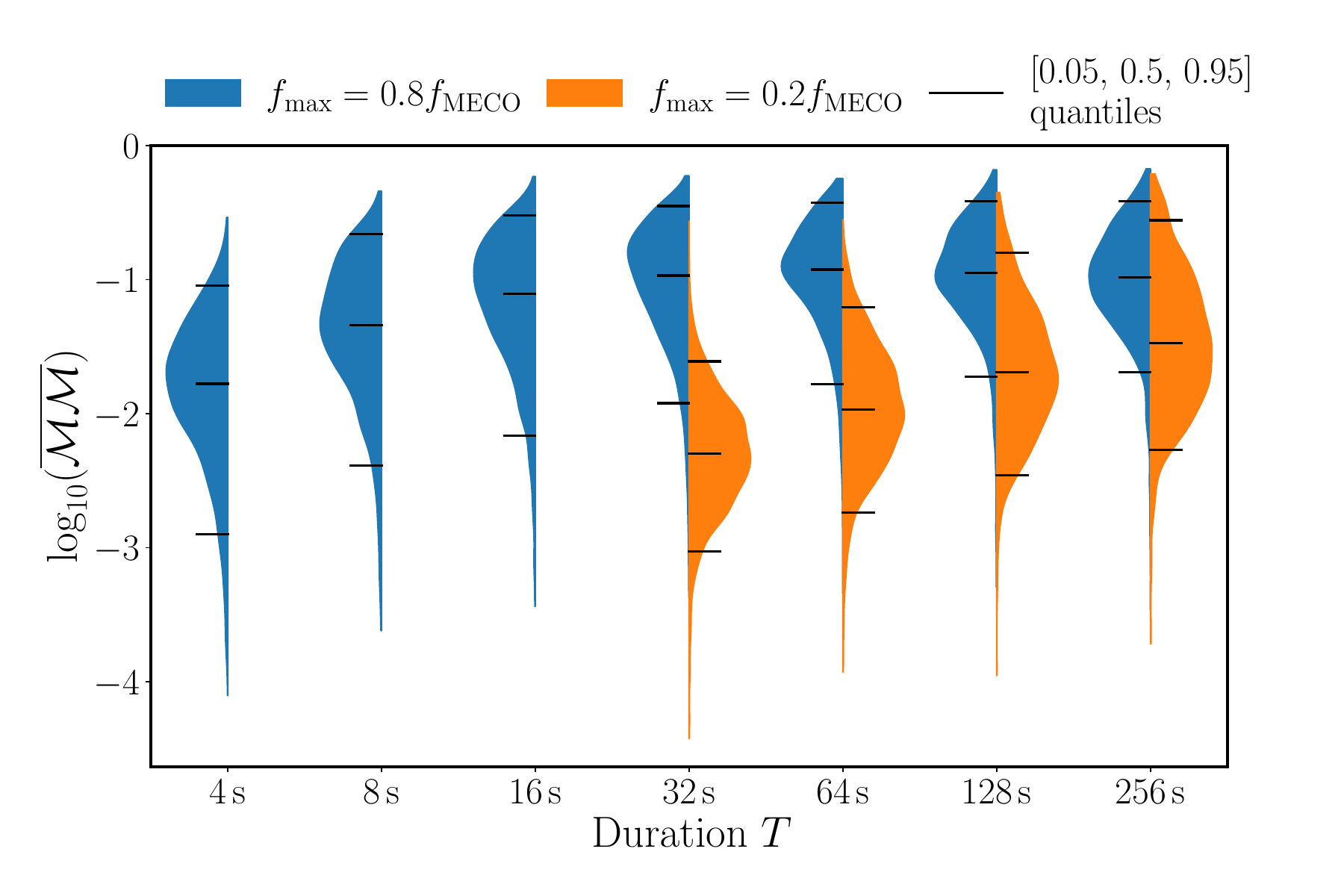}
\caption{\justifying Violin plots showing the distribution of the mismatch $\overline{\mathcal{MM}}$ (Eq.~\eqref{eq:MinMisMatchDef}) between \pyEFPE and \XP in the precessing spin case as a function of the segment duration $T$, or equivalently the chirp mass $\mathcal{M}_c$ range as specified in table~\ref{table:ChirpMassRanges}. In the left part of the violins we compute the mismatch (Eq.~\eqref{eq:InnerProdDef}) up to $f_\mathrm{max} = 0.8 f_\mathrm{MECO}$, while in the right side we only compute it up to $f_\mathrm{max} = 0.2 f_\mathrm{MECO}$. The MECO frequency is obtained with the phenomenological fit introduced in Ref.~\cite{Pratten:2020fqn}. When $f_\mathrm{max} = 0.2 f_\mathrm{MECO}$ we do not show mismatches for $T \leq 16\, \mathrm{s}$ (i.e. $\mathcal{M}_c \leq 8 M_\odot$), since some samples have $f_\mathrm{max} < f_\mathrm{min}$.
Each distribution shown consists of 2000 random samples drawn from the distributions described in the text.}
\label{fig:MM_violins_IMRPhenomXP_fmax_0.8_0.2MECO}
\end{figure}

In Fig.~\ref{fig:MM_violins_IMRPhenomXP_fmax_0.8_0.2MECO} we compare \pyEFPE against \XP~\cite{Pratten:2020ceb,lalsuite_code}, one of the most widely used models in GW data analysis. \XP is a precessing frequency-domain phenomenological inspiral-merger-ringdown (IMR) model obtained by twisting up the aligned-spin model \texttt{IMRPhenomXAS}~\cite{Pratten:2020fqn}. The inspiral part of \texttt{IMRPhenomXAS}, used to generate the waveform at frequencies below the MECO ($f \lesssim f_\mathrm{MECO}$) is based on \Ftwo at 3.5PN, augmented adding pseudo-PN coefficients to pseudo-6PN in the phase and pseudo-4.5PN in the amplitude. The pseudo-PN coefficients mimic the expected analytical structure of higher PN orders and are fitted to \texttt{SEOBNRv4}~\cite{Bohe:2016gbl} and NR~\cite{Boyle:2019kee}. 
The Euler angles to rotate \XP are derived using a PN expansion within the MSA framework described in Ref.~\cite{Chatziioannou:2017tdw} for the quasi-circular limit of the 2PN precessing equations used in \pyEFPE (Eq.~\eqref{eq:raw_prec_eqs}).

Therefore, the comparison with \XP is specially relevant, as it is fitted to EOB and NR, more closely representing the exact GW emission of a binary than the PN approximants. 
Since \pyEFPE is an inspiral model, we only compare it with the inspiral part of \XP. To take as much of the inspiral as possible while avoiding any effects from the merger-ringdown, we first use a maximum frequency to compute mismatches of $f_\mathrm{max} = 0.8 f_\mathrm{MECO}$, where $f_\mathrm{MECO}$ is computed with the phenomenological fit introduced in Ref.~\cite{Pratten:2020fqn}. The parameters are drawn from the same distribution as in Fig.~\ref{fig:MM_violins_prec_SpinTaylorT4_PN_spin_4_6}. However, we observe that in this case the mismatches are significantly bigger, specially for longer signals. This is mostly due to the much higher pseudo-PN-order information contained in \XP. Nonetheless, as we will see in Sec.~\ref{sec:validate:PE:XP}, \pyEFPE still describes this waveform with enough accuracy to obtain compatible parameter estimation results. 
To reduce the effect of these higher pseudo-PN-order corrections, that are most relevant at high frequencies, we also compute mismatches to a very conservative lower maximum frequency of $f_\mathrm{max} = 0.2 f_\mathrm{MECO}$. In this case, we observe much smaller mismatches. The results of Fig.~\ref{fig:MM_violins_IMRPhenomXP_fmax_0.8_0.2MECO} highlight the importance of adding more PN (or pseudo-PN) information to the phasing and amplitude of \pyEFPE, especially if we want to describe high SNR systems all the way to the beginning of the merger-ringdown phase. This is left as future work.

\subsubsection{Comparison with eccentric-spin-aligned models}
\label{sec:validate:comp:SAeccentric}

\begin{figure}[t!]
\centering  
\includegraphics[width=0.5\textwidth]{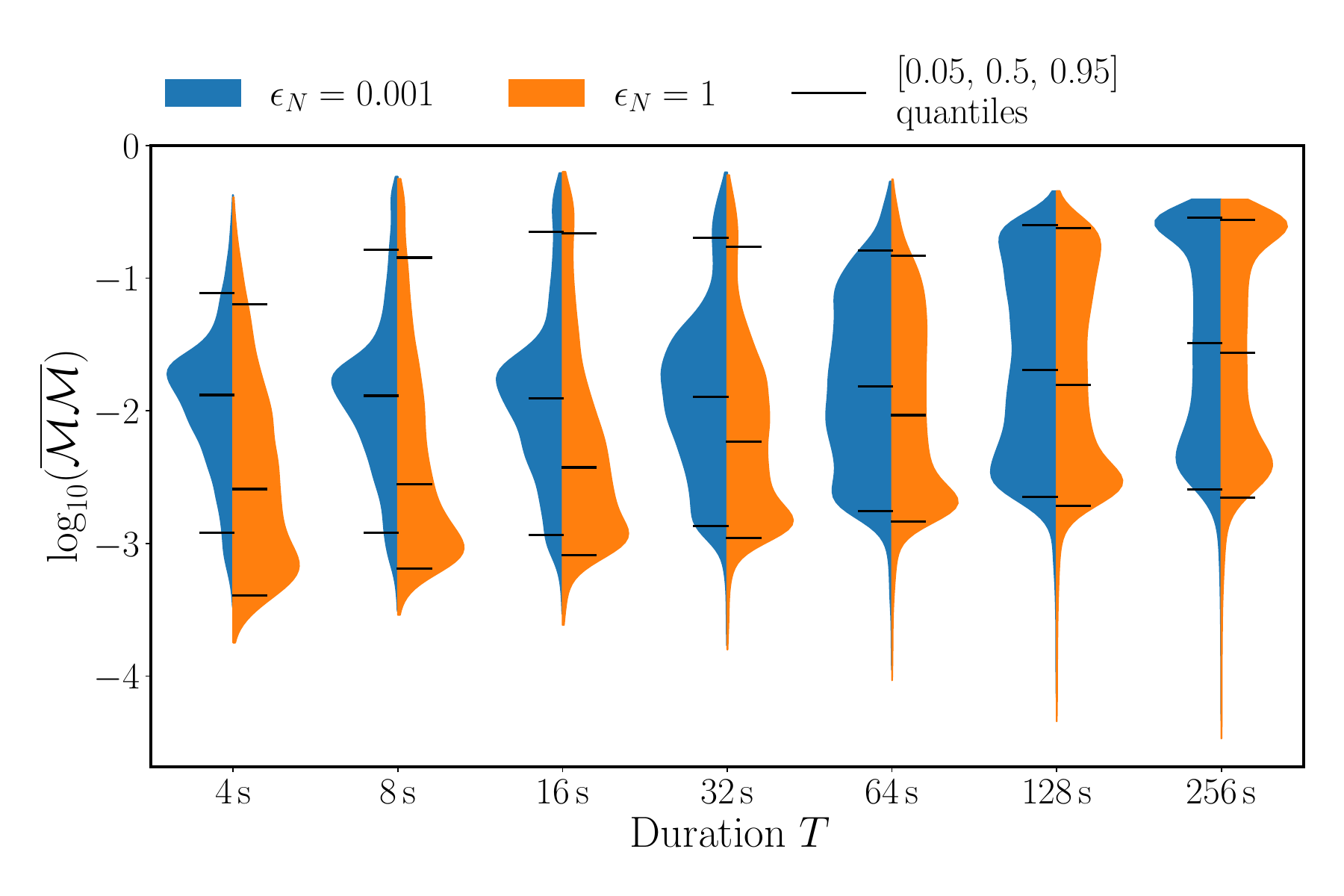}
\caption{\justifying Violin plots showing the distribution of the mismatch $\overline{\mathcal{MM}}$ (Eq.~\eqref{eq:MinMisMatchDef}) between \pyEFPE and \FtwoEcc in the aligned spin eccentric case as a function of the segment duration $T$, or equivalently the chirp mass $\mathcal{M}_c$ range as specified in table~\ref{table:ChirpMassRanges}. In the left part of the violins we show the mismatches with the default configuration of \pyEFPE, that has $\epsilon_N = 10^{-3}$, including as many Fourier modes as necessary to reach that amplitude tolerance. In the right part of the violins, we show the mismatches with \pyEFPE\ using $\epsilon_N = 1$, considering only the leading-order Fourier mode. For the eccentricities $e \leq 0.15$ considered here, this corresponds to the $l=m=n=2$ mode modeled by \FtwoEcc. Each distribution consists of 10,000 random samples drawn from the distributions described in the text.}
\label{fig:MM_violins_TaylorF2Ecc_Atol_1e-3_1}
\end{figure}

In this section we restrict ourselves to comparisons with spin-aligned eccentric waveforms. In particular, we compare \pyEFPE with \FtwoEcc~\cite{Moore:2016qxz,lalsuite_code}, a frequency-domain spin-aligned inspiral model that in the quasi-circular limit (i.e. $e \to 0$) reduces to \Ftwo. This model has a more limited description of eccentricity than \pyEFPE, since it only considers the leading order eccentricity contribution to the phasing equations of Eq.~\eqref{eq:RR_eqs_no_SP}, which are $\ord{e^2}$ and it neglects all eccentricity corrections to the amplitude, which start at $\ord{e}$. Therefore, \FtwoEcc is only valid for small eccentricities, in contrast with \pyEFPE, that is valid for arbitrarily large eccentricities. Because of this, for the mismatches shown in Fig.~\ref{fig:MM_violins_TaylorF2Ecc_Atol_1e-3_1} we consider small initial eccentricities, uniformly distributed with $e_0 \in [0, 0.15]$. As in Sec.~\ref{sec:validate:comp:SAQC} we use $f_\mathrm{max} = 0.8 f_\mathrm{ISCO}$. However, to avoid mismatches being dominated by the PN differences appearing at large spins that were discussed in that section, we use a smaller spin range of $\chi_{i z,0} \in [-0.3, 0.3]$. We test two different tolerances for the \pyEFPE amplitudes, the default one used in the rest of this work, of $\epsilon_N = 10^{-3}$, and a value of $\epsilon_N = 1$, that guarantees we are only including the leading Fourier mode, that for the low eccentricities considered corresponds to the same $l=m=n=2$ mode modeled by \FtwoEcc. 

\begin{figure}[t!]
\centering  
\includegraphics[width=0.5\textwidth]{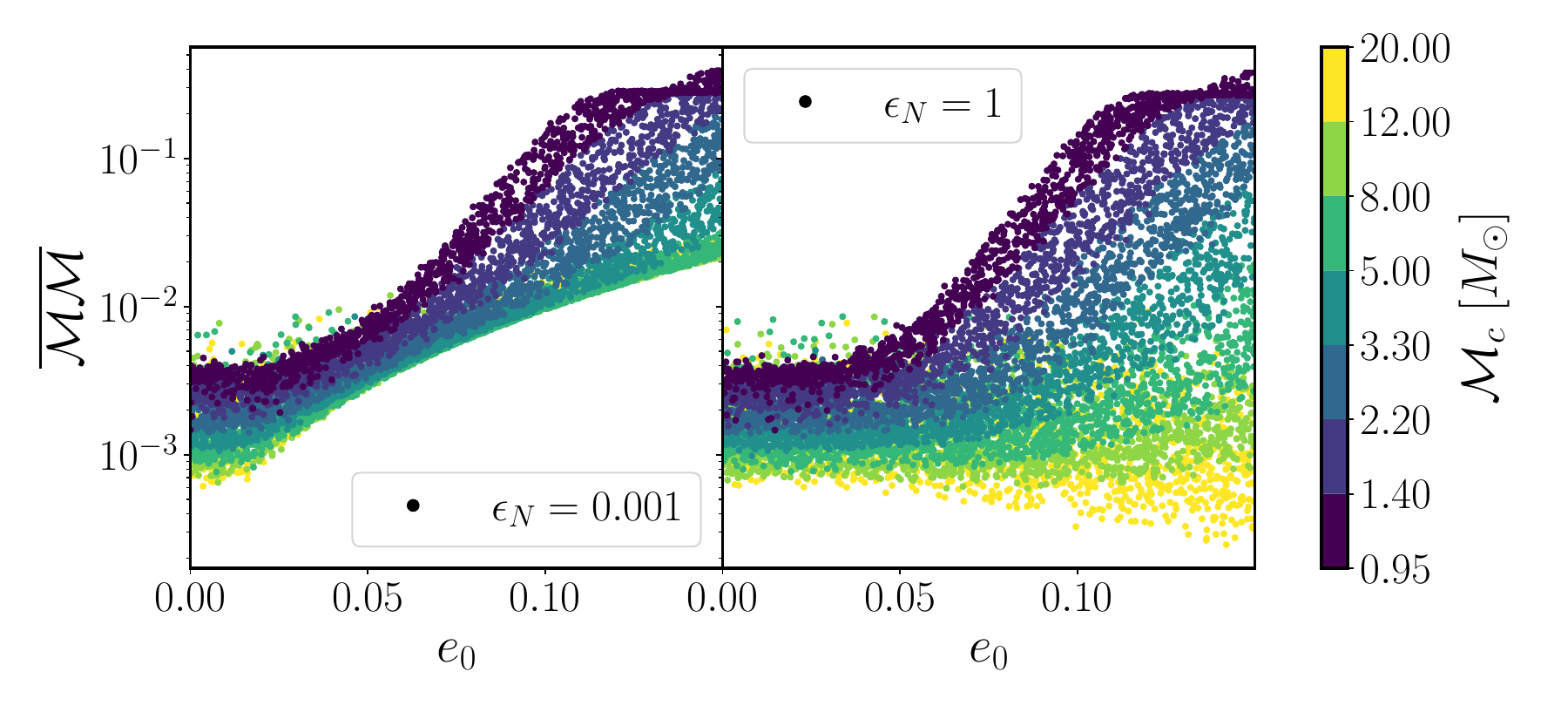}
\caption{\justifying Scatter plot of the mismatches between \pyEFPE and \FtwoEcc as a function of the initial eccentricity $e_0$ and chirp mass $\mathcal{M}_c$. We plot a subset of the samples shown in Fig.~\ref{fig:MM_violins_TaylorF2Ecc_Atol_1e-3_1} satisfying $q \in [0.5, 1]$ and $\chi_{iz,0} \in [-0.15, 0.15]$. In the left panel we show the same $\epsilon_N = 10^{-3}$ \pyEFPE configuration shown in the left violins of Fig.~\ref{fig:MM_violins_TaylorF2Ecc_Atol_1e-3_1}, while in the right panel we show the $\epsilon_N = 1$ configuration shown in the right violins. }
\label{fig:ecc_scatter}
\end{figure}

In Fig.~\ref{fig:MM_violins_TaylorF2Ecc_Atol_1e-3_1} we see that the mismatches are generally small. However, they grow as the waveforms become longer. Also, while the case with $\epsilon_N = 10^{-3}$ generally has larger mismatches than the $\epsilon_N = 1$ case, the difference is more important for small mass systems.
To better understand these trends, in Fig.~\ref{fig:ecc_scatter} we show some of the mismatches of Fig.~\ref{fig:MM_violins_TaylorF2Ecc_Atol_1e-3_1} as a function of initial eccentricity $e_0$ and chirp mass $\mathcal{M}_c$. To reduce the scatter not due to these two variables, we take the samples of Fig.~\ref{fig:MM_violins_TaylorF2Ecc_Atol_1e-3_1} that satisfy $q \in [0.5, 1]$ and $\chi_{i z, 0} \in [-0.15, 0.15]$. 

If, as in the right panel of Fig.~\ref{fig:ecc_scatter} where $\epsilon_N = 1$, we compare the same Fourier mode of \pyEFPE and \FtwoEcc, we observe that for smaller chirp masses (i.e. longer signals), the mismatches start growing at smaller eccentricities and are generally larger. The reason being that, when \FtwoEcc neglects $\ord{e^4}$ contributions to the phasing of the binary, it is ignoring phase terms starting at 0PN, that rapidly grow as $M f_\mathrm{min}$ decreases. At fixed eccentricity, the $n$-PN the contribution to the phase scales as $(M f_\mathrm{min})^{-(5 - 2 n)/3}$, so neglecting any eccentricity term up to 2PN order leads to problems for very low mass systems. 

In the left panel of Fig.~\ref{fig:ecc_scatter}, where $\epsilon_N = 10^{-3}$, we again observe that including the necessary Fourier modes in \pyEFPE, leads to larger mismatches with \FtwoEcc. Differences between the two waveforms appear at lower eccentricities than when $\epsilon_N = 1$, since the leading higher order Fourier modes are of order $\ord{e}$, compared to the $\ord{e^4}$ corrections neglected in the phase. We observe that this effect is specially relevant for systems with larger masses, since they have shorter durations and thus less time to get rid of the eccentricity, having a sizable eccentricity for a bigger fraction of the inspiral than smaller mass systems.

\subsection{Parameter Estimation}
\label{sec:validate:PE}
We now directly test \pyEFPE in GW parameter estimation (PE). Given a GW event candidate, identified by a modeled~\cite{Usman:2015kfa,Zackay:2019kkv,Aubin:2020goo,Nitz:2021zwj,Sakon:2022ibh,Phukon:2024amh} or unmodeled~\cite{Klimenko:2008fu,Lynch:2015yin,Drago:2020kic} search, we want to determine the properties of the system that could have generated the observed strain $d(t)$ in the detector, assuming some signal model with parameters $\bm{\theta}$. We use Bayesian inference~\cite{Bayes:1764vd, Veitch:2014wba} to compute the posterior probability of the signal parameters given the data and a signal model, $p(\bm{\theta}|d)$, using Bayes' theorem 

\begin{equation}
    p(\bm{\theta}|d) = \frac{\mathcal{L}(d|\bm{\theta}) \pi(\bm{\theta})}{\mathcal{Z}}
    \label{eq:BayesTheorem}
\end{equation}

\noindent where $\pi(\bm{\theta})$ is the prior on the parameters $\bm{\theta}$, $\mathcal{L}(d|\bm{\theta})$ is the likelihood, and $\mathcal{Z}$ the evidence or marginalized likelihood. Assuming stationary Gaussian noise, the likelihood reduces to the Whittle likelihood~\cite{Veitch:2014wba, Thrane:2018qnx}

\begin{align}
    \mathcal{L}(d|\bm{\theta}) & \propto \exp{\left\{ -\frac{1}{2} \sum_{i=1}^N  \langle h_i(\bm{\theta}) - d_i | h_i(\bm{\theta}) - d_i \rangle_i \right\}} \nonumber \\
    & \propto \exp{\left\{  \sum_{i=1}^N\left( \langle h_i(\bm{\theta}) | d_i \rangle_i -\frac{1}{2} \langle h_i(\bm{\theta}) | h_i(\bm{\theta}) \rangle_i \right) \right\}} \, ,
    \label{eq:Likelihood_def}    
\end{align}

\noindent where $d_i$ and $h_i$ represent the measured data (including noise) and the GW signal in the $i$-th detector. Similarly, $\langle \cdot | \cdot \rangle_i$ is the noise weighted inner product, already introduced in Eqs.~(\ref{eq:InnerProdDef},\ref{eq:InnerProdDiscrete}), with the $i$ subscript denoting that the PSD of the $i$-th detector is used. Finally, in Eq.~\eqref{eq:BayesTheorem} we have introduced the evidence $\mathcal{Z}$, which is a constant which ensures that the posterior probability $p(\bm{\theta}|d)$ is normalized, i.e.,

\begin{equation}
    \mathcal{Z} = \int \mathcal{L}(d|\bm{\theta}) \pi(\bm{\theta}) \d\bm{\theta} \, .
    \label{eq:Evidence_def}
\end{equation}

The evidence $\mathcal{Z}$ measures the expected value of the likelihood in the prior and is used for Bayesian model comparison~\cite{Kass:1995bf}. 

Due to the complexity of the integral in Eq.~\eqref{eq:Evidence_def}, we use Markov chain Monte Carlo (MCMC) methods to obtain the posterior samples. Here, we perform PE using \texttt{bilby}~\cite{Ashton:2018jfp,Smith:2019ucc,Romero-Shaw:2020owr}, and in particular, its implementation of the \texttt{dynesty}~\cite{Speagle:2020dqf} nested sampling algorithm. 

To test the PE performance of \pyEFPE and avoid confounding factors from random noise realizations, we perform zero-noise injections, equivalent to averaging the posterior over infinite Gaussian noise realizations~\cite{Rodriguez:2013oaa}. Therefore, we set the detector strain in Eq.~\eqref{eq:Likelihood_def} to 
\begin{equation}
    d_i = h_i^\mathrm{inj}(\bm{\theta}_\mathrm{inj}) \, ,
    \label{eq:d_inj}
\end{equation}
where $h^\mathrm{inj}$ and $\bm{\theta}_\mathrm{inj}$ represent the injected model and parameters respectively. We inject our mock signals into the LIGO-Hanford (H1), LIGO-Livingston (L1) and the Virgo detectors, assuming the LIGO A+ and Virgo AdV+ sensitivities projected for O5~\cite{Abbott_2020,ObservingScenariosPSDs}. We analyze data from a minimum frequency of $f_\mathrm{min} = 20\,\mathrm{Hz}$, which we also use as the initial frequency for \pyEFPE (i.e. $f_0^\mathrm{GW,22}=f_\mathrm{min} = 20\,\mathrm{Hz}$), meaning that the initial eccentricity $e_0$ is measured at 20$\mathrm{Hz}$. Finally, for the internal configuration of \pyEFPE we use the default values, also used in the mismatch studies in Sec.~\ref{sec:validate:comp}. 

We choose broad uninformative priors that are uniform in component masses, component spin magnitudes, coalescence time, and isotropic in location in the sky and binary and spin orientations. The distance prior is set assuming uniform probability in source volume, using the \textit{Planck15} cosmology~\cite{Planck:2015fie}. Finally, when the system is eccentric we use flat priors in the initial eccentricity $e_0 \in [0, 0.4]$ and mean anomaly $\ell_0 \in [0, 2\pi]$.

\subsubsection{\textnormal{\pyEFPE} injection -- recovery}
\label{sec:validate:PE:pyEFPE}

\begin{figure*}[t!]
\centering  
\includegraphics[width=0.75\textwidth]{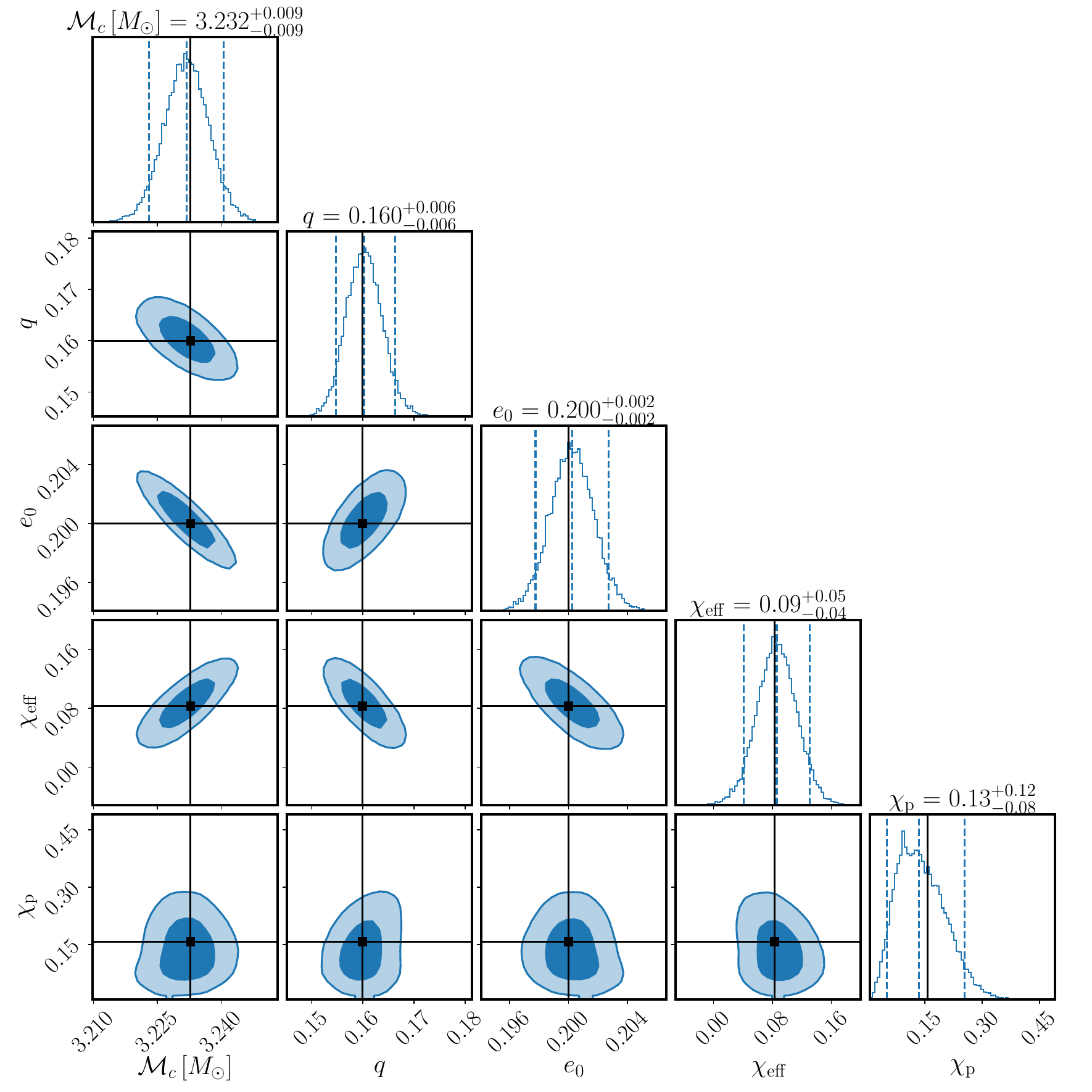}
\caption{\justifying Corner plot showing the joint posterior distributions of the most important intrinsic parameters of the \pyEFPE injection-recovery study. Specifically, the plot displays the chirp mass $\mathcal{M}_c$, mass ratio $q$, initial eccentricity $e_0$, effective inspiral spin parameter $\chi_\mathrm{eff}$ and effective precession spin parameter $\chi_\mathrm{p}$. The diagonal panels display the marginal distributions for each parameter, along with the median and 90\% confidence interval. The off-diagonal panels show the bivariate correlations between pairs of parameters, with the contours representing the $50\%$ and $90\%$ confidence regions. The black lines mark the values of the injected parameters.}
\label{fig:inj_EFPE_rec_EFPE_intrinsic_important}
\end{figure*}

To validate the reliability of PE with the \pyEFPE approximant, we start by testing whether an injection of \pyEFPE with specific parameters results in posterior distributions that are compatible with the injected values. 
To this end, we consider a precessing NSBH-like signal with detector-frame component masses $m_1 = 10 M_\odot$ and $m_2 = 1.6 M_\odot$ with moderate initial eccentricity $e_0 = 0.2$ defined at a reference frequency of 20 Hz, and recover the signal parameters with \pyEFPE. 
We analyze 32 s of data up to a maximum frequency of 300 Hz, taken to be below the Schwarzschild ISCO frequency, i.e., $f_\mathrm{max} \approx 0.8 f_\mathrm{ISCO}$. 
In Fig.~\ref{fig:inj_EFPE_rec_EFPE_intrinsic_important} we show the posterior distributions obtained for the most important intrinsic parameters, together with their injected values. 
More details on the injected signal and posterior distributions for all parameters are given in App.~\ref{sec:appendix:PE_extra}. In Fig.~\ref{fig:inj_EFPE_rec_EFPE_intrinsic_important}, we quantify the presence of spin-induced orbital precession in terms of the widely used effective precession spin parameter~\cite{Schmidt:2014iyl}, defined as 

\begin{equation}
    \chi\sub{p} = \max\left\{ \chi_{1\perp,0}, \frac{q (4 q + 3)}{4 + 3 q} \chi_{2\perp,0} \right\} \, ,
    \label{eq:chi_p_def}
\end{equation}

\noindent where $\chi_{i \perp,0} = \sqrt{\chi_{i x,0}^2 + \chi_{i y,0}^2}$ is the initial magnitude of the in-plane spin.
Firstly, we note that the estimated posteriors are in excellent agreement with the injected values. 
Secondly, despite the moderate SNR $(\approx 19.7)$ of the injected signal, the intrinsic parameters are very well constrained.
While a more detailed investigation into the behavior of PE for eccentric precessing inspirals is outside of the scope of this paper, and left for future work, we suggest a simple, intuitive explanation for this: 
As discussed in Refs.~\cite{Cutler:1992tc,Cutler:1994ys} for quasi-circular binaries, the signal and template waveform must remain in phase for the mismatch to be small. Consequently, each parameter $\theta$ can be inferred with an accuracy approximately corresponding to the change $\Delta\theta$ that shifts the number of GW cycles in the sensitivity band by order unity. 
A similar argument extends to each Fourier mode of an eccentric signal. From Eq.~\eqref{eq:hpc_decomp} it follows that in order to maintain a high match, parameters that cause the phase $n \lambda + (m - n)\delta\lambda$ to vary by more than one cycle need to be excluded. 
This effectively ``measures'' the accumulated orbital and periastron advance phases, $\lambda$ and $\delta\lambda$, with $\ord{1}$ precision. Therefore, if these phases are large, the parameters they depend on can be inferred with high accuracy. 
At leading PN order, the evolutions of $\lambda$ and $\delta\lambda$ are given by (see App.~\ref{sec:appendix:PN_formulas})
\begin{subequations}
    \label{eq:dphases_dy_LO}
    \begin{align}
        \frac{\d\lambda}{\d y} & = \frac{5}{32 \nu} \frac{y^{-6}}{1 + 7 e^2/8} \, , \label{eq:dphases_dy_LO:lambda} \\
        \frac{\d\delta\lambda}{\d y} & = \frac{15}{32 \nu} \frac{y^{-4}}{1 + 7 e^2/8} \, . \label{eq:dphases_dy_LO:delta_lambda}
    \end{align}
\end{subequations}
\noindent Neglecting $\ord{e^2}$ terms, and integrating from $y_0$ to coalescence, the accumulated phases are
\begin{subequations}
    \label{eq:phases_y_LO}
    \begin{align}
        \lambda & = \frac{1}{32 \nu} y_0^{-5} \left(1 + \ord{e_0^2} \right) \\& \approx \frac{1}{32} \left(\frac{G \mathcal{M}_c \pi f_\mathrm{min}}{c^3} \right)^{-5/3} \, , \label{eq:phases_y_LO:lambda} \\
        \delta\lambda & = \frac{5}{32 \nu} y_0^{-3} \left(1 + \ord{e_0^2} \right) \\& \approx \frac{5}{32} \nu^{-2/5} \left(\frac{G \mathcal{M}_c \pi f_\mathrm{min}}{c^3} \right)^{-1} \, . \label{eq:phases_y_LO:delta_lambda}
    \end{align}
\end{subequations}
\noindent For typical parameter values the approximate number of accumulated orbital cycles $\mathcal{N}_\mathrm{orb}$ and periastron advance cycles $\mathcal{N}_\mathrm{ecc}$ are given by
\begin{subequations}
    \label{eq:cycles_f_LO_numbers}
    \begin{align}
        \mathcal{N}_\mathrm{orb} = \frac{\lambda}{2 \pi} & \approx 3500 \left( \frac{\mathcal{M}_c}{M_\odot} \right)^{-5/3} \left( \frac{f_\mathrm{min}}{20\,\mathrm{Hz}} \right)^{-5/3}  , \label{eq:cycles_f_LO_numbers:lambda} \\
        \mathcal{N}_\mathrm{ecc} = \frac{\delta\lambda}{2 \pi} & \approx 140 \left( \frac{\nu}{1/4} \right)^{-2/5} \left( \frac{\mathcal{M}_c}{M_\odot} \right)^{-1} \left( \frac{f_\mathrm{min}}{20\,\mathrm{Hz}} \right)^{-1} .\label{eq:cycles_f_LO_numbers:delta_lambda}
    \end{align}
\end{subequations}
\noindent For the injected binary system, this yields $\mathcal{N}_\mathrm{orb} = 497$ and $\mathcal{N}_\mathrm{ecc} = 58$.
Therefore, from Eq.~\eqref{eq:cycles_f_LO_numbers:lambda}, $\mathcal{M}_c$ can be determined with sub-percent level accuracy from tracking the orbital phase $\lambda$ with $\ord{1}$ precision. 
Likewise, since $\delta\lambda$ is proportional to $\nu^{-2/5}$ (Eq.~\eqref{eq:cycles_f_LO_numbers:delta_lambda}), the symmetric mass ratio can be determined with percent level accuracy if we measure $\delta\lambda$ with $\ord{1}$ precision, as is the case when higher-order Fourier modes are observed.

In Fig.~\ref{fig:inj_EFPE_rec_EFPE_intrinsic_important} we also find that $\mathcal{M}_c$, $q$, $e_0$ and $\chi_\mathrm{eff}$ are correlated as they dominate the phase evolution of the binary during the inspiral. 
On the other hand, $\chi\sub{p}$ seems to be largely uncorrelated with the other parameters, since the effect of precession approximately decouples from the inspiral rate~\cite{Schmidt:2012}, modulating the amplitude with little secular impact on the phase.
It is particularly noteworthy that $e_0$ and $\chi_p$ do not seem to be strongly correlated, suggesting that eccentricity and precession can be independently constrained. 
Both effects produce amplitude modulations (see e.g. Fig.~\ref{fig:waveform_example_td}), with eccentricity modulations occurring on the orbital timescale. For high-mass systems, where only the final inspiral cycles and the merger-ringdown are observed, the orbital and spin-precession time-scales are of similar order, and the effects of eccentricity and precession might be more difficult to disentangle~\cite{Romero-Shaw:2022fbf}. However, for low-mass binaries, we find that the orbital and spin-precession time-scales are cleanly separated, suggesting that the two effects are distinguishable. Furthermore, as previously argued, for low-mass systems with many cycles in the detector sensitivity band, eccentricity is primarily constrained by its impact on the binary phase, while precession leaves the phase largely unaffected~\cite{Schmidt:2012}. 

\subsubsection{Precessing quasi-circular injection}
\label{sec:validate:PE:XP}

\begin{figure*}[t!]
\centering  
    \begin{subfigure}{0.495\textwidth}
        \centering
        \includegraphics[width=\textwidth]{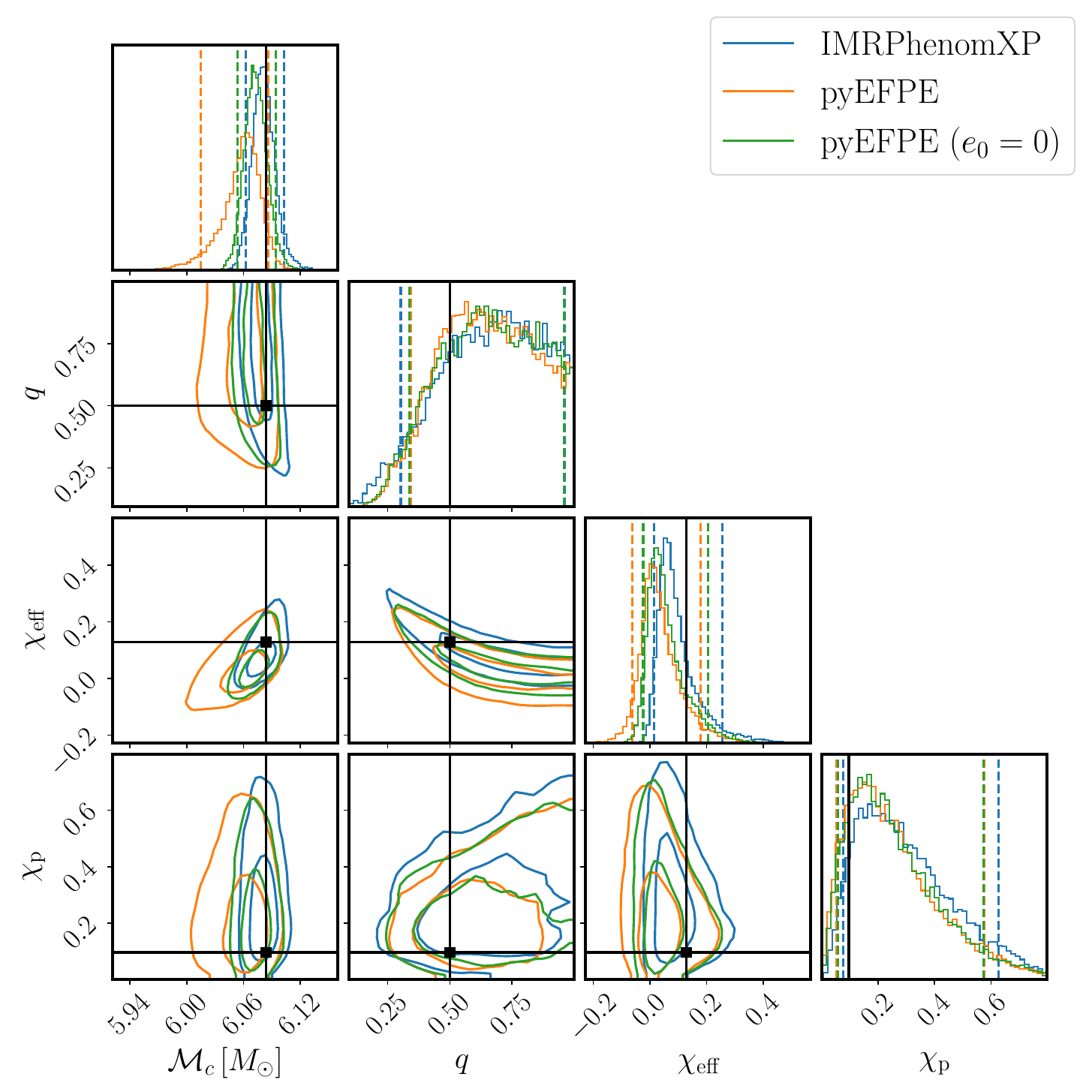} 
        \caption{Low SNR (SNR=$12.7$).}
        \label{fig:inj_XP_16s_intrinsic_important:LowSNR}
    \end{subfigure}
    \begin{subfigure}{0.495\textwidth}
        \centering
        \includegraphics[width=\textwidth]{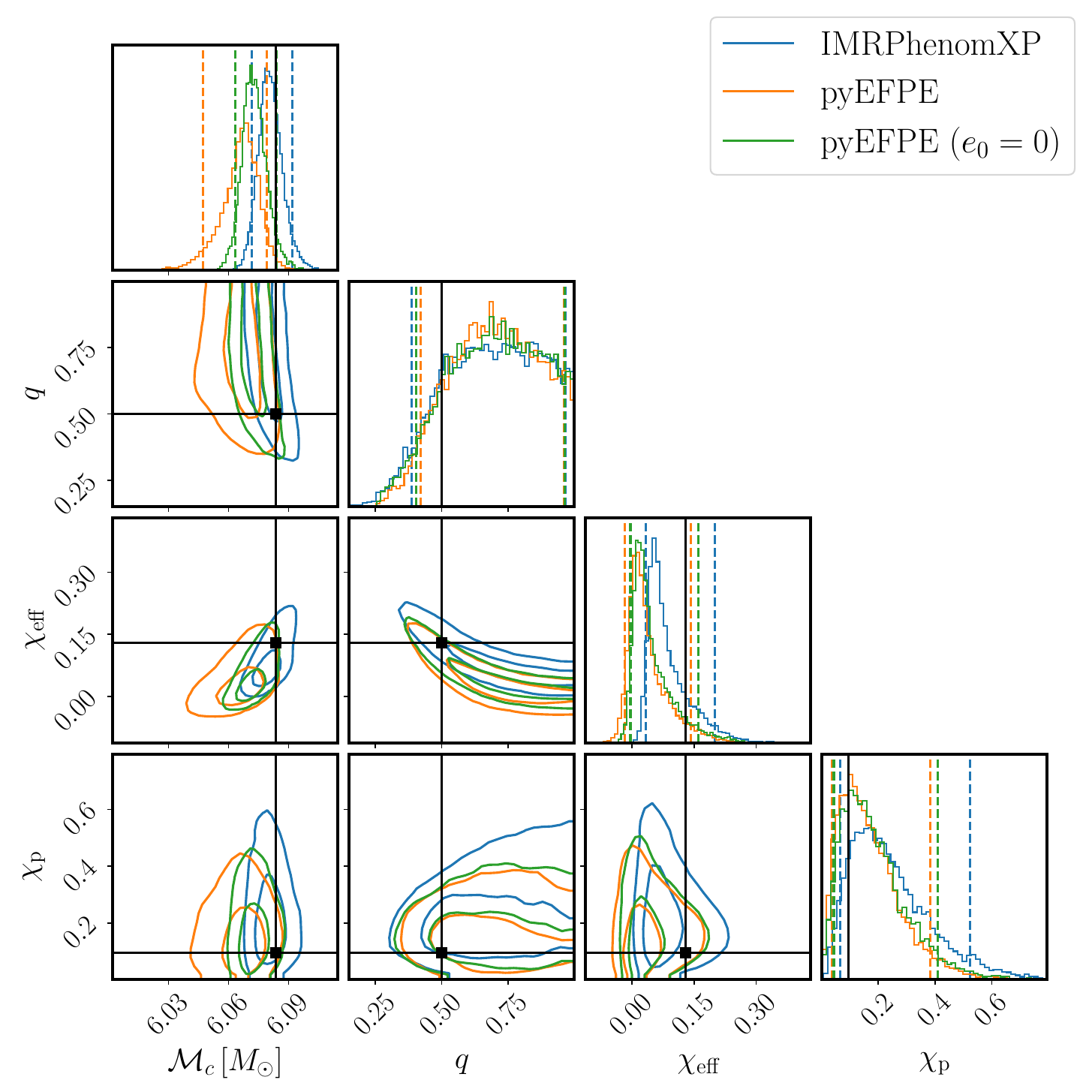} 
        \caption{High SNR (SNR=$25.5$).}
        \label{fig:inj_XP_16s_intrinsic_important:HighSNR}
    \end{subfigure}
\caption{\justifying Corner plot showing the joint posterior distributions of the most important intrinsic parameters of the \XP injection recovered with \XP, \pyEFPE and \pyEFPE with $e_0 = 0$ for the low SNR case (left panel) and high SNR case (right panel). Specifically, the plot displays the chirp mass $\mathcal{M}_c$, mass ratio $q$ effective inspiral spin parameter $\chi_\mathrm{eff}$ and effective precession spin parameter $\chi_\mathrm{p}$. In each corner plot, the diagonal panels display the marginal distributions for each parameter, along with the 90\% confidence interval. The off-diagonal panels show the bivariate correlations between pairs of parameters, with the contours representing the $50\%$ and $90\%$ confidence regions. The black lines mark the values of the injected parameters.}
\label{fig:inj_XP_16s_intrinsic_important}
\end{figure*}

Next, we assess whether \pyEFPE accurately recovers the parameters of a precessing, quasi-circular binary, in particular $e_0=0$.
For the injection we use the inspiral-merger-ringdown waveform model \XP~\cite{Pratten:2020fqn,Pratten:2020ceb} described in Sec.~\ref{sec:validate:comp:PrecesingQC}

In particular, we perform two binary black hole (BBH) \XP injections with detector-frame masses of $m_1 = 10M_\odot$ and $m_2 = 5 M_\odot$, consistent with the low mass end of the observed BBH population~\cite{KAGRA:2021duu}. We analyze 16 s of data up to a maximum frequency of 250 Hz, i.e. $f_\mathrm{max} \approx 0.85 f_\mathrm{ISCO}$. All parameters of the two injections are identical except for the luminosity distance, allowing us to explore signals with different SNRs. The ``low SNR'' injection has $d_L = 2000$~Mpc and an SNR of $12.7$, typical of the bulk of events observed by GW detectors~\cite{KAGRA:2021vkt}, while the ``high SNR''  has $d_L = 1000$~Mpc and SNR of $25.5$. We recover each zero-noise injection with three different signal models: 

\begin{itemize}
    \item \XP, which is the same model as used for the injection, and therefore gives the baseline result.
    \item \pyEFPE with a flat prior for the initial eccentricity $e_0 \in [0,0.4]$. 
    \item \pyEFPE, where we fix the initial eccentricity $e_0 = 0$, to more closely resemble the configuration of the \XP run and to isolate differences pertaining to differences in the non-eccentric sector.
\end{itemize}

Figure~\ref{fig:inj_XP_16s_intrinsic_important} shows the posterior distributions of the most important intrinsic parameters, together with the injected values, for both SNRs and all three waveform models. We refer the reader to appendix~\ref{sec:appendix:PE_extra} for the complete results. 
We find that the posterior distributions generally agree with the injected values. 
However, even in the \XP recovery, the posteriors are not Gaussian distributions centered at the injected values, and exhibit noticeable biases. 
These deviations arise from prior effects and correlations between parameters~\cite{Cutler:1994ys,Chatziioannou:2014coa,Pratten:2020igi}. 
Nevertheless, the overall shape of the posteriors is broadly consistent between the different approximants, with small differences in the inferred values of $\mathcal{M}_c$ and $\chi_\mathrm{eff}$ that become more noticeable at higher SNR (right panel).

\begin{figure}[t!]
\centering  
\includegraphics[width=0.5\textwidth]{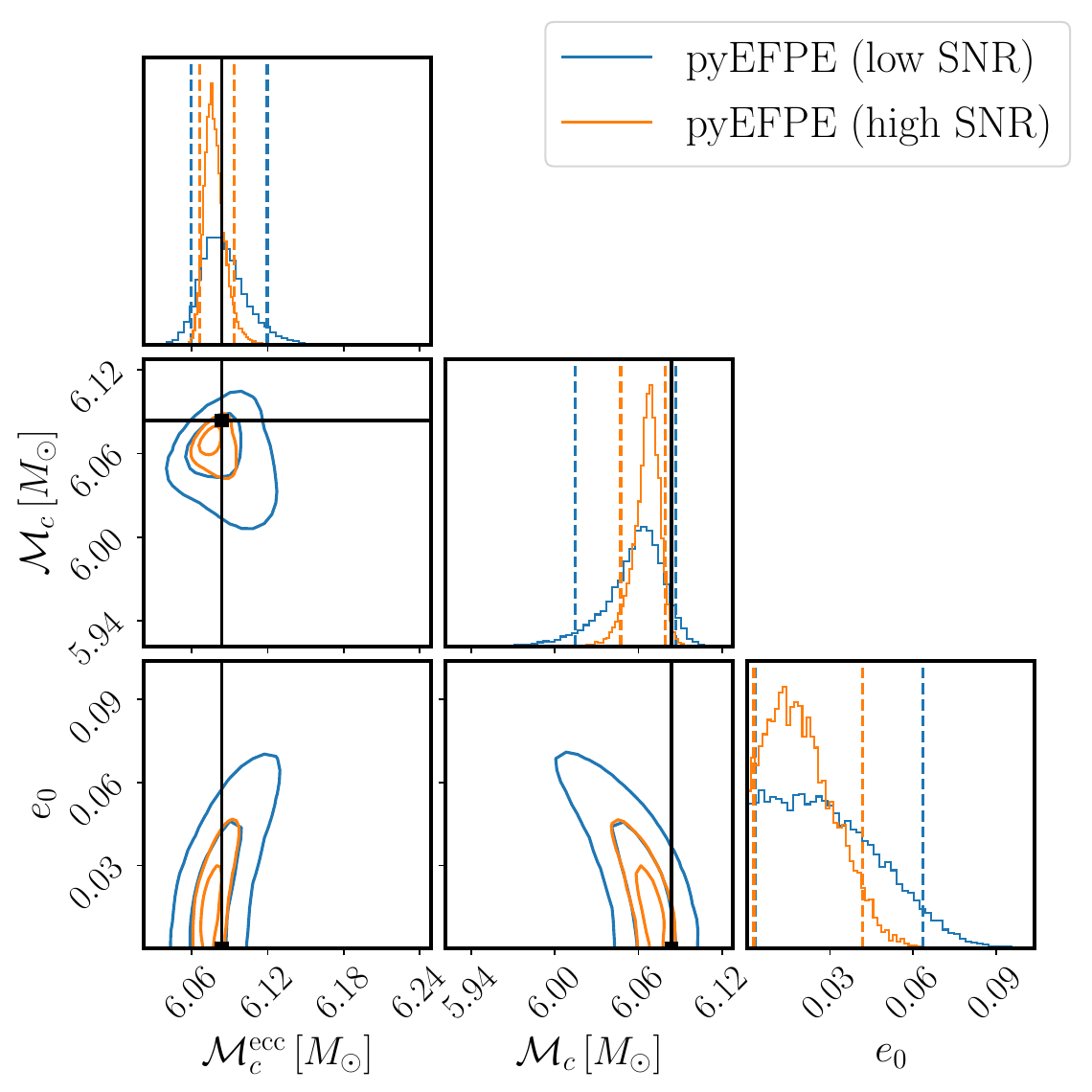}
\caption{\justifying Corner plot showing the joint and marginalised 1D posterior distributions of the eccentric chirp mass $\mathcal{M}_c^\mathrm{ecc}$, chirp mass $\mathcal{M}_c$ and initial eccentricity $e_0$ for the \XP injection recovered with \pyEFPE for the low SNR (blue) and high SNR (orange) cases. The dashed vertical lines mark the 90\% confidence interval. The contours for the 2D posteriors represent the $50\%$ and $90\%$ confidence regions. The black lines mark the injected values.}
\label{fig:inj_XP_16s_joint_EFPE_McEcc}
\end{figure}

The most striking difference we observe concerns the chirp mass recovered by \pyEFPE, which is significantly lower than the injected value. 
However, this is expected since the chirp mass and eccentricity are degenerate, both appearing at 0PN order in the phase.
This degeneracy can be captured by an eccentric chirp mass $\mathcal{M}_c^\mathrm{ecc}$~\cite{Favata:2021vhw}, which at $\ord{e_0^2}$ can be computed as
\footnote{Note that Ref.~\cite{Favata:2021vhw} defines $\mathcal{M}_c^\mathrm{ecc} = \mathcal{M}_c/\left( 1 - \frac{157}{24} e_0^2\right)^{3/5}$. This definition is problematic, since it has a divergence at $e_0 = \sqrt{24/157} \approx 0.39$. Given that it was derived using formulas accurate up to $\ord{e^2}$, it is preferable to use the $\ord{e_0^2}$ expansion of Eq.~\eqref{eq:Mc_ecc_def}.
}

\begin{equation}
    \mathcal{M}_c^\mathrm{ecc} = \left(1 + \frac{157}{40} e_0^2 \right) \mathcal{M}_c.
    \label{eq:Mc_ecc_def}
\end{equation}
Figure~\ref{fig:inj_XP_16s_joint_EFPE_McEcc} shows the posteriors for the eccentric chirp mass $\mathcal{M}_c^\mathrm{ecc}$, the regular chirp mass $\mathcal{M}_c$ and the initial eccentricity $e_0$ for the \pyEFPE low and high SNR injections. We observe that the eccentricity posteriors for both runs strongly support a quasi-circular binary ($e_0 = 0$), and have tails extending to $e_0 <0.09$ and $e_0 < 0.06$ for the low and high SNR case, respectively. 
The aforementioned anti-correlation between $e_0$ and $\mathcal{M}_c$ is evident in the 2D posterior: To fit the injected signal, non-zero eccentricity is compensated by a lower chirp mass, ultimately leaving the injected chirp mass outside of the 1D 90\% credible interval. On the other hand, when using the eccentric chirp mass of Eq.~\eqref{eq:Mc_ecc_def}, this degeneracy is significantly reduced and the recovered median value of $\mathcal{M}_c^\mathrm{ecc}$ agrees with the injected value.

\begin{figure*}[t!]
\centering  
    \begin{subfigure}{0.45\textwidth}
        \centering
        \includegraphics[width=\textwidth]{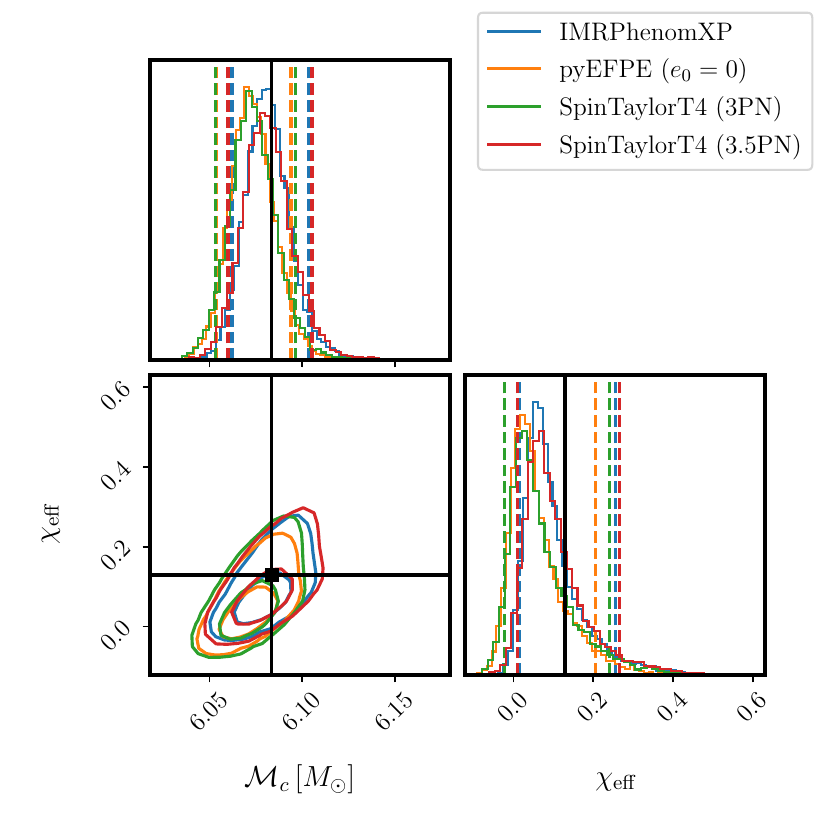} 
        \caption{Low SNR (SNR=$12.7$).}
        \label{fig:inj_XP_16s_PN_comparison:LowSNR}
    \end{subfigure}
    \begin{subfigure}{0.45\textwidth}
        \centering
        \includegraphics[width=\textwidth]{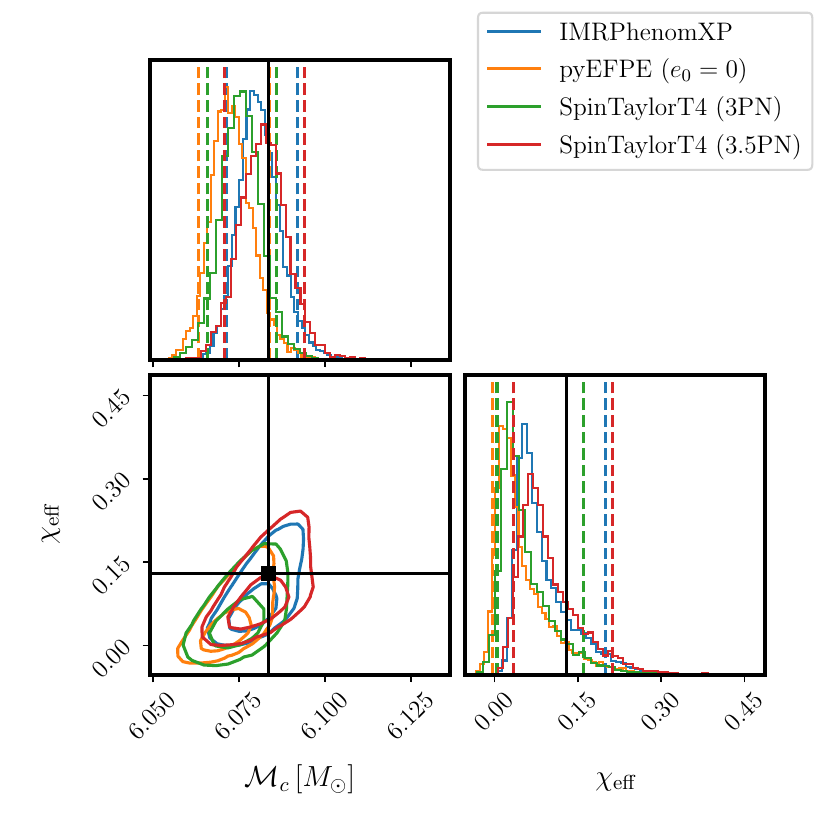} 
        \caption{High SNR (SNR=$25.5$).}
        \label{fig:inj_XP_16s_PN_comparison:HighSNR}
    \end{subfigure}
\caption{\justifying Corner plot showing the joint posterior distributions of the chirp mass $\mathcal{M}_c$ and effective inspiral spin parameter $\chi_\mathrm{eff}$ for the \XP injection recovered with \XP, \pyEFPE with $e_0 = 0$, and \STfour taking up to 3PN and 3.5PN terms in the non-spinning part of the evolution equations. In the left panel we plot the low SNR case and in the right panel the high SNR case. In each corner plot, the diagonal panels display the marginal distributions for each parameter, along with the 90\% confidence interval. The off-diagonal panel shows the bivariate correlations between the two parameters, with the contours representing the $50\%$ and $90\%$ confidence regions. The black lines mark the values of the injected parameters.}
\label{fig:inj_XP_16s_PN_comparison}
\end{figure*}

Going back to Fig.~\ref{fig:inj_XP_16s_intrinsic_important}, we again observe that, even in the case where the initial eccentricity is fixed to $e_0 = 0$, the \pyEFPE $\mathcal{M}_c$ and $\chi_\mathrm{eff}$ posteriors still differ slightly from the \XP ones, specially in the high SNR case. These disagreements are due to differences in the models that, as discussed in Sec.~\ref{sec:validate:comp:PrecesingQC}, lead to differences in their predictions. To explore these waveform systematics in more detail, we perform two more analyses on both injections, using the \STfour waveform with different amounts of PN information. In both analyses the amplitude of the GWs is modeled at 0PN and the spinning terms of the equations of motion are taken to 3PN, the maximum available in the \texttt{lalsuite}~\cite{lalsuite_code} implementation used. However, in one run we limit the non-spinning terms to 3PN, the same order used in \pyEFPE, and in the other to the maximum 3.5PN allowed in \texttt{lalsuite}.

In Fig.~\ref{fig:inj_XP_16s_PN_comparison} we show the posteriors for $\mathcal{M}_c$ and $\chi_\mathrm{eff}$ for the four quasi-circular analyses. Looking at the low SNR case in Fig.~\ref{fig:inj_XP_16s_PN_comparison:LowSNR}, the \pyEFPE run with $e_0 = 0$ and the \STfour run at 3PN agree very well with each other, while the \XP run and the \STfour run at 3.5PN are also in very good agreement with each other. This highlights the importance of including higher-order PN terms in the waveform. For the high SNR case shown in Fig.~\ref{fig:inj_XP_16s_PN_comparison:HighSNR}, the \pyEFPE run with $e_0 = 0$ and the \STfour run at 3PN, and the \XP run and the \STfour run at 3.5PN are still clustered into pairs, however, we note that in this case there are bigger differences due to small differences in the models discussed in Sec.~\ref{sec:validate:comp:PrecesingQC}. The most important differences are: i) the different PN information included in the precession equations of each model (3PN for \STfour and 2PN for \pyEFPE), ii) the difference between the SUA and FFT to compute the frequency-domain waveforms and iii) the residual eccentricity of Eq.~\eqref{eq:emin} taken into account by \pyEFPE.

\section{Conclusion}
\label{sec:conclusion}

In this paper we have introduced \pyEFPE, a post-Newtonian waveform model for the inspiral of spin-precessing compact binaries on eccentric orbits. \pyEFPE improves upon previous EFPE models available in the literature~\cite{Klein:2018ybm,Klein:2021jtd,Arredondo:2024nsl} by i) introducing closed-form analytical expressions for the Newtonian Fourier mode amplitudes (Sec.~\ref{sec:NewtAmp}), ii) improving the numerical stability of the MSA approximation (Sec.~\ref{sec:MSA}), iii) extending the PN information in the equations of motion (Sec.~\ref{sec:RR} and appendix~\ref{sec:appendix:PN_formulas}) and iv) providing an optimized and simple numerical implementation (Sec.~\ref{sec:waveform}). 
In addition, we introduced a scheme to interpolate the slow-varying waveform amplitudes, enabling a speedup of the waveform generation by a factor up to $\sim \ord{15}$ with minimal loss of accuracy, making the model well-suited for real data analysis applications. The \pyEFPE waveform is publicly available in~\cite{pyEFPE_repo}, with a \texttt{Python} implementation designed to be easy to modify and use to make it as useful as possible for the scientific community. 

In Sec.~\ref{sec:validate}, we thoroughly validated the \pyEFPE waveform: First,  we checked the phenomenology of inspiral waveforms with spin-precession and orbital eccentricity. As no other widely available frequency-domain waveform model includes both effects, we then performed mismatch comparisons between \pyEFPE and other waveforms in the quasi-circular limit (\STfour, \XP, and \Ftwo) and the eccentric, spin-aligned limit (\FtwoEcc), finding good agreement overall. Finally, we demonstrated the efficacy of \pyEFPE to perform full Bayesian inference on simulated GW data in current ground-based GW detector networks, and to recover the parameters of signals described by \pyEFPE and \XP.
This makes \pyEFPE a powerful tool for GW science, enabling the study of inspiral signals with both spin-precession and orbital eccentricity, which are considered critical features for understanding the formation and evolution of compact binaries~\cite{Rodriguez:2016vmx,Zevin:2021rtf,Gerosa:2021mno}. 
Integrating \pyEFPE into LISA data analysis pipelines~\cite{Roebber:2020hso,Buscicchio:2021dph,Klein:2022rbf,Pratten:2023krc} will be a key goal, with the inference of stellar mass binaries and intermediate mass ratio inspirals being core components of the global fit~\cite{LISA:2024hlh}. Whilst only a handful of stellar mass binaries are expected~\cite{Buscicchio:2024asl}, LISA may be able to resolve a handful of low-redshift sources with high precision, providing valuable insight into the degree of orbital eccentricity and spin-precession in the population at low frequencies.

While \pyEFPE represents a significant advancement in waveform modeling, it still omits several physical effects present in the inspiral of compact binaries, such as higher-order modes~\cite{Pratten:2020ceb}, mode asymmetries~\cite{Boyle:2014ioa, Ghosh:2023mhc} or tidal effects~\cite{Flanagan:2007ix,Yang:2018bzx,Schmidt:2019wrl,LaHaye:2022yxa,Dones:2024odv}. Additionally, as highlighted in Sec.~\ref{sec:validate}, higher-order PN terms in the equations of motion are required to accurately describe very high SNR signals, particularly for binaries with highly unequal mass ratios or large anti-aligned spins. We leave the incorporation of these effects into \pyEFPE for future work. 

The final major component missing in \pyEFPE is the merger-ringdown phase of the waveform, where the PN expansion is no longer applicable. 
As demonstrated in recent phenomenological models (see e.g. Refs.~\cite{Pratten:2020fqn,Pratten:2020ceb}), the merger-ringdown regime can be accurately modeled by fitting physically motivated ans\"{a}tze to numerical relativity simulations.
Integrating such a calibrated model into \pyEFPE’s description of the inspiral phase would represent a significant step toward constructing a computationally efficient inspiral-merger-ringdown waveform model for spin-precessing, eccentric binaries.
In this regard, \pyEFPE offers a new avenue towards the construction of complete and computationally efficient inspiral-merger-ringdown waveforms for compact binaries with orbital eccentricity and spin precession.

\section*{Code Availability}

The repository containing the waveform model, along with scripts to reproduce many of the figures in this paper, is available at Ref.~\cite{pyEFPE_repo}. 

\section*{Acknowledgments}
We thank Antoine Klein for helpful discussions and code comparisons. 
We thank Antoni Ramos-Buades for his helpful comments and suggestions on the manuscript.
G.M. acknowledges support from the Ministerio de Universidades through Grants No. FPU20/02857 and No. EST24/00099, and from the Agencia Estatal de Investigaci\'on through the Grant IFT Centro de Excelencia Severo Ochoa No. CEX2020-001007-S, funded by MCIN/AEI/10.13039/501100011033.
G.P. is very grateful for support from a Royal Society University Research Fellowship URF{\textbackslash}R1{\textbackslash}221500 and RF{\textbackslash}ERE{\textbackslash}221015, and gratefully acknowledges support from an NVIDIA Academic Hardware Grant.
G.P. and P.S. acknowledge support from STFC grants ST/V005677/1 and ST/Y00423X/1, and UKSA grants supporting the UK's contribution to LISA Ground Segment activities ST/Y004922/1.
P.S. also acknowledges support from a Royal Society Research Grant RG{\textbackslash}R1{\textbackslash}241327.
The authors are grateful for computational resources provided by the LIGO Laboratory (CIT) and supported by the National Science Foundation Grants PHY-0757058 and PHY-0823459, the University of Birmingham's BlueBEAR HPC service, which provides a High Performance Computing service to the University's research community, as well as resources provided by Supercomputing Wales, funded by STFC grants ST/I006285/1 and ST/V001167/1 supporting the UK Involvement in the Operation of Advanced LIGO.
This manuscript has the LIGO document number P2500025.


\onecolumngrid
\appendix

\section{Comparison of Newtonian Fourier Mode Amplitudes}
\label{sec:appendix:mode_amp_comparison}
\begin{figure}[h!]
\centering  
\includegraphics[width=0.5\textwidth]{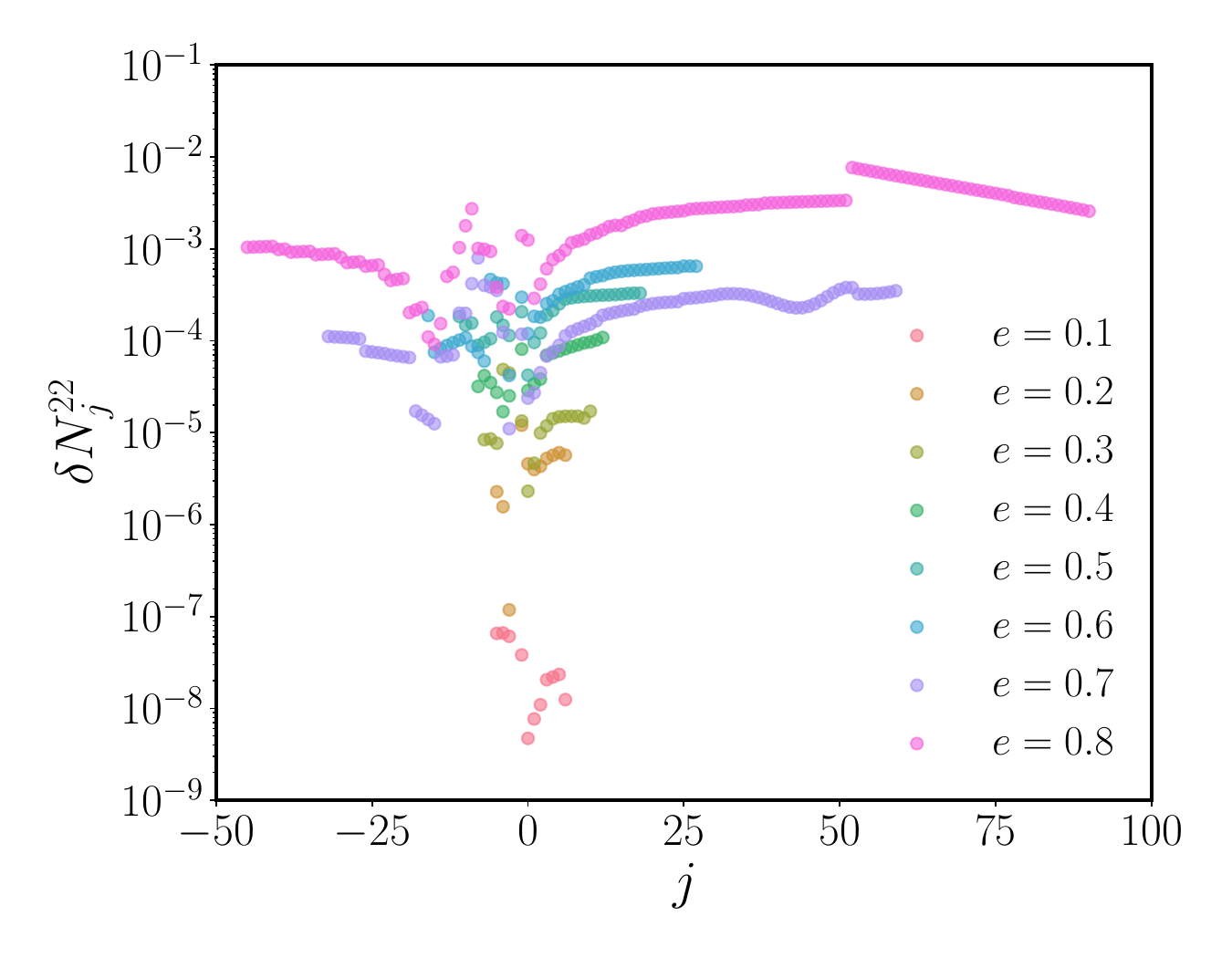}
\caption{\justifying Here we show the relative error between the exact solution $N^{22,\rm ex.}_j$, corresponding to Eq.~\eqref{eq:N22_j}, and the perturbative solution $N^{22,\rm pert.}_j$ derived in~\cite{Arredondo:2024nsl}, where $\delta N^{22}_j = |(N^{22,\rm ex.}_j - N^{22,\rm pert.}_j) / N^{22,\rm ex.}_j|$. We explore the approximation error over the range of $j$ required to accurately reconstruct the Fourier mode amplitudes for a range of eccentricities. 
}
\label{fig:diff_N2jj_plot}
\end{figure}

In Sec.~\ref{sec:NewtAmp}, we outlined a derivation of the Newtonian Fourier mode amplitudes in terms of exact closed-form analytical expressions involving Bessel functions, denoted $N^{22,\rm ex.}_j$. A solution for the Fourier mode amplitudes was previously derived in~\cite{Arredondo:2024nsl}, based on nested series expansions that build on earlier work in~\cite{Boetzel:2017zza,Klein:2021jtd}. We denote the amplitudes estimated this way as $N^{22,\rm pert.}_j$, where we explicitly follow the recommended truncation of the summation as outlined in~\cite{Arredondo:2024nsl}. 
In Fig.~\ref{fig:diff_N2jj_plot} we demonstrate agreement between the two methods, with the maximum relative errors being $\sim \mathcal{O}(10^{-2})$. As expected, the errors increase with eccentricity as accurately reproducing the exact solution requires an increasing number of terms in the series.
Likewise, the apparent discontinuities as a function of $j$ are due to the finite truncation of the nested summation operations. 

\section{Alternative interpretation of amplitude tolerance $\epsilon_N$}
\label{sec:appendix:epsilonN_logL}
In this appendix we provide an alternative interpretation of the amplitude tolerance $\epsilon_N$, introduced in Sec.~\ref{sec:NewtAmp:Number}, that is more directly connected to data analysis applications and can guide the choice of an appropriate $\epsilon_N$ for a given use case. Specifically, we note that selecting only a subset of the Fourier modes effectively decomposes the exact waveform $h$ into two components, the selected part $h_\mathrm{sel}$ and the unselected residual $\delta h$, such that
\begin{subequations}
\label{eq:h_hseldh}
\begin{align}
    h & = h_\mathrm{sel} + \delta h \, , \label{eq:h_hseldh:h} \\
    h_\mathrm{sel} & = \sum_{l,m} \sum_{n \in n^{\mathrm{sel}}_{l,m}} h_{l,m,n} \, , \label{eq:h_hseldh:hsel} \\
    \delta h & =  \sum_{l,m} \sum_{n \not\in n^{\mathrm{sel}}_{l,m}} h_{l,m,n} \, , \label{eq:h_hseldh:dh} \\
    h_{l,m,n}(t) & = \mathcal{A}_{l,m,n}(t) \rme^{-\rmi(n\lambda(t) + (m-n)\delta\lambda(t))} \ ,  \label{eq:h_hseldh:hlmn}
\end{align}
\end{subequations}
\noindent where we have used the expansion of Eq.~\eqref{eq:hpc_decomp}, dropping the $+,\times$ polarization subscripts by assuming the waveform has been projected onto a detector. When using the approximated waveform $h_\mathrm{sel}$ to analyze the exact signal $h$, the error induced in the log-likelihood (Eq.~\eqref{eq:Likelihood_def}) is
\begin{equation}
    \Delta \log\mathcal{L}_\mathrm{sel} \equiv \log\mathcal{L}(h|h) - \log\mathcal{L}(h|h_\mathrm{sel}) = \frac{1}{2} \langle \delta h | \delta h \rangle \, .
    \label{eq:DeltalogLsel}
\end{equation}
To relate $\epsilon_N$ to this log-likelihood error, we note that Fourier modes with different values of $n$ and/or $m$ are approximately orthogonal under the noise weighted inner product $\langle \cdot | \cdot \rangle$, as the integral in Eq.~\eqref{eq:InnerProdDef} becomes rapidly oscillatory when the phases of the modes in Eq.~\eqref{eq:h_hseldh:hlmn} differ, leading to destructive interference. That is,
\begin{equation}
 \langle h_{l,m,n} | h_{l',m',n'} \rangle 
 = \langle  h_{l,m,n} | h_{l',m,n} \rangle \delta_{n n'}\delta_{m m'} \, ,
 \label{eq:mode_ortogonality}
\end{equation}
where $\delta_{ij}$ denotes the Kronecker delta. 
Using this together with Eq.~\eqref{eq:h_hseldh} we then find that
\begin{subequations}
\label{eq:hh_dhdh_modes}
\begin{align}
    \langle h | h \rangle & = \sum_{l,l',m, n}  \langle  h_{l,m,n} | h_{l',m,n} \rangle \, , \\
    \langle \delta h | \delta h \rangle & = \sum_{l,l',m} \sum_{n \not\in (n^{\mathrm{sel}}_{l,m} \cup n^{\mathrm{sel}}_{l',m})} \langle  h_{l,m,n} | h_{l',m,n} \rangle \, .
\end{align}    
\end{subequations}
Therefore, the fraction $\langle \delta h | \delta h \rangle/\langle h | h \rangle$ is closely related to the error introduced in Eq.~\eqref{eq:Newtonian_orders_needed}, since it measures the squared modulus of the neglected Fourier modes. Nonetheless, $\langle \delta h | \delta h \rangle/\langle h | h \rangle$ is significantly more complicated, given that it will depend on the detector PSD $S_n(f)$ through the noise weighted inner product and on the orientation of the binary through the angular dependence of the amplitudes $\mathcal{A}_{l,m,n}$. Nonetheless, if we average over angles (as was done in Sec.~\ref{sec:NewtAmp:Number}) and ignore the frequency dependence of the PSD ($S_n(f) = S_0$), we recover a result equivalent to Eq.~\eqref{eq:Newtonian_orders_needed}, i.e.
\begin{subequations}
\begin{align}
   \langle h | h \rangle & = C \sum_{l,m, n} | N_{n -m}^{l m} |^2 \, , \\
    \langle \delta h | \delta h \rangle & = C \sum_{l,m} \sum_{n \not\in n^{\mathrm{sel}}_{l,m}} | N_{n -m}^{l m} |^2 \, , \\
    \Delta \log\mathcal{L}_\mathrm{sel} & < \frac{1}{2} \langle h | h \rangle \epsilon_N \, , \label{eq:DeltalogLsel_assump}
\end{align}    
\end{subequations}
\noindent where $C$ is detector dependent constant. 

\begin{figure}[h!]
\centering  
\includegraphics[width=0.6\textwidth]{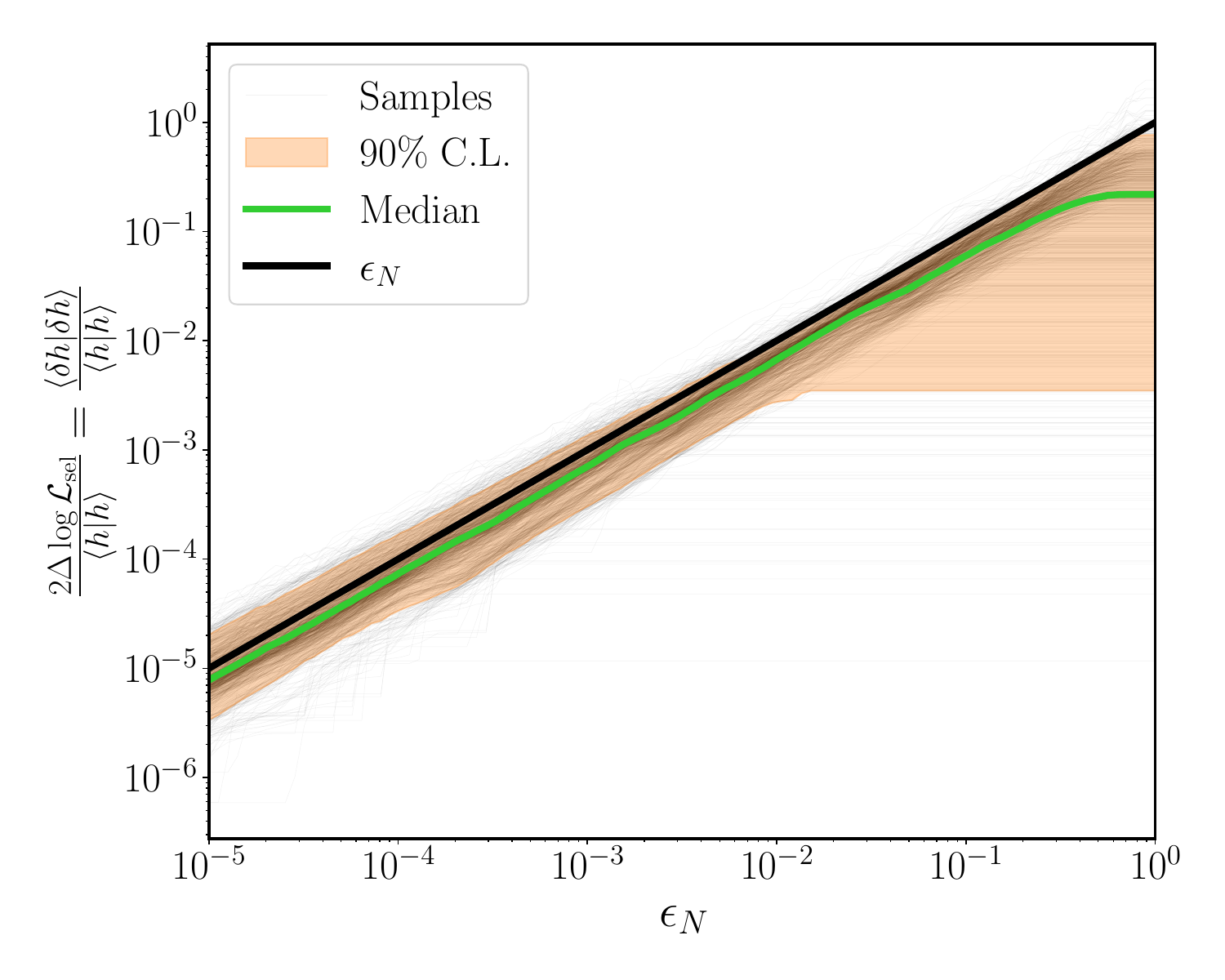}
\caption{\justifying Relative error in the log-likelihood, $2 \log\mathcal{L}_\mathrm{sel}/\langle h | h \rangle$, as a function of the amplitude tolerance $\epsilon_N$ for 1000 random \pyEFPE samples, distributed as in Sec.~\ref{sec:validate:comp:interp} (with $T=256\,\mathrm{s}$, $\mathcal{M}_c \in [0.95, 20] M_\odot$, and $e_0 \in [0 , 0.7]$). The ``exact'' waveform $h$ is generated using $\epsilon_N=10^{-9}$ in \pyEFPE and compared to waveforms $h_\mathrm{sel}$ with varying $\epsilon_N \in [10^{-5},1]$. Inner products are computed using the Advanced LIGO A+ PSD~\cite{Abbott_2020,ObservingScenariosPSDs}. In addition to showing the log-likelihood error for each sample, we display the median and 90\% confidence interval. For comparison, we also plot the reference line $\epsilon_N$, which we expect the errors to be comparable to.
}
\label{fig:logLrelerr_atol}
\end{figure}

For a given signal, the specific angular configuration influences the relative contribution of different modes in Eq.~\eqref{eq:hh_dhdh_modes}. Additionally, assuming a frequency independent PSD is far from realistic. Nevertheless, we argue that Eq.~\eqref{eq:DeltalogLsel_assump} should approximately hold. To assess this, Fig.~\ref{fig:logLrelerr_atol} shows the values of $2 \Delta \log\mathcal{L}_\mathrm{sel}/\langle h | h \rangle$ as a function of the amplitude tolerance $\epsilon_N$ for 1000 random \pyEFPE samples, distributed as in Sec.~\ref{sec:validate:comp:interp} (with $T=256\,\mathrm{s}$, $\mathcal{M}_c \in [0.95, 20] M_\odot$, and $e_0 \in [0 , 0.7]$). The ``exact'' waveform $h$ is generated using $\epsilon_N=10^{-9}$ in \pyEFPE and compared to waveforms $h_\mathrm{sel}$ with varying $\epsilon_N \in [10^{-5},1]$.  Despite employing the Advanced LIGO A+ PSD~\cite{Abbott_2020,ObservingScenariosPSDs} in the inner products and analyzing specific samples (i.e., without averaging over angles), Fig.~\ref{fig:logLrelerr_atol} confirms that Eq.~\eqref{eq:DeltalogLsel_assump} approximately holds, with the median of the $2 \Delta \log\mathcal{L}_\mathrm{sel}/\rho_\mathrm{opt}^2$ distribution consistently below $\epsilon_N$. However, some samples exhibit errors slightly exceeding the prediction of Eq.~\eqref{eq:DeltalogLsel_assump}, with the largest deviations reaching up to a factor of $\sim 3$ times $\epsilon_N$. Therefore, we expect the following inequality to approximately hold
\begin{equation}
    \Delta \log\mathcal{L}_\mathrm{sel} \lesssim \rho_\mathrm{opt}^2 \epsilon_N \, , \label{eq:DeltalogLsel_approx}
\end{equation}
\noindent corresponding to Eq.~\eqref{eq:DeltalogLsel_assump}, relaxing the factor $1/2$ and identifying $\langle h | h \rangle$ with the squared optimal SNR $\rho_\mathrm{opt}^2$.
Using Eq.~\eqref{eq:DeltalogLsel_approx}, we can estimate an acceptable value of $\epsilon_N$ for studying a signal with optimal SNR $\rho_\mathrm{opt}$, noting that log-likelihood errors smaller than $\sim 1$ do not significantly impact data analysis applications~\cite{Jaranowski:2005hz}, having only a minimal effect on event significance in searches and PE posteriors~\cite{Morras:2023pug} and evidences (Eq.~\eqref{eq:Evidence_def}). By imposing this condition in Eq.~\eqref{eq:DeltalogLsel_approx} (i.e. $\Delta \log\mathcal{L}_\mathrm{sel} \lesssim 1$), we find that an appropriate choice for $\epsilon_N$ is
\begin{equation}
    \epsilon_N \lesssim \frac{1}{\rho_\mathrm{opt}^2}  \, , \label{eq:sensible_epsilonN}
\end{equation}
which suggests that the default value of $\epsilon_N = 10^{-3}$ in \pyEFPE should be valid for $\rho_\mathrm{opt} \lesssim 32$.

\section{Expressions for the evolution equations}
\label{sec:appendix:PN_formulas}

In this work we use harmonic coordinates (HC) and the covariant Tulczyjew-Dixon spin-supplementary condition \cite{Tulczyjew:1959zza,Dixon:1970zza}.

\subsection{Evolution of $y$ and $e^2$}
\label{sec:appendix:PN_formulas:ye2}

We obtain the non-spinning evolution of $y$ and $e^2$ from the evolution of $x = (1-e^2) y^2 = (M \omega)^{2/3}$ and $e$, given in Ref.~\cite{Arun:2009mc}. As was noted in Ref.~\cite{Ebersold:2019kdc}, Ref.~\cite{Arun:2009mc} has an error in the expression for $\d e/\d t$, coming from the fact that they obtained it by just transforming the eccentricity from ADM to harmonic coordinates (i.e. $e^\mathrm{ADM} \to e^\mathrm{HC}$), but one also has to transform  $de^\mathrm{ADM}/dt \to de^\mathrm{HC}/dt$, since the derivative is not gauge invariant. 
Note that we do not take the formulas for $\d x/\d t$ from Ref.~\cite{Ebersold:2019kdc}, as although they indicate these are given in harmonic coordinates, there appears to be a discrepancy, potentially involving the use of ADM expressions instead. 
When using the updated formulae, the expressions for $a_4$ and $a_6$ simplify, since there are now no terms proportional to $1/(1-e^2)$.

We use the approximation of the enhancement functions presented in Ref.~\cite{Klein:2018ybm}, written in a way that improves convergence, due to the inclusion of factors of $\sqrt{1-e^2}$. Similarly, as in Ref.~\cite{Klein:2018ybm}, we include in the 3PN enhancement functions $\kappa_i$ only the terms proportional to $\log(n)$, as the other ones are in finite number and can be combined with non-tail terms.

For the spinning contributions, we use the 1.5PN and 2PN fully-spinning terms from Ref.~\cite{Klein:2018ybm}, correcting a typo found when comparing with Ref.~\cite{Henry:2023tka}, where we note that in the coefficient of $b_4^\mathrm{SS}$ multiplying $(\uvec{L} \cdot \bm{s}_2)^2$, an $111/4$ should be replaced by an $111/2$. For eccentric-spin-aligned binaries, the spin-spin and spin-orbit phasing coefficients to 3PN were recently found in Ref.~\cite{Henry:2023tka}. We include these coefficients in our waveform model, making it accurate up to 3PN in spin, ignoring the perpendicular-in-spin contributions to the 3PN spin-spin terms. 

We write the evolution equations for $y$ and $e^2$ as

\begin{subequations}
\label{eq:RR_eqs_appendix}
\begin{align}
 \D y &= \nu y^9 \left(a_0 + \sum_{n=2}^{6} a_n  y^n \right), \\
 \D e^2 &= - \nu y^8 \left(b_0 + \sum_{n=2}^{6} b_n  y^n \right),
\end{align}
\end{subequations}

At 3PN order, the non-spinning (NS) part of the coefficients is given by:

\begin{subequations}
\begin{align}
	a_0 =&\; \frac{32}{5} + \frac{28}{5} e^2 \,, \\
	a_2 =&\; -\frac{1486}{105} - \frac{88}{5} \nu + \left( \frac{12296}{105} - \frac{5258}{45}\nu 
		\right) e^2 + \left( \frac{3007}{84} - \frac{244}{9} \nu \right) e^4 \,, \\
	a_3^\mathrm{NS} =&\; \frac{128 \pi}{5} \phi_y \,, \\
	a_4^\mathrm{NS} =&\; \frac{34103}{2835} + \frac{13661}{315}\nu + \frac{944}{45}\nu^2 + \left(-\frac{489191}{1890} - \frac{209729}{630}\nu + \frac{147443}{270}\nu^2\right) e^2 +  \left(\frac{2098919}{7560} - \frac{2928257}{2520}\nu + \frac{34679}{45}\nu^2\right)e^4 \nonumber\\ 
        & + \left(\frac{53881}{2520}-\frac{7357}{90}\nu+\frac{9392}{135}\nu^2\right)e^6  + \frac{1 - \sqrt{1 
		- e^2}}{\sqrt{1 - e^2}} \bigg[ 16 - \frac{32}{5} \nu + \left(266 - \frac{532}{5}\nu \right) 
		e^2 + \left( -\frac{859}{2} + \frac{859}{5} \nu \right) e^4 \nonumber\\
		& + \left(-65 + 26 \nu \right) e^6 
		\bigg] \,, \\
	a_5^\mathrm{NS} =&\; \pi \left( -\frac{4159}{105} \psi_y - \frac{756}{5} \nu \zeta_y \right) \,, \\
	a_6^\mathrm{NS} =&\; \frac{16447322263}{21829500} - \frac{54784}{525}\gamma_E + \frac{512}{15}\pi^2 +\left(-\frac{56198689}{34020} + \frac{902}{15}\pi^2\right)\nu + \frac{541}{140}\nu^2-\frac{1121}{81}\nu^3 + \bigg[\frac{33232226053}{10914750}  \nonumber\\
        & - \frac{392048}{525}\gamma_E + \frac{3664}{15}\pi^2 + \left(-\frac{588778}{1701} + \frac{2747}{40}\pi^2\right) \nu -\frac{846121}{1260}\nu^2 - \frac{392945}{324} \nu^3 \bigg] e^2 + \bigg[-\frac{227539553251}{58212000} - \frac{93304}{175}\gamma_E \nonumber\\
        & + \frac{872}{5}\pi^2 + \left(\frac{124929721}{12960}-\frac{41287}{960} \pi^2 \right) \nu +\frac{148514441}{30240}\nu^2 - \frac{2198212}{405}\nu^3 \bigg] e^4  + \bigg[-\frac{300856627}{67375} - \frac{4922}{175}\gamma_E + \frac{46}{5}\pi^2 \nonumber \\
        & + \left(\frac{1588607}{432}-\frac{369}{80}\pi^2\right)\nu + \frac{12594313}{3780}\nu^2 - \frac{44338}{15}\nu^3\bigg] e^6 + \left(-\frac{243511057}{887040} +\frac{4179523}{15120}\nu + \frac{83701}{3780}\nu^2 - \frac{1876}{15}\nu^3\right) e^8
 \nonumber \\
		&+ \frac{1 - \sqrt{1 - e^2}}{\sqrt{1 - e^2}} \bigg\{-\frac{616471}{1575} + \left( 
		\frac{9874}{315} - \frac{41}{30} \pi^2 \right) \nu + \frac{632}{15} \nu^2 + \bigg[ 
		\frac{2385427}{1050} + \left( -\frac{274234}{45} + \frac{4223}{240} \pi^2 \right) \nu + 
		\frac{70946}{45} \nu^2 \bigg] e^2 \nonumber\\
		&+ \bigg[\frac{8364697}{4200} + \bigg( \frac{1900517}{630} - \frac{32267}{960} \pi^2 
		\bigg) \nu - \frac{47443}{90} \nu^2 \bigg] e^4 + \bigg[-\frac{167385119}{25200} + \left( 
		\frac{4272491}{504} - \frac{123}{160} \pi^2 \right) \nu - \frac{43607}{18} \nu^2 \bigg] 
		e^6 \nonumber\\
		&+ \bigg( -\frac{65279}{168} + \frac{510361}{1260} \nu - \frac{5623}{45} \nu^2 \bigg) e^8
		\bigg\} + \frac{1284}{175} \kappa_y \nonumber\\
		&+ \left( \frac{54784}{525} + \frac{392048}{525} e^2 + \frac{93304}{175} e^4 + 
		\frac{4922}{175} e^6 \right) \log\left[ \frac{1 + \sqrt{1 - e^2}}{8 y \left(1 - e^2 
		\right)^{3/2}} \right] \,,\\
	b_0 =&\; \frac{608}{15} e^2 + \frac{242}{15} e^4 \,, \\
	b_2 =&\; \left( -\frac{1878}{35} - \frac{8168}{45} \nu \right) e^2 + \left( \frac{59834}{105} - 
		\frac{7753}{15} \nu \right) e^4 + \left( \frac{13929}{140} - \frac{3328}{45} \nu \right) 
		e^6 \,, \\
	b_3^\mathrm{NS} =&\; \frac{788 \pi e^2}{3} \phi_e \,, \\
	b_4^\mathrm{NS} =&\; \left(-\frac{949877}{945} + \frac{18763}{21}\nu + \frac{1504}{5}\nu^2\right) e^2 + \left(-\frac{3082783}{1260} - \frac{988423}{420}\nu + \frac{64433}{20}\nu^2\right) e^4 + \bigg(\frac{23289859}{7560} -\frac{13018711}{2520}\nu \nonumber\\
        & + \frac{127411}{45}\nu^2\bigg) e^6  +  \left(\frac{420727}{1680} - \frac{362071}{1260}\nu + \frac{1642}{9}\nu^2\right) e^8 +
        \sqrt{1 - e^2}\bigg[ \left(\frac{2672}{3} - \frac{5344}{15} \nu \right) e^2 + \left( 2321 - \frac{4642}{5} \nu \right) e^4 \nonumber\\
		&+ \left( \frac{565}{3} - \frac{226}{3} \nu \right) e^6\bigg] 
		\,, \\
	b_5^\mathrm{NS} =&\; \pi\left( -\frac{55691}{105} \psi_e - \frac{610144}{315} \nu \zeta_e \right) e^2 \,, \\
	b_6^\mathrm{NS} =&\; \left[\frac{61669369961}{4365900} - \frac{2633056}{1575}\gamma_E + \frac{24608}{45}\pi^2 +\left(\frac{50099023}{56700}+\frac{779}{5}\pi^2\right)\nu - \frac{4088921}{1260}\nu^2 - \frac{61001}{243}\nu^3 \right] e^2 \nonumber\\
		&+ \left[\frac{66319591307}{21829500} - \frac{9525568}{1575}\gamma_E + \frac{89024}{45}\pi^2 +\left(\frac{28141879}{450}-\frac{139031}{480}\pi^2\right)\nu - \frac{21283907}{1512}\nu^2 -\frac{86910509}{9720}\nu^3  \right] e^4 \nonumber\\
		&+ \left[-\frac{1149383987023}{58212000} - \frac{4588588}{1575}\gamma_E +\frac{42884}{45}\pi^2 + \left(\frac{11499615139}{453600}-\frac{271871}{960}\pi^2\right)\nu + \frac{61093675}{2016}\nu^2  - \frac{2223241}{90}\nu^3\right] e^6 \nonumber\\
		&+ \left[\frac{40262284807}{4312000} - \frac{20437}{175}\gamma_E + \frac{191}{5}\pi^2 + \left(-\frac{5028323}{280}-\frac{6519}{320}\pi^2\right)\nu +\frac{24757667}{1260}\nu^2 -\frac{11792069}{1215}\nu^3 \right] e^8 \nonumber\\
		&+ \bigg(\frac{302322169}{887040} - \frac{1921387}{5040}\nu +\frac{41179}{108}\nu^2 - \frac{386792}{1215}\nu^3 \bigg) e^{10} + \sqrt{1 - e^2} \bigg\{\bigg[ 
		-\frac{22713049}{7875} + \left( -\frac{11053982}{945} + \frac{8323}{90} \pi^2 \right) \nu 
		\nonumber\\
		&+\frac{108664}{45} \nu^2 \bigg] e^2 +\left[\frac{178791374}{7875} + \left( 
		-\frac{38295557}{630} + \frac{94177}{480} \pi^2 \right) \nu + \frac{681989}{45} \nu^2 
		\right] e^4 + \bigg[\frac{5321445613}{189000} \nonumber\\
		&+ \left( -\frac{26478311}{756} + \frac{2501}{1440} \pi^2 \right) \nu + \frac{450212}{45} 
		\nu^2 \bigg] e^6 + \left[\frac{186961}{168} - \frac{289691}{252}\nu + \frac{3197}{9} \nu^2 
		\right] e^8 \bigg\} + \frac{1460336}{23625} \left( 1 - \sqrt{1 - e^2} \right) \nonumber\\ 
		&+ \frac{428}{1575} e^2 \kappa_e + \left( \frac{2633056}{1575} e^2 + \frac{9525568}{1575} 
		e^4 + \frac{4588588}{1575} e^6 + \frac{20437}{175} e^8 \right) \log\left[ \frac{1 + \sqrt{1 
		- e^2}}{8 y \left(1 - e^2 \right)^{3/2}} \right] \,,
\end{align}
\end{subequations}
\noindent where the tail terms are given by~\cite{Klein:2018ybm}
\begin{subequations}
\begin{align}
	\phi_y =& \left(1 - e^2 \right)^{7/2} \tilde{\phi} \nonumber \\ 
 =&  1 + \frac{97}{32} e^2 + \frac{49}{128} e^4 - \frac{49}{18432} e^6 - 
		\frac{109}{147456} e^8 - \frac{2567}{58982400} e^{10} + \ord{e^{12}} \,,\\
	\phi_e =& \frac{192 \left(1 - e^2 \right)^{9/2}}{985 e^2} \left( \sqrt{1- e^2} \phi - 
		\tilde{\phi} \right) \nonumber \\ 
 =& 1 + \frac{5969}{3940} e^2 + \frac{24217}{189120} e^4 + \frac{623}{4538880} e^6 - 
		\frac{96811}{363110400} e^8 - \frac{5971}{4357324800} e^{10} + \ord{e^{12}} \,, \\
	\psi_y =& \left( 1 - e^2 \right)^{9/2} \left( -\frac{8064}{4159} \sqrt{1 - e^2} \phi + 
		\frac{4032}{4159} \tilde{\phi} + \frac{8191}{4159} \tilde{\psi} \right) \nonumber \\ 
 =& 1 - \frac{207671}{8318} e^2 - \frac{8382869}{266176} e^4 - \frac{8437609}{4791168} 
		e^6 + \frac{10075915}{306634752} e^8 - \frac{38077159}{15331737600} e^{10} + \ord{e^{12}} \,, \\
	\zeta_y =& \left(1 - e^2 \right)^{7/2} \left[ \frac{160 \left(1 - e^2 \right)^{3/2}}{567} \phi 
		+ \left( -\frac{176}{567} + \frac{80}{567} e^2 \right) \tilde{\phi} + \frac{583 \left(1 - 
		e^2 \right)}{567} \tilde{\zeta} \right] \nonumber \\ 
 =& 1 + \frac{113002}{11907} e^2 + \frac{6035543}{762048} e^4 + \frac{253177}{571536} 
		e^6 - \frac{850489}{877879296} e^8 - \frac{1888651}{10973491200} e^{10} + \ord{e^{12}} \,, \\
	\psi_e =& \frac{16382 \left(1 - e^2\right)^{9/2}}{55691 e^2} \left[ \left( \frac{9408}{8191} - 
		\frac{14784}{8191} e^2 \right) \sqrt{1 - e^2} \phi+\left(-\frac{9408}{8191} + 
		\frac{4032}{8191} e^2 \right) \tilde{\phi} + \left(1 - e^2 \right) \left( \sqrt{1 - e^2} 
		\psi - \tilde{\psi} \right)\right] \nonumber \\ 
 =& 1 - \frac{9904271}{891056} e^2 - \frac{101704075}{10692672} e^4 - 
		\frac{217413779}{513248256} e^6 + \frac{35703577}{6843310080} e^8 - 
		\frac{3311197679}{9854366515200} e^{10} + \ord{e^{12}} \,, \\
	\zeta_e =& \frac{12243 \left(1 - e^2\right)^{9/2}}{76268 e^2} \left[-\frac{16 \left(1 - e^2 
		\right)^{3/2}}{53} \phi	+ \left( \frac{16}{53} - \frac{80}{583} e^2 \right) \tilde{\phi} 
		+ \left(1 - e^2 \right) \left( \sqrt{1 - e^2} \zeta - \tilde{\zeta} \right)\right] \nonumber \\ 
 =& 1 + \frac{11228233}{2440576} e^2 + \frac{37095275}{14643456} e^4 + 
		\frac{151238443}{1405771776} e^6 - \frac{118111}{611205120} e^8 - 		
		\frac{407523451}{26990818099300} e^{10} + \ord{e^{12}} \,, \\
	\kappa_y =& -\frac{ 934088 \left(1 - e^2 \right)^5}{33705} \left( \tilde{\kappa} - \tilde{F} 
		\right)	\nonumber \\ 
 =& 244 \log 2 \left( e^2 - \frac{18881}{1098} e^4 + \frac{6159821}{39528} e^6 - 
		\frac{16811095}{19764} e^8 + \frac{446132351}{123525} e^{10} \right) - 243 \log 3 \bigg(e^2 
		- \frac{39}{4} e^4 + \frac{2735}{64}e^6 \nonumber\\
		&+ \frac{25959}{512} e^8 - \frac{638032239}{409600} e^{10}\bigg) - \frac{48828125 \log 
		5}{5184} \left( e^6 - \frac{83}{8} e^8 + \frac{12637}{256} e^{10} \right) - 
		\frac{4747561509943 \log 7}{33177600} e^{10} + \ord{e^{12}} \,, \\
	\kappa_e =& -\frac{5604528 \left(1 - e^2 \right)^6}{3745 e^2} \left[ \sqrt{1 - e^2} \left( 
		\kappa - F \right) - \left(\tilde{\kappa} - \tilde{F} \right) \right] \nonumber \\ 
 =& 6536 \log 2 \left(1 - \frac{22314}{817} e^2 + \frac{7170067}{19608}e^4 - 
		\frac{10943033}{4128} e^6 + \frac{230370959}{15480} e^8 - \frac{866124466133}{8823600} 
		e^{10} \right) \nonumber\\
		&- 6561 \log 3 \left( 1 - \frac{49}{4} e^2 + \frac{4369}{64} e^4 + \frac{214449}{512} e^6 - 
		\frac{623830739}{81920} e^8 + \frac{76513915569}{1638400} e^{10} \right) \nonumber\\
		&- \frac{48828125 \log 5}{64} \left( e^4 - \frac{293}{24} e^6 + \frac{159007}{2304} e^8 - 
		\frac{6631171}{27648} e^{10} \right) - \frac{4747561509943 \log 7}{245760} \left( e^8 - 
		\frac{259}{20} e^{10} \right) + \ord{e^{12}} \,.
\end{align}
\end{subequations}

The spin-orbit (SO) terms can be found in Refs.~\cite{Klein:2018ybm,Henry:2023tka}, and in terms of $\chi_\mathrm{eff}$ (Eq.~\ref{eq:chi_eff_0}) and $\delta \chi$ (Eq.~\ref{eq:dchi_def}) they are given by:

\begin{subequations}
\begin{align}
a_3^\mathrm{SO} =& 
    \left(-\frac{752}{15} -138 e^2 -\frac{611}{30} e^4\right) \chi_\mathrm{eff} + \left(-\frac{152}{15} -\frac{154}{15} e^2 + \frac{17}{30} e^4\right) \delta\mu \delta\chi \, , \\
a_5^\mathrm{SO} =&
    \bigg[-\frac{5861}{45}+\frac{4004}{15}\nu+\left(-\frac{968539}{630}+\frac{259643}{135}\nu\right) e^2+\left(-\frac{4856917}{2520}+\frac{943721}{540}\nu\right) e^4+\left(-\frac{64903}{560}+\frac{5081}{45}\nu\right) e^6 \nonumber \\
    &+\frac{e^2}{\sqrt{1-e^2}} \left(-\frac{1416}{5}+\frac{1652}{15}\nu+\left(\frac{2469}{5}-\frac{5761}{30}\nu\right) e^2+\left(\frac{222}{5}-\frac{259}{15}\nu\right)e^4\right)\bigg] \chi_\mathrm{eff} \nonumber\\
    &+\bigg[-\frac{21611}{315}+\frac{632}{15}\nu+\left(-\frac{55415}{126}+\frac{36239}{135}\nu\right) e^2+\left(-\frac{72631}{360}+\frac{12151}{108}\nu\right) e^4 +\left(\frac{909}{560}-\frac{143}{45}\nu\right) e^6 \nonumber\\
    &+\frac{e^2}{\sqrt{1-e^2}} \left(-\frac{472}{5}+\frac{236}{15}\nu+\left(\frac{823}{5}-\frac{823}{30}\nu\right) e^2+\left(\frac{74}{5}-\frac{37}{15}\nu\right) e^4\right)\bigg] \delta \mu  \delta \chi \, , \\
a_6^\mathrm{SO} =& 
    -\frac{3008}{15} \pi  \theta_{y\chi} \chi_\mathrm{eff} -\frac{592}{15} \pi  \theta_{y\delta} \delta \mu  \delta \chi \, , \\
b_3^\mathrm{SO} =& 
    e^2 \left(-\frac{3272}{9}-\frac{26263}{45}e^2 - \frac{812}{15}e^4\right) \chi_\mathrm{eff} +e^2 \left(-\frac{3328}{45}-\frac{1993}{45}e^2+\frac{23}{15}e^4\right) \delta \mu  \delta \chi \, , \\
b_5^\mathrm{SO} =& 
    e^2 \bigg[-\frac{13103}{35}+\frac{289208}{135}\nu+\left(-\frac{548929}{63}+\frac{61355}{6}\nu\right) e^2+\left(-\frac{6215453}{840}+\frac{1725437}{270}\nu\right) e^4+\left(-\frac{87873}{280}+\frac{13177}{45}\nu\right) e^6 \nonumber\\
    & + \sqrt{1-e^2} \left(-1184+\frac{4144}{9}\nu+\left(-\frac{13854}{5}+\frac{16163}{15}\nu\right) e^2+\left(-\frac{626}{5}+\frac{2191}{45}\nu\right) e^4\right)\bigg] \chi_\mathrm{eff} \nonumber \\ 
    & + e^2 \bigg[-\frac{32857}{105}+\frac{52916}{135}\nu+\left(-\frac{1396159}{630}+\frac{126833}{90}\nu\right) e^2+\left(-\frac{203999}{280}+\frac{56368}{135}\nu\right) e^4+\left(\frac{5681}{1120}-\frac{376}{45}\nu\right) e^6\nonumber\\ 
    & + \sqrt{1-e^2} \left(-\frac{1184}{3}+\frac{592}{9}\nu+\left(-\frac{4618}{5}+\frac{2309}{15}\nu\right) e^2+\left(-\frac{626}{15}+\frac{313}{45}\nu\right) e^4\right)\bigg] \delta \mu  \delta \chi \, , \\
b_6^\mathrm{SO} =& 
    e^2 \left(-\frac{92444}{45} \pi  \theta_{e\chi} \chi_\mathrm{eff} -\frac{19748}{45} \pi  \theta_{e\delta} \delta \mu  \delta \chi \right) \, ,
\end{align}
\end{subequations}

\noindent where we have introduced the following enhancement functions, related to the spin-orbit tail terms,

\begin{subequations}
    \begin{align}
    \theta_{y\chi} = & 1 + \frac{21263}{3008}e^2 + \frac{52387}{12032}e^4 + \frac{253973}{1732608}e^6 - \frac{82103}{13860864}e^8  + \ord{e^{10}} \, , \\
    \theta_{y\delta} = & 1+\frac{1897}{592}e^2 - \frac{461}{2368}e^4 - \frac{42581}{340992}e^6 - \frac{3803}{1363968}e^8 + \ord{e^{10}} \, , \\
    \theta_{e\chi} = & 1 + \frac{377077}{92444}e^2 + \frac{7978379}{4437312}e^4 + \frac{5258749}{106495488}e^6 + \ord{e^8} \, , \\
    \theta_{e\delta} = & 1 + \frac{37477}{19748}e^2+\frac{95561}{947904}e^4 - \frac{631523}{22749696}e^6 + \ord{e^8} \, . 
    \end{align}
\end{subequations}

For the spin-spin (SS) part, we take the 2PN fully-spinning contributions from Ref.~\cite{Klein:2018ybm}, given in terms of the following function

\begin{align}
	\sigma(a, b, c, a_1 + a_2 q, b_1 + b_2 q, c_1 + c_2 q) =&\; a \bm{s}^2 + b \left( \uvec{L} 
		\cdot \bm{s} \right)^2 + c \left| \uvec{L} \times \bm{s} \right|^2 \cos 2 \psi \nonumber\\
		&+ \sum_{i=1}^2 \left[ \left(a_1 + a_2 q_i \right) \bm{s}_i^2 + \left( b_1 + b_2 q_i 
		\right) \left( \uvec{L} \cdot \bm{s}_i \right)^2 + \left(c_1 + c_2 q_i \right) \left| 
		\uvec{L} \times \bm{s}_i \right|^2 \cos 2 \psi_i \right] \, ,
\end{align}

\noindent where $q_i$ is the quadrupole parameter, defined in such a way that $q_i = 1$ for black holes. The 2PN fully-spinning contribution to the evolution equations of $y$ and $e^2$ are

\begin{subequations}
    \label{eq:ab4SS_full}
    \begin{align}
a_4^\mathrm{SS} = &
    \sigma \bigg( - \frac{84}{5} - \frac{228}{5} e^2 - \frac{33}{5} e^4, \frac{242}{5} 
	+ \frac{654}{5} e^2 + \frac{381}{20} e^4, - \frac{447}{10} e^2 - \frac{93}{10} e^4, \frac{88}{5} - 16 q + \left(48 - \frac{216}{5} q \right) e^2 + \left( \frac{69}{10} - \frac{63}{10} q \right) e^4, \nonumber\\
	&  - \frac{244}{5} + 48 q + \left(-132 + \frac{648}{5} q \right) 
	e^2 + \left(- \frac{96}{5} + \frac{189}{10} q \right) e^4, \left(1 - q\right) \left( \frac{447}{10} e^2 + \frac{93}{10} e^4 \right) \bigg) \, , \\
b_4^\mathrm{SS} = &
    \sigma \bigg( \frac{2}{3} 
	-\frac{1961}{15} e^2 - \frac{2527}{12}e^4 - \frac{157}{8} e^6, - \frac{2}{3} + 
	\frac{5623}{15} e^2 + \frac{2393}{4} e^4 + \frac{447}{8} e^6, -\frac{5527}{30} e^2 - 
	\frac{10117}{30} e^4 - \frac{5507}{160} e^6, \nonumber\\
	& -\frac{4}{3} + \left( \frac{682}{5} - \frac{1876}{15} q \right) 
	e^2 + \left(\frac{1337}{6} - \frac{595}{3} q \right) e^4 + \left( \frac{83}{4} - 
	\frac{37}{2} q \right) e^6,	\frac{4}{3} + \left( - \frac{5618}{15} + \frac{1876}{5} q 
	\right) e^2 \nonumber\\
	&+ \left(- \frac{1203}{2} + 595 q \right) e^4 + \left( - \frac{225}{4} + \frac{111}{2} q 
	\right) e^6 , \left( \frac{2764}{15} - \frac{921}{5} q \right) e^2 + \left(\frac{1687}{5} - 
	\frac{5056}{15} q \right) e^4 + \left( \frac{551}{16} - \frac{172}{5} q \right) e^6 \bigg)  \, .
    \end{align}
\end{subequations}

The next spin-spin contribution enters at 3PN, for which we consider only the aligned-spin (AS) part, derived in Ref.~\cite{Henry:2023tka}. To simplify the equations we introduce the following parameters.

\begin{subequations}
    \begin{align}
        \delta q_S  = q_1 + q_2 - 2 \, , \\
        \delta q_A  = q_1 - q_2 \, ,
    \end{align}
\end{subequations}

\noindent which measure the symmetric and anti-symetric deviations from the quadrupole parameters of a binary black hole. In terms of these new variables, the 3PN aligned spin-spin coefficients are

\begin{subequations}
\begin{align}
a_6^\mathrm{SS,AS} = &
    \bigg\{\frac{30596}{105}+\frac{2539 \delta q_S}{105}+\frac{443 \delta q_A \delta \mu }{30}+\left(-\frac{688}{5}-\frac{172 \delta q_S}{5}\right) \nu +\bigg[\frac{115078}{45}+\frac{21317 \delta q_S}{60}+\frac{3253 \delta q_A \delta \mu }{60} \nonumber\\
    & +\left(-\frac{3962}{3}-\frac{1981 \delta q_S}{6}\right) \nu \bigg] e^2+\bigg[\frac{4476649}{2520}+\frac{133703 \delta q_S}{420}+\frac{481 \delta q_A \delta \mu }{48}+\left(-\frac{53267}{45}-\frac{53267 \delta q_S}{180}\right) \nu \bigg] e^4 \nonumber \\
    & + \bigg[\frac{17019}{140}+\frac{29831 \delta q_S}{1120}+\frac{29 \delta q_A \delta \mu }{160}+\left(-\frac{1343}{15}-\frac{1343 \delta q_S}{60}\right) \nu \bigg] e^6+\frac{1-\sqrt{1-e^2}}{\sqrt{1-e^2}} \bigg(-\frac{244}{15}-\frac{52 \delta q_S}{15}-\frac{4 \delta q_A \delta \mu }{15} \nonumber\\
    & + \left(\frac{16}{5}+\frac{4 \delta q_S}{5}\right) \nu +\bigg[\frac{6283}{30}+\frac{1339 \delta q_S}{30}+\frac{103 \delta q_A \delta \mu }{30}+\left(-\frac{206}{5}-\frac{103 \delta q_S}{10}\right) \nu \bigg] e^2 + \bigg[-\frac{48007}{120}-\frac{10231 \delta q_S}{120} \nonumber\\
    & -\frac{787 \delta q_A \delta \mu }{120}+\left(\frac{787}{10}+\frac{787 \delta q_S}{40}\right) \nu \bigg] e^4+\bigg[-\frac{183}{20}-\frac{39 \delta q_S}{20}-\frac{3 \delta q_A \delta \mu }{20}+\left(\frac{9}{5}+\frac{9 \delta q_S}{20}\right) \nu \bigg] e^6 \bigg) \bigg\} \chi_\mathrm{eff}^2 \nonumber\\
    & + \bigg\{\left(\frac{3134}{15}+\frac{443 \delta q_S}{15}\right) \delta \mu +\left(\frac{5078}{105}-\frac{344 \nu }{5}\right) \delta q_A+\bigg[\left(\frac{30421}{45}+\frac{3253 \delta q_S}{30}\right) \delta \mu +\left(\frac{21317}{30}-\frac{1981 \nu }{3}\right) \delta q_A\bigg] e^2 \nonumber\\
    & + \bigg[\left(-\frac{111}{5}+\frac{481 \delta q_S}{24}\right) \delta \mu +\left(\frac{133703}{210}-\frac{53267 \nu }{90}\right) \delta q_A\bigg] e^4+\bigg[\left(-\frac{149}{40}+\frac{29 \delta q_S}{80}\right) \delta \mu +\left(\frac{29831}{560}-\frac{1343 \nu }{30}\right) \delta q_A\bigg] e^6 \nonumber\\
    & + \frac{1-\sqrt{1-e^2}}{\sqrt{1-e^2}} \bigg(\left(-\frac{104}{15}-\frac{8 \delta q_S}{15}\right) \delta \mu +\left(-\frac{104}{15}+\frac{8 \nu }{5}\right) \delta q_A+\bigg[\left(\frac{1339}{15}+\frac{103 \delta q_S}{15}\right) \delta \mu +\left(\frac{1339}{15}-\frac{103 \nu }{5}\right) \delta q_A\bigg] e^2 \nonumber\\
    & + \bigg[\left(-\frac{10231}{60}-\frac{787 \delta q_S}{60}\right) \delta \mu +\left(-\frac{10231}{60}+\frac{787 \nu }{20}\right) \delta q_A\bigg] e^4+ \bigg[\left(-\frac{39}{10}-\frac{3 \delta q_S}{10}\right) \delta \mu +\left(-\frac{39}{10}+\frac{9 \nu }{10}\right) \delta q_A\bigg]e^6\bigg)\bigg\} \chi_\mathrm{eff}  \delta \chi \nonumber\\
    & +\bigg\{\frac{39}{5}+\frac{2539 \delta q_S}{105}+\frac{443 \delta q_A \delta \mu }{30}+\left(-\frac{1163}{15}-\frac{172 \delta q_S}{5}\right) \nu +\bigg[\frac{659}{15}+\frac{21317 \delta q_S}{60}+\frac{3253 \delta q_A \delta \mu }{60} \nonumber\\
    & + \left(-\frac{2399}{15}-\frac{1981 \delta q_S}{6}\right) \nu \bigg] e^2 + \bigg[\frac{1769}{90}+\frac{133703 \delta q_S}{420}+\frac{481 \delta q_A \delta \mu }{48}+\left(\frac{2021}{72}-\frac{53267 \delta q_S}{180}\right) \nu \bigg] e^4 + \bigg[\frac{19}{10}\nonumber\\
    & + \frac{29831 \delta q_S}{1120} + \frac{29 \delta q_A \delta \mu }{160}+\left(-\frac{3}{10}-\frac{1343 \delta q_S}{60}\right) \nu \bigg]e^6 + \frac{1-\sqrt{1-e^2}}{\sqrt{1-e^2}} \bigg(-\frac{4}{15}-\frac{52 \delta q_S}{15}-\frac{4 \delta q_A \delta \mu }{15} \nonumber\\
    & + \left(\frac{32}{15}+\frac{4 \delta q_S}{5}\right) \nu + \bigg[\frac{103}{30}+\frac{1339 \delta q_S}{30}+\frac{103 \delta q_A \delta \mu }{30}+\left(-\frac{412}{15}-\frac{103 \delta q_S}{10}\right) \nu \bigg] e^2+\bigg[-\frac{787}{120}-\frac{10231 \delta q_S}{120} \nonumber\\
    &- \frac{787 \delta q_A \delta \mu }{120} + \left(\frac{787}{15}+\frac{787 \delta q_S}{40}\right) \nu \bigg]e^4 + \bigg[-\frac{3}{20}-\frac{39 \delta q_S}{20}-\frac{3 \delta q_A \delta \mu }{20}+\left(\frac{6}{5}+\frac{9 \delta q_S}{20}\right) \nu \bigg] e^6\bigg)\bigg\} \delta \chi^2 \, , \\
b_6^\mathrm{SS,AS} = &
    e^2 \bigg\{\frac{1468414}{945}+\frac{2852 \delta q_S}{105}+\frac{3461 \delta q_A \delta \mu }{30}+\left(-\frac{57844}{45}-\frac{14461 \delta q_S}{45}\right) \nu +\bigg[\frac{47715853}{3780}+\frac{1464091 \delta q_S}{840}+\frac{11007 \delta q_A \delta \mu }{40}\nonumber\\
    & + \left(-\frac{21865}{3}-\frac{21865 \delta q_S}{12}\right) \nu \bigg] e^2 + \bigg[\frac{4255831}{504}+\frac{166844 \delta q_S}{105}+\frac{2941 \delta q_A \delta \mu }{48}+\left(-\frac{222533}{45}-\frac{222533 \delta q_S}{180}\right) \nu \bigg] e^4 \nonumber\\
    & + \bigg[\frac{414027}{1120}+\frac{365363 \delta q_S}{4480}+\frac{511 \delta q_A \delta \mu }{640}+\left(-\frac{1287}{5}-\frac{1287 \delta q_S}{20}\right) \nu \bigg] e^6 + \sqrt{1-e^2} \bigg(\frac{49532}{45}+\frac{10556 \delta q_S}{45}+\frac{812 \delta q_A \delta \mu }{45} \nonumber\\
    & + \left(-\frac{3248}{15}-\frac{812 \delta q_S}{15}\right) \nu +\bigg[\frac{140117}{60}+\frac{29861 \delta q_S}{60}+\frac{2297 \delta q_A \delta \mu }{60}+\left(-\frac{2297}{5}-\frac{2297 \delta q_S}{20}\right) \nu \bigg] e^2 \nonumber\\
    & + \bigg[\frac{3721}{180}+\frac{793 \delta q_S}{180}+\frac{61 \delta q_A \delta \mu }{180}+\left(-\frac{61}{15}-\frac{61 \delta q_S}{60}\right) \nu \bigg] e^4\bigg)\bigg\} \chi_\mathrm{eff}^2 \nonumber\\
    & + e^2 \bigg\{\left(\frac{176426}{135}+\frac{3461 \delta q_S}{15}\right) \delta \mu +\left(\frac{5704}{105}-\frac{28922 \nu }{45}\right) \delta q_A+\bigg[\left(\frac{387212}{135}+\frac{11007 \delta q_S}{20}\right) \delta \mu \nonumber\\
    & + \left(\frac{1464091}{420}-\frac{21865 \nu }{6}\right) \delta q_A\bigg] e^2 + \bigg[\left(\frac{2562}{5}+\frac{2941 \delta q_S}{24}\right) \delta \mu +\left(\frac{333688}{105}-\frac{222533 \nu }{90}\right) \delta q_A\bigg] e^4 \nonumber\\
    & + \bigg[\left(-\frac{33}{32}+\frac{511 \delta q_S}{320}\right) \delta \mu +\left(\frac{365363}{2240}-\frac{1287 \nu }{10}\right) \delta q_A\bigg] e^6+\sqrt{1-e^2} \bigg(\left(\frac{21112}{45}+\frac{1624 \delta q_S}{45}\right) \delta \mu \nonumber\\
    & +\left(\frac{21112}{45}-\frac{1624 \nu }{15}\right) \delta q_A+\bigg[\left(\frac{29861}{30}+\frac{2297 \delta q_S}{30}\right) \delta \mu + \left(\frac{29861}{30}-\frac{2297 \nu }{10}\right)\delta q_A \bigg] e^2 \nonumber\\
    & + \bigg[\left(\frac{793}{90}+\frac{61 \delta q_S}{90}\right) \delta \mu + \left(\frac{793}{90}-\frac{61 \nu }{30}\right)\delta q_A\bigg] e^4\bigg)\bigg\} \chi_\mathrm{eff}  \delta \chi \nonumber\\
    & + e^2 \bigg\{\frac{8887}{135}+\frac{2852 \delta q_S}{105}+\frac{3461 \delta q_A \delta \mu }{30}+\left(-\frac{13127}{27}-\frac{14461 \delta q_S}{45}\right) \nu + \bigg[\frac{161077}{540}+\frac{1464091 \delta q_S}{840}+\frac{11007 \delta q_A \delta \mu }{40} \nonumber\\
    & + \left(-\frac{185723}{270}-\frac{21865 \delta q_S}{12}\right) \nu \bigg] e^2+\bigg[\frac{14827}{90}+\frac{166844 \delta q_S}{105}+\frac{2941 \delta q_A \delta \mu }{48}+\left(-\frac{45373}{360}-\frac{222533 \delta q_S}{180}\right) \nu \bigg] e^4 \nonumber\\
    & + \bigg[\frac{283}{32}+\frac{365363 \delta q_S}{4480}+\frac{511 \delta q_A \delta \mu }{640}+\left(-\frac{117}{20}-\frac{1287 \delta q_S}{20}\right) \nu \bigg] e^6+\sqrt{1-e^2} \bigg(\frac{812}{45}+\frac{10556 \delta q_S}{45}+\frac{812 \delta q_A \delta \mu }{45}\nonumber\\
    & + \left(-\frac{6496}{45}-\frac{812 \delta q_S}{15}\right) \nu + \bigg[\frac{2297}{60}+\frac{29861 \delta q_S}{60}+\frac{2297 \delta q_A \delta \mu }{60}+\left(-\frac{4594}{15}-\frac{2297 \delta q_S}{20}\right) \nu \bigg] e^2\nonumber\\
    & + \bigg[\frac{61}{180}+\frac{793 \delta q_S}{180}+\frac{61 \delta q_A \delta \mu }{180}+\left(-\frac{122}{45}-\frac{61 \delta q_S}{60}\right) \nu \bigg] e^4\bigg)\bigg\} \delta \chi^2 \, .
\end{align}
\end{subequations}

\subsection{Evolution of the argument of periastron $\delta\lambda$}
\label{sec:appendix:PN_formulas:k}

For the evolution equation of the argument of periastron $\delta\lambda$ (Eq.~\eqref{eq:RR_eqs_no_SP:dl}), we need an expression for the periastron advance $k$, which we express as the following PN series:

\begin{equation}
    k = y^2\sum_{n=0}^{4} k_n y^n.
    \label{eq:k_PN_series}
\end{equation}

The non-spinning part of $k_n$ can be obtained from Ref.~\cite{Arun:2009mc}, by taking the expression of the periastron advance $k$ in terms of the dimensionless energy $\varepsilon$ and angular momentum $j$, substituting the expressions for $\varepsilon(x,e^\mathrm{ADM})$ and $j(x,e^\mathrm{ADM})$, and converting the eccentricity from ADM to harmonic coordinates using the expressions also provided in that reference. Doing this we obtain
\begin{subequations}
\begin{align}
    k_0 =& 3  \, , \\
    k_2^\mathrm{NS} =& \frac{27}{2}-7 \nu +\left(\frac{51}{4}-\frac{13}{2}\nu\right) e^2   \, , \\
    k_4^\mathrm{NS} =& \frac{105}{2}+\left(-\frac{625}{4}+\frac{123}{32}\pi^2\right)\nu + 7 \nu^2+\left(\frac{573}{4}+\left(-\frac{357}{2}+\frac{123}{128}\pi^2\right)\nu +40 \nu^2\right) e^2 \nonumber\\
    &+\left(\frac{39}{2}-\frac{55}{4}\nu+\frac{65}{8}\nu^2\right) e^4  +\sqrt{1-e^2} \left(15-6 \nu +(30-12 \nu ) e^2\right)  \, . 
\end{align}
\end{subequations}

The fully-spinning part of $k_n$ can be found to 2PN order in Ref.~\cite{Klein:2018ybm}

\begin{subequations}
\label{eq:kS_full}
\begin{align}
    k_1 =& -\frac{7}{2} \chi_\mathrm{eff} - \frac{1}{2} \delta\mu \delta\chi \, , \\
    k_2^\mathrm{SS} =& \frac{3}{4}\left\{ 3 \left(\uvec{L} \cdot \bm{s} \right)^2 - \bm{s}^2 + \sum_{i=1}^2 \left(q_i - 1\right) \left[ 3 \left(\uvec{L} \cdot \bm{s}_i \right)^2 - \bm{s}_i^2 \right] \right\}  \, . \label{eq:kS_full:SS}
\end{align}
\end{subequations}

\noindent Finally, the aligned-spin contributions to the periastron advance $k$ can be found to 3PN in the supplementary material of Ref.~\cite{Henry:2023tka}. Using our notation, these coefficients are

\begin{subequations}
\begin{align}
    k_3 =& \left[-26+8 \nu +\left(-\frac{105}{4}+\frac{49 \nu }{4}\right) e^2\right] \chi_\mathrm{eff} +\left[-8+\frac{\nu }{2}+\left(-\frac{15}{4}+\frac{7 \nu }{4}\right) e^2\right] \delta \mu  \delta \chi \, , \\
    k_4^\mathrm{SS,AS} =& \left\{\frac{181}{8}+\frac{33 \delta q_S}{8}+\frac{3 \delta q_A \delta \mu }{4}+\left(-\frac{5}{2}-\frac{5 \delta q_S}{8}\right) \nu +\left[\frac{369}{16}+\frac{75 \delta q_S}{16}+\frac{3 \delta q_A \delta \mu }{16}+\left(-\frac{29}{4}-\frac{29 \delta q_S}{16}\right) \nu \right] e^2\right\} \chi_\mathrm{eff}^2 \nonumber\\
    & + \left\{\left(\frac{43}{4}+\frac{3 \delta q_S}{2}\right) \delta \mu +\left(\frac{33}{4}-\frac{5 \nu }{4}\right) \delta q_A+\left[\left(\frac{21}{8}+\frac{3 \delta q_S}{8}\right) \delta \mu +\left(\frac{75}{8}-\frac{29 \nu }{8}\right) \delta q_A\right] e^2\right\} \chi_\mathrm{eff} \delta \chi \nonumber\\
    & + \left\{\frac{1}{8}+\frac{33 \delta q_S}{8}+\frac{3 \delta q_A \delta \mu }{4}+\left(-\frac{7}{2}-\frac{5 \delta q_S}{8}\right) \nu +\left[-\frac{3}{16}+\frac{75 \delta q_S}{16}+\frac{3 \delta q_A \delta \mu }{16}-\frac{29 \delta q_S \nu }{16}\right] e^2\right\} \delta \chi ^2 \, . 
\end{align}
\end{subequations}

\subsection{Precession averaged fully spinning spin-spin coefficients}
\label{sec:appendix:PN_formulas:SpinAvg-SS}

The precession averages for most of the PN coefficients given in Appendices~\ref{sec:appendix:PN_formulas:ye2} and \ref{sec:appendix:PN_formulas:k} are straight-forward, since they are written in terms of $\delta\chi$ and $\delta\chi^2$. Therefore, to compute their averages it will suffice to do $\delta \chi \to \av{\delta \chi}$ and $\delta\chi^2 \to \av{\delta \chi^2}$, which can be computed with Eq.~\eqref{eq:dchi_prec_avg} and Eq.~\eqref{eq:dchi2_prec_avg} respectively. In contrast, the fully-spinning coefficients entering at 2PN, as written in Eq.~\eqref{eq:ab4SS_full} and Eq.~\eqref{eq:kS_full:SS}, depend on different quadratic functions of the full spins of the component objects, whose averages are given in Sec.~\ref{sec:RR:spin_avg}. Substituting these averages in the fully spinning spin-spin coefficients we obtain

\begin{subequations}
    \begin{align}
        \av{a_4^\mathrm{SS}} = & \left[-\frac{84}{5}-\frac{228}{5} e^2-\frac{33}{5} e^4\right] \av{s_\perp^2} + \left[\frac{8}{5}-8 \delta q_S+\left(\frac{24}{5}-\frac{108 \delta q_S}{5}\right) e^2+\left(\frac{3}{5}-\frac{63 \delta q_S}{20}\right) e^4\right] \left(s_1^2+s_2^2\right)\nonumber\\
        & + \left[-8-\frac{108}{5} e^2-\frac{63}{20} e^4\right] \delta q_A \left(s_1^2-s_2^2\right)+\left[\frac{156}{5}+12 \delta q_S+\left(84+\frac{162 \delta q_S}{5}\right) e^2+\left(\frac{123}{10}+\frac{189 \delta q_S}{40}\right) e^4\right] \chi_\mathrm{eff}^2 \nonumber\\
        & + \left[24+\frac{324}{5} e^2+\frac{189}{20} e^4\right] \delta q_A \chi_\mathrm{eff} \av{\delta \chi}  +\left[-\frac{2}{5}+12 \delta q_S+\left(-\frac{6}{5}+\frac{162 \delta q_S}{5}\right) e^2+\left(-\frac{3}{20}+\frac{189 \delta q_S}{40}\right) e^4\right] \av{\delta \chi^2} \, , \\
        \av{b_4^\mathrm{SS}} = & \left[\frac{2}{3}-\frac{1961}{15} e^2-\frac{2527}{12} e^4 - \frac{157}{8} e^6\right] \av{s_\perp^2}+\bigg[-\frac{4}{3}+\left(\frac{34}{3}-\frac{938 \delta q_S}{15}\right) e^2+\left(\frac{49}{2}-\frac{595 \delta q_S}{6}\right) e^4\nonumber\\
        & + \left(\frac{9}{4}-\frac{37 \delta q_S}{4}\right) e^6\bigg] \left(s_1^2+s_2^2\right)+\left[-\frac{938}{15}-\frac{595}{6} e^2-\frac{37}{4} e^4\right] e^2 \left(s_1^2-s_2^2\right) \delta q_A+\bigg[\frac{2}{3}+\left(\frac{3667}{15}+\frac{469 \delta q_S}{5}\right) e^2 \nonumber\\
        & + \left(\frac{4613}{12}+\frac{595 \delta q_S}{4}\right) e^4+\left(\frac{287}{8}+\frac{111 \delta q_S}{8}\right) e^6\bigg] \chi_\mathrm{eff}^2+\left[\frac{938}{5}+\frac{595}{2} e^2+\frac{111}{4} e^4\right] e^2 \delta q_A \chi_\mathrm{eff} \av{\delta \chi} \nonumber\\
        & + \left[\frac{2}{3}+\left(\frac{1}{3}+\frac{469 \delta q_S}{5}\right) e^2+\left(-\frac{13}{4}+\frac{595 \delta q_S}{4}\right) e^4+\left(-\frac{3}{8}+\frac{111 \delta q_S}{8}\right) e^6\right] \av{\delta \chi^2} \, , \\
        \av{k_2^\mathrm{SS}} = & -\frac{3}{4} \av{s_\perp^2} - \frac{3}{8} \delta q_S \left(s_1^2+s_2^2\right)-\frac{3}{8} \delta q_A \left(s_1^2-s_2^2\right)+\left(\frac{3}{2}+\frac{9 \delta q_S}{16}\right) \chi_\mathrm{eff}^2+\frac{9 \delta q_A }{8} \chi_\mathrm{eff} \av{\delta \chi} +\frac{9 \delta q_S}{16}  \av{\delta \chi^2} \, ,
    \end{align}
\end{subequations}

\noindent where we have introduced the variable

\begin{equation}
    \av{s_\perp^2} = \av{s^2 - \chi_\mathrm{eff}^2}=  \av{\sigma_0^{(1)}} - \chi_\mathrm{eff}^2  =  s_{1\perp,0}^2 + s_{2\perp,0}^2 + \Delta_{J^2}^2 + \delta\mu \left(\frac{\delta\chi_0 - \av{\delta\chi}}{y}\right) \, .
    \label{eq:sperp_prec_avg}
\end{equation}

\section{Leading order time to coalescence}
\label{sec:appendix:tLO}

\begin{figure}[t!]
\centering  
\includegraphics[width=0.5\textwidth]{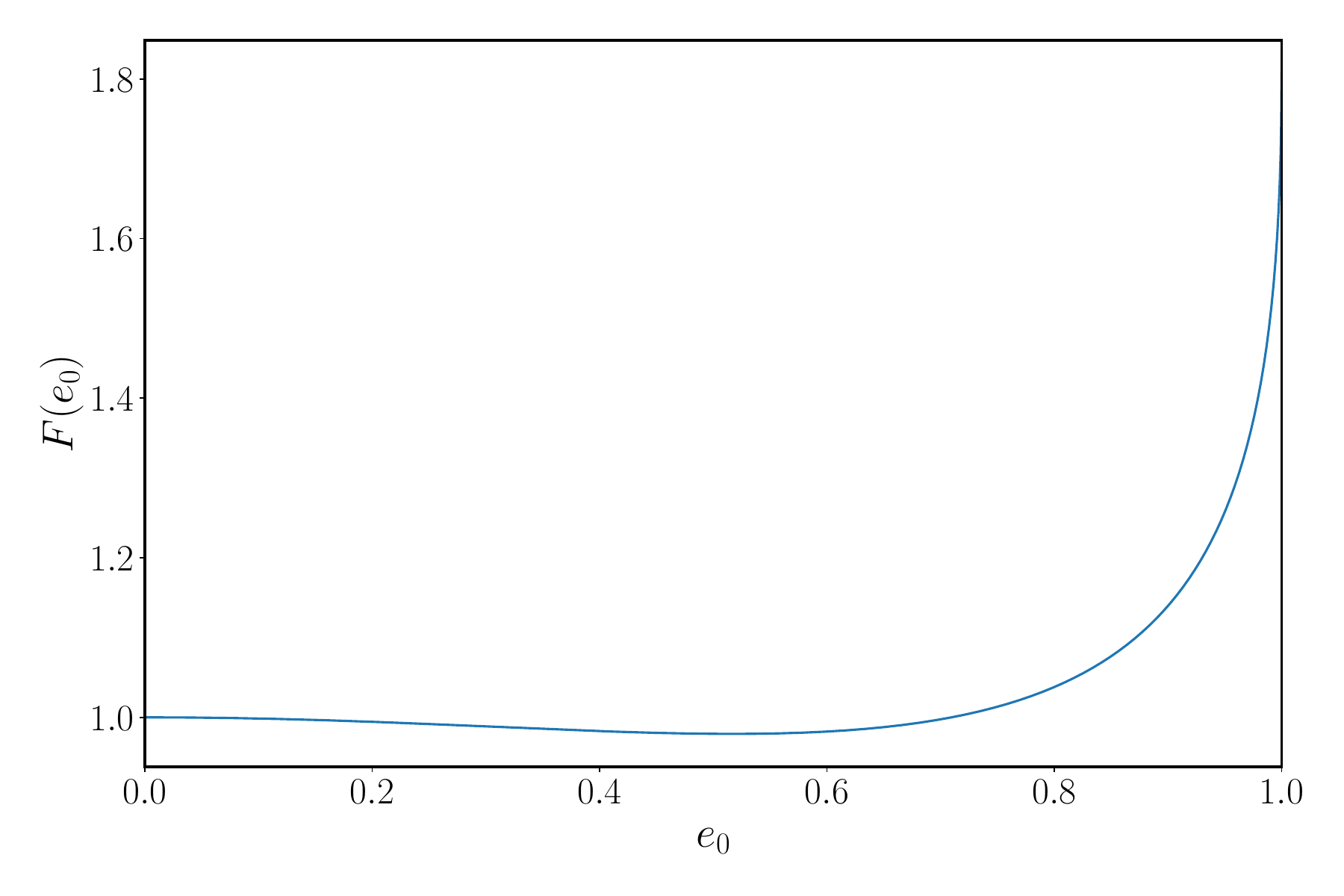}
\caption{\justifying Plot of the function $F(e_0)$, defined in Eq.~\eqref{eq:Fe_def}, as a function of the initial eccentricity $e_0$. At $e_0=0$, we have that $F(0) = 1$, while at $e_0 = 1$ we have that $F(1) = 768/425 \approx 1.81$.}
\label{fig:Fe_plot}
\end{figure}

To estimate the time to coalescence at leading order in PN, we use the 0PN expressions for $\D y$ and $\D e^2$, given in appendix~\ref{sec:appendix:PN_formulas:ye2}, i.e.
\begin{subequations}
    \label{eq:DyDe_LO}
    \begin{align}
        \D y & = \nu y^9 \left( \frac{32}{5} + \frac{28}{5} e^2 \right)\, , \label{eq:DyDe_LO:Dy} \\
        \D e^2 & = -\nu y^8 \left( \frac{608}{15} e^2 + \frac{242}{15} e^4 \right) \, . \label{eq:DyDe_LO:De} 
        \end{align}
\end{subequations}
Dividing Eq.~\eqref{eq:DyDe_LO:Dy} by Eq.~\eqref{eq:DyDe_LO:De} we obtain 

\begin{equation}
    \frac{\d y}{\d e^2} = - y \frac{48 + 42 e^2}{e^2 (304 + 121 e^2)} \, ,
    \label{eq:dy_de2_eq}
\end{equation}

\noindent which can be integrated to obtain $y$ as a function of the eccentricity $e$, i.e.
\begin{equation}
    y(e) = y_0 \frac{h(e)}{h(e_0)} \, ,
    \label{eq:y_of_e_LO}
\end{equation}
\noindent where we have defined 
\begin{equation}
    h(e) = e^{-6/19} \left( 1+ \frac{121}{304} e^2 \right)^{-435/2299} \, .
    \label{eq:he_def}
\end{equation}
From Eq.~\eqref{eq:he_def} and Eq.~\eqref{eq:y_of_e_LO}, we observe that, at coalescence ($y \to \infty$), the eccentricity becomes 0 (i.e. $e \to 0$). Therefore, we can obtain the coalescence time substituting $y(e)$ (Eq.~\eqref{eq:y_of_e_LO}) in $\D e^2$ (Eq.~\eqref{eq:DyDe_LO:De}) and integrating $\d t/\d e^2$ from $e^2=e_0^2$ to $e^2=0$, obtaining
\begin{align}
    \tau_c & = \frac{5}{256} \frac{M}{\nu y_0^8} \frac{F(e_0)}{\sqrt{1 - e_0^2}} \, ,
    \label{eq:tcoal_0PN_y0}    
\end{align}
\noindent where $F(e_0)$ is defined as~\cite{Maggiore_Vol1}
\begin{align}
    F(e_0)  & =  \frac{24}{19} h^8(e_0) \sqrt{1 - e_0^2}   \int_0^{e_0^2}  \frac{x^{5/19}  \left(1  +  \frac{121}{304} x \right)^{1181/2299}}{(1 - x)^{3/2}} \d x ,\label{eq:Fe_def} \\ 
    & = \frac{48}{19} h^8(e_0) \sqrt{1 - e_0^2} \int_1^{\frac{1}{\sqrt{1 -e_0^2}}} \left(1 - \frac{1}{u^2}\right)^{5/19} \left( \frac{425}{304} - \frac{121}{304} \frac{1}{u^2} \right)^{1181/2299} \d u \, . \label{eq:Fe_u}
\end{align}
This function, shown in Fig.~\ref{fig:Fe_plot}, is always of order one and can therefore easily be approximated. In our case, for $e_0^2 \leq 0.4$ we take the Maclaurin seies of Eq.~\eqref{eq:Fe_def} with respect to $x_0=e_0^2$, while for $e_0^2 > 0.4$ we take the Maclaurin seies of Eq.~\eqref{eq:Fe_u} with respect to $w_0=\sqrt{1 - e_0^2}$. 
Substituting the definition of $y$, given in Eq.~\eqref{eq:raw_prec_eqs_defs:y}, we can compute the time to coalescence in terms of the initial frequency and eccentricity as
\begin{equation}
    \tau_c = \frac{5}{256} \left( \frac{G \mathcal{M}_c}{c^3} \right)^{-5/3} \left(\pi f_0^\mathrm{GW,22} \right)^{-8/3} (1 - e_0^2)^{7/2} F(e_0) \, ,
    \label{eq:tcoal_0PN_f}
\end{equation}
\noindent where we have reintroduced the universal constants $G$ and $c$ and we have defined the chirp mass as
\begin{equation}
    \mathcal{M}_c = \nu^{3/5} M = \frac{(m_1 m_2)^{3/5}}{(m_1 + m_2)^{1/5}}.
    \label{eq:Mchirp_derived}
\end{equation}

\section{Series reversion}
\label{sec:appendix:SeriesReversion}
Given a function $y(x)$ with Maclaurin series
\begin{equation}
    y(x) = a_1 x + a_2 x^2 + a_3 x^3 + \ldots \, ,
    \label{eq:series_general}
\end{equation}
we want to find the Maclaurin series of the inverse function $x(y)$, which we denote 
\begin{equation}
    x(y) = A_1 y + A_2 y^2 + A_3 y^3 + \ldots \, ,
    \label{eq:inverse_series_general}
\end{equation}
such that 
\begin{equation}
    y(x(y)) = a_1 A_1 y + (a_2 A_1^2 + a_1 A_2) y ^2 + (a_3 A_1^3 + 2 a_2 A_1 A_2 + a_1 A_3) y^3 + \ldots = y \, .
    \label{eq:series_inverse_series}
\end{equation}
Solving Eq.~\eqref{eq:series_inverse_series} order by order, we can obtain the constants $A_j$, which are well known and tabulated in the literature (e.g. in Ref.~\cite{Abramowitz_and_Stegun}). These are given as:
\begin{subequations}
\begin{align}
    a_1 A_1 & = 1 \, , \label{eq:Ajs_series_reversion:A1} \\
    a_1^2 A_2 & = -\tilde{a}_2 \, , \label{eq:Ajs_series_reversion:A2} \\
    a_1^3 A_3 & = 2 \tilde{a}_2^2 - \tilde{a}_3 \, , \label{eq:Ajs_series_reversion:A3} \\
    a_1^4 A_4 & = 5 \tilde{a}_2 \tilde{a}_3 - \tilde{a}_4 - 5 \tilde{a}_2^3 \, , \label{eq:Ajs_series_reversion:A4} \\
    a_1^5 A_5 & = 6\tilde{a}_2\tilde{a}_4 + 3 \tilde{a}_3^2 + 14 \tilde{a}_2^4 - \tilde{a}_5 - 21 \tilde{a}_2^2 \tilde{a}_3 \, , \label{eq:Ajs_series_reversion:A5}
\end{align}
\label{eq:Ajs_series_reversion}
\end{subequations}    
\noindent where we have defined
\begin{equation}
    \tilde{a}_j = \frac{a_j}{a_1} \, .
    \label{eq:atilde_series_reversion}
\end{equation}
In the context of inverting Eq.~\eqref{eq:omega_modes_RK} to obtain the expression $t^\mathrm{SPA}_i (\omega)$ in Eq.~\eqref{eq:tSPA_modes_RK}. We can use the above formulas identifying 
\begin{subequations}
\begin{align}
x & \to t_i^\mathrm{SPA} - t_j \, , \\
y & \to \omega - \omega_i(t_j) \ , \\
a_k & \to 
\begin{cases}
Q_{\omega,i,j,k} &, 1 \leq k  \leq 3 \\
0 &, k \geq 4
\end{cases}\, , \\
A_k & \to Q_{t,i,j,k} \, .
\end{align}
\end{subequations}

\section{Wigner $D$-matrices and spin-weighted spherical harmonics used}
\label{sec:appendix:WignerD}

The Wigner $D$-matrix can be written in terms of the Wigner small $d$-matrix as

\begin{equation}
    D^l_{m',m}(\alpha, \beta, \gamma) = \rme^{-\rmi m' \alpha}  d^l_{m',m}(\beta) \rme^{-\rmi m \gamma} \, , 
    \label{eq:WigerD_of_Wignerd}
\end{equation}

\noindent where with our convention, the Wigner small $d$-matrix is given by

\begin{align}
d^l_{m',m}(\beta) &= (-1)^{m' - m} \sqrt{(l+m')!(l-m')!(l+m)!(l-m)!} \sum_{k = \max(0, m-m')}^{\min(l+m, l-m')} \frac{(-1)^k \sin^{2k + m' - m}\left(\frac{\beta}{2}\right) \cos^{2(l-k)+m - m'}\left(\frac{\beta}{2}\right)}{k! (k - m + m')! (l + m - k)! (l - m' - k)!} \, ,
\end{align}

\noindent which agrees with the convention of Ref.~\cite{Klein:2021jtd} for integer $m$ and $m'$. In this work, we only need to compute the Wigner matrices for $l = 2$ with $m = 0, \pm 2$. The required values of $d^l_{m',m}(\theta)$ are given by
\begin{subequations}
\begin{align}
    d^2_{2 , 2} & = \left(\frac{1 + \cos\theta}{2} \right)^2 \, , \\
    d^2_{1 , 2} & = \frac{1}{2} \sin\theta (1 + \cos\theta) \, , \\
    d^2_{0 , 2} & = \sqrt{\frac{3}{8}} \sin^2\theta \, , \\
    d^2_{-1 , 2} & = \frac{1}{2} \sin\theta (1 - \cos\theta) \, , \\
    d^2_{-2 , 2} & = \left(\frac{1 - \cos\theta}{2} \right)^2 \, , \\
    d^2_{1 , 0} & = -\sqrt{\frac{3}{2}} \sin\theta \cos\theta \, , \\
    d^2_{0 , 0} & = \frac{3 \cos^2\theta - 1}{2} \, .
\end{align}
\end{subequations}

\noindent The rest of the elements of the Wigner small $d$-matrix can be obtained by using that

\begin{equation}
    d^l_{m',m} = (-1)^{m-m'} d^l_{m,m'} = d^l_{-m,-m'} \, .
    \label{eq:dlmpm_dlmmp}
\end{equation}
For the spin-weighted spherical harmonics we use the following convention

\begin{align}
{}_s Y_{lm}(\theta, \phi) &= (-1 )^m \sqrt{\frac{2l + 1}{4 \pi}} e^{-i s \psi}
D^l_{-m, s}(\phi, \theta, - \psi) \nonumber \\ &= (-1 )^m \sqrt{\frac{2l + 1}{4 \pi}}
\rme^{\rmi m \phi} d^l_{-m, s}(\theta) \, .
\label{eq:Ylm_def}
\end{align}

With the $d$-matrix property of Eq.~\eqref{eq:dlmpm_dlmmp}, we can simplify the spin $-2$ spherical harmonics, used for decomposing GW emission, as

\begin{equation}
    {}_{-2} Y_{lm}(\theta, \phi) = \sqrt{\frac{2l + 1}{4 \pi}}
\rme^{\rmi m \phi} d^l_{m, 2}(\theta) \, .
\end{equation}

\section{Extra parameter estimation information and results}
\label{sec:appendix:PE_extra}

As mentioned in Sec.~\ref{sec:validate:PE}, for the PE analyses performed in this paper, we use \texttt{bilby}. In line with the standard \texttt{bilby} conventions, we specify quasi-circular injections via 15 parameters, namely the detector-frame component masses $m_1$ and $m_2$, the dimensionless spin magnitudes $a_1$ and $a_2$, the angle between the component spins and the angular momentum vector $\theta_1$ and $\theta_2$, the azimuthal angle between spin vectors $\phi_{12}$, the azimuthal angle between the total and orbital angular momenta $\phi_{JL}$, the angle between the total angular momentum and the vector from the binary to the observer $\theta_{JN}$, the reference binary phase $\phi_\mathrm{ref} = \lambda_0$, the luminosity distance $d_L$, the right ascension $\mathrm{ra}$, the declination $\mathrm{dec}$, the polarization $\psi$ and the reference geocent time $t^\mathrm{geo} = t_c$. For eccentric injections we need two more parameters, which we choose to be the initial eccentricity $e_0$ and initial mean anomaly $\ell_0$. In table~\ref{tab:injected_params} we show the parameters of the three injections analyzed in Sec.~\ref{sec:validate:PE}, together with the SNR at each detector and the total network SNR.

For sampling, we use \texttt{bilby}'s~\cite{Ashton:2018jfp,Smith:2019ucc,Romero-Shaw:2020owr} \texttt{acceptance-walk} configuration of \texttt{dynesty}, running three parallel chains with $n_\mathrm{live} = 1000$ and $n_\mathrm{accept} = 60$. With these settings, each PE analysis required the number of likelihood evaluations listed in table~\ref{tab:Nlikelihood}. As expected, the number of calls increases with SNR and is higher in the eccentric case, due to the additional two parameters and the more constrained posteriors. Nonetheless, we observe that all these numbers are $\ord{10^8}$, in agreement with what was discussed in Eq.~\eqref{eq:T_PE_estimation}. Consistently with the waveform timings of Fig.~\ref{fig:pyEFPE_comparison_runtime_violins} and Eq.~\eqref{eq:T_PE_estimation}, the PE analyses took $\ord{1~\mathrm{week}}$ to run in 16 cores.

\begin{figure}[t!]
\centering  
\includegraphics[width=\textwidth]{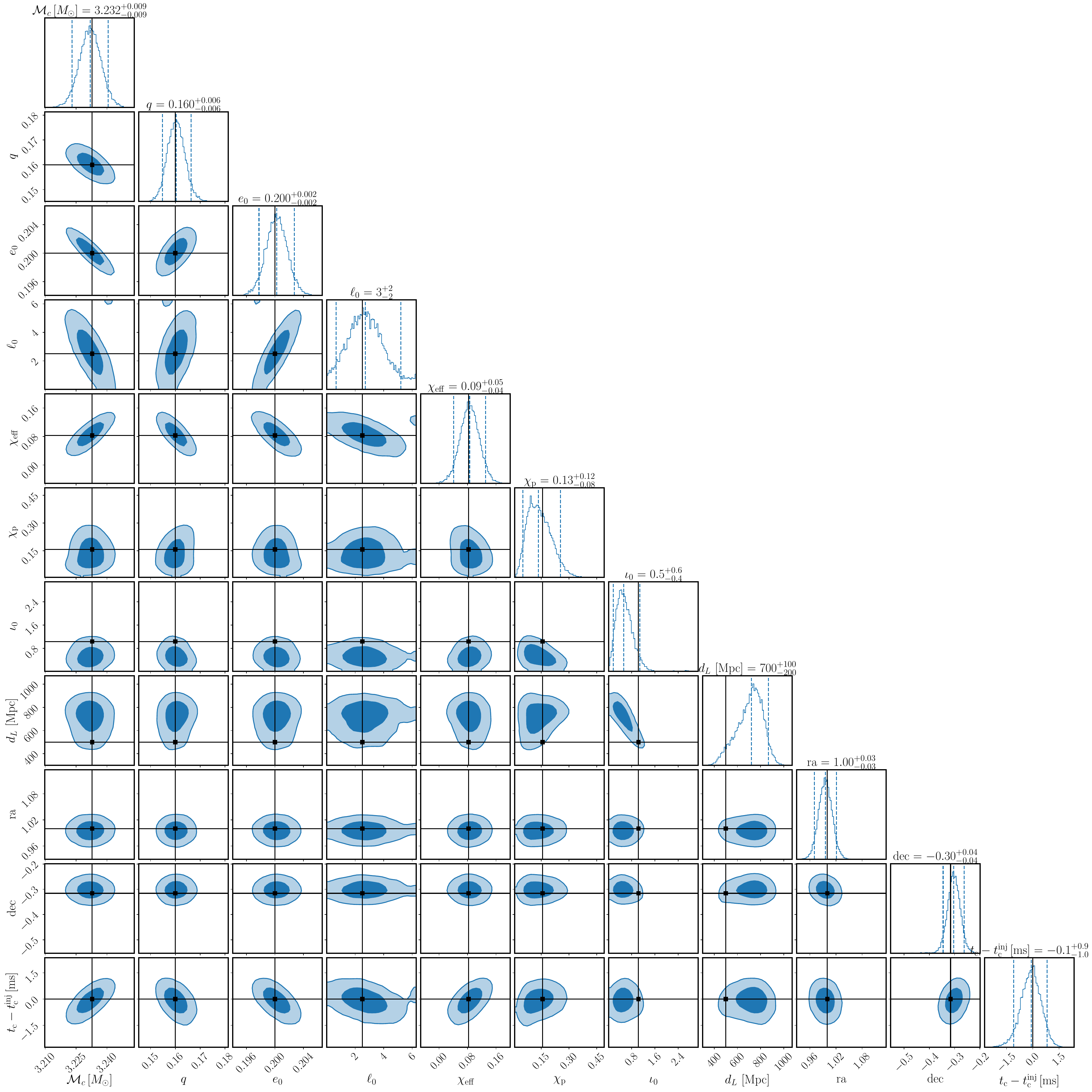}
\caption{\justifying Corner plot showing the joint posterior distributions of the most important parameters of the \pyEFPE injection-recovery study. Specifically, the plot displays the chirp mass $\mathcal{M}_c$, mass ratio $q$, initial eccentricity $e_0$, initial mean anomaly $\ell_0$, effective inspiral spin parameter $\chi_\mathrm{eff}$, effective precession spin parameter $\chi_\mathrm{p}$, initial inclination $\iota_0$, luminosity distance $d_L$, right ascension $\mathrm{ra}$, declination $\mathrm{dec}$ and difference between measured and injected coalescence times $t_c - t_c^\mathrm{inj}$. The diagonal panels display the marginal distributions for each parameter, along with the median and 90\% confidence interval. The off-diagonal panels show the bivariate correlations between pairs of parameters, with the contours representing the $50\%$ and $90\%$ confidence regions. The black lines mark the values of the injected parameters.}
\label{fig:inj_EFPE_rec_EFPE_all_important}
\end{figure}

\begin{figure}[t!]
\centering  
\includegraphics[width=\textwidth]{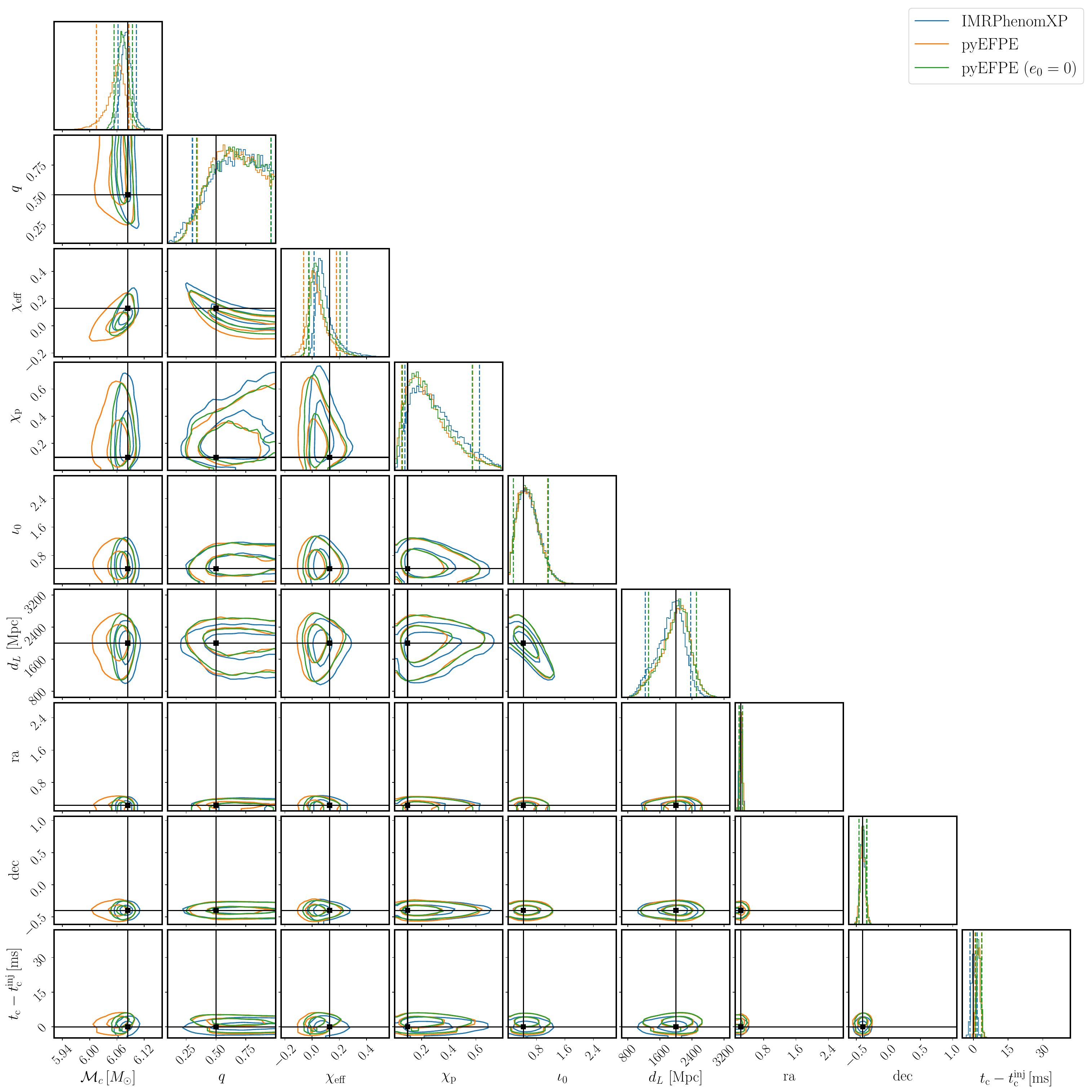}
\caption{\justifying Corner plot showing the joint posterior distributions of the most important parameters of the \XP injection recovered with \XP, \pyEFPE and \pyEFPE with $e_0 = 0$ for the low SNR case. Specifically, the plot displays the chirp mass $\mathcal{M}_c$, mass ratio $q$, effective inspiral spin parameter $\chi_\mathrm{eff}$, effective precession spin parameter $\chi_\mathrm{p}$, initial inclination $\iota_0$, luminosity distance $d_L$, right ascension $\mathrm{ra}$, declination $\mathrm{dec}$ and difference between measured and injected coalescence times $t_c - t_c^\mathrm{inj}$. In each corner plot, the diagonal panels display the marginal distributions for each parameter, along with the 90\% confidence interval. The off-diagonal panels show the bivariate correlations between pairs of parameters, with the contours representing the $50\%$ and $90\%$ confidence regions. The black lines mark the values of the injected parameters.}
\label{fig:inj_XP_16s_lowSNR_all_important}
\end{figure}

\begin{figure}[t!]
\centering  
\includegraphics[width=\textwidth]{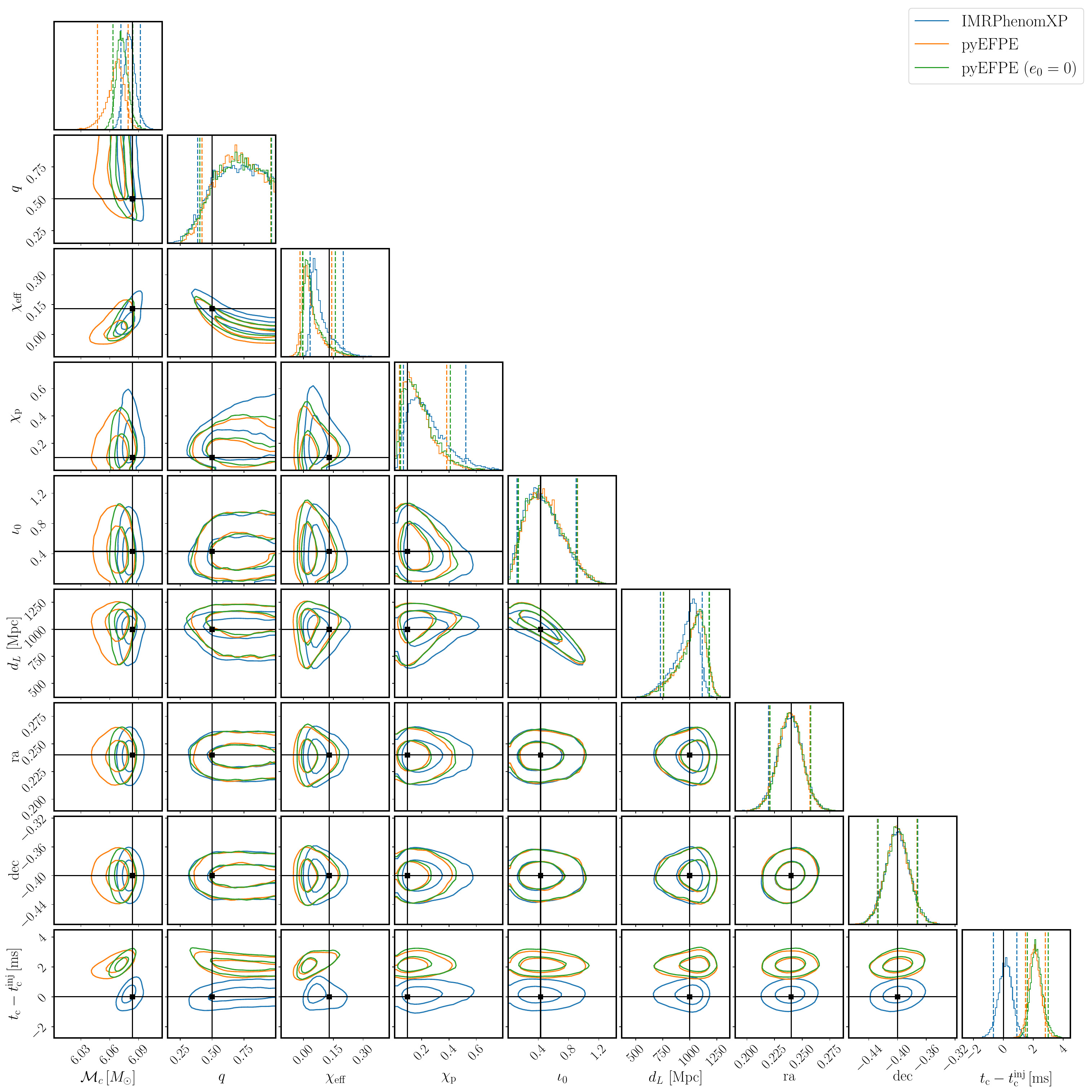}
\caption{\justifying Corner plot showing the joint posterior distributions of the most important parameters of the \XP injection recovered with \XP, \pyEFPE and \pyEFPE with $e_0 = 0$ for the high SNR case. Specifically, the plot displays the chirp mass $\mathcal{M}_c$, mass ratio $q$, effective inspiral spin parameter $\chi_\mathrm{eff}$, effective precession spin parameter $\chi_\mathrm{p}$, initial inclination $\iota_0$, luminosity distance $d_L$, right ascension $\mathrm{ra}$, declination $\mathrm{dec}$ and difference between measured and injected coalescence times $t_c - t_c^\mathrm{inj}$. In each corner plot, the diagonal panels display the marginal distributions for each parameter, along with the 90\% confidence interval. The off-diagonal panels show the bivariate correlations between pairs of parameters, with the contours representing the $50\%$ and $90\%$ confidence regions. The black lines mark the values of the injected parameters.}
\label{fig:inj_XP_16s_highSNR_all_important}
\end{figure}

In Figs.~\ref{fig:inj_EFPE_rec_EFPE_all_important}, \ref{fig:inj_XP_16s_lowSNR_all_important} and \ref{fig:inj_XP_16s_highSNR_all_important} we show the corner plots of the posteriors for all important parameters, recovered by the PEs on the \pyEFPE and Low and High SNR \XP injections, respectively. Note that in both Fig.~\ref{fig:inj_XP_16s_lowSNR_all_important} and Fig.~\ref{fig:inj_XP_16s_highSNR_all_important}, the time posteriors for the \pyEFPE and \XP analyses are very different due to the reference time of the waveforms being different. For \pyEFPE it is the coalescence time described in Eq.~\eqref{eq:RK_final_tshift} while for \XP, it is the peak amplitude time~\cite{Pratten:2020fqn}.

\begin{table}[h]
    \centering
    \begin{tabular}{c|c|c|c|}
         & \pyEFPE & \XP  & \XP  \\
         &  & (Low SNR) & (High SNR) \\
         \hline
         $m_1 \, [M_\odot]$ & 10 & 10 & 10 \\
         $m_2 \, [M_\odot]$ & 1.6 & 5 & 5 \\
         $a_1$ & 0.2 & 0.2 & 0.2 \\
         $a_2$ & 0.3 & 0.1 & 0.1 \\
         $\theta_1 \, [\mathrm{rad}]$ & 0.9 & 0.5 & 0.5 \\
         $\theta_2 \, [\mathrm{rad}]$ & 2.2 & 1.2 & 1.2 \\
         $\phi_{12} \, [\mathrm{rad}]$ & 3.0 & 4.0 & 4.0 \\
         $\phi_{JL} \, [\mathrm{rad}]$ & 3.3 & 5.0 & 5.0 \\
         $\theta_{JN} \, [\mathrm{rad}]$ & 1.0 & 0.4 & 0.4 \\
         $d_L  \, [\mathrm{Mpc}]$ & 500 & 2000 & 1000 \\
         $\phi_\mathrm{ref} \, [\mathrm{rad}]$ & 0.9 & 1.6 & 1.6 \\
         $\mathrm{ra} \, [\mathrm{rad}]$ & 1.0 & 0.24 & 0.24 \\
         $\mathrm{dec} \, [\mathrm{rad}]$ & -0.316 & -0.4 & -0.4 \\
         $\psi \, [\mathrm{rad}]$ & 0.6 & 2.7 & 2.7 \\
         $t_c^\mathrm{GPS}\, [\mathrm{s}]$ & 1262276684 & 1249852257 & 1249852257 \\
         $e_0$ & 0.2 & -- & -- \\
         $\ell_0 \, [\mathrm{rad}]$ & 2.5 & -- & -- \\
         \hline
         H1 SNR & 13.25 & 7.76 & 15.52 \\
         L1 SNR & 13.61 & 9.31 & 18.61 \\
         V1 SNR & 5.06 & 3.91 & 7.82 \\
         Network SNR & 19.7 & 12.7 & 25.5 \\
         \hline
    \end{tabular}
    \caption{Values of the parameters for the three injections analyzed in Sec.~\ref{sec:validate:PE}. We also list the SNR in each detector (H1, L1 and V1), computed with their projected O5 sensitivities~\cite{Abbott_2020,ObservingScenariosPSDs}, also used in the PEs of Sec.~\ref{sec:validate:PE}. Finally, we show the total network SNR, obtained by combining in quadrature the individual detector SNRs.}
    \label{tab:injected_params}
\end{table}

\begin{table}[h]
    \centering
    \begin{tabular}{c|c|c|c|}
         & \pyEFPE & \XP  & \XP  \\
         &  & (Low SNR) & (High SNR) \\
         \hline
         \pyEFPE & $1.87 \cdot 10^8$ & $0.81 \cdot 10^8$ & $1.11 \cdot 10^8$ \\
         \pyEFPE ($e_0 = 0 $) & ------ &  $0.81 \cdot 10^8$ &  $1.09 \cdot 10^8$ \\
         \XP & ------ & $0.81 \cdot 10^8$ & $1.06 \cdot 10^8$ \\
         \hline
    \end{tabular}
    \caption{Number of likelihood evaluations required in each PE analysis of Sec.~\ref{sec:validate:PE}. Columns correspond to different injections, and rows to different recovery cases.}
    \label{tab:Nlikelihood}
\end{table}

\twocolumngrid

\FloatBarrier

\bibliography{Refs}

\end{document}